\documentclass{aastex62}
\usepackage{amsmath}
\usepackage{amsfonts}
\usepackage{amssymb}
\usepackage{bm}
\usepackage{multirow}
\usepackage{graphicx}
\usepackage{booktabs}
\usepackage{hyperref}
\usepackage{cleveref}
\usepackage{pifont}% http://ctan.org/pkg/pifont
\newcommand{\cmark}{\ding{51}}%
\newcommand{\xmark}{\ding{55}}%

\def \apjs {{\rm {ApJS}}}

\def \aap {{\rm {A\&A}}}

\def \apjl{\rm {ApJL}}

\date{\today}
\received{xxx}
\revised{xxx}
\accepted{xxx}
\submitjournal{xx}

\shorttitle{PEXO}
\shortauthors{Feng et al.}

\begin{document}
\title{PEXO: a global modeling framework for nanosecond timing, microsecond astrometry, and $\mu$m/s radial velocities}
\author[0000-0001-6039-0555]{Fabo Feng}
\affiliation{Department of Terrestrial Magnetism, Carnegie Institution of Washington, Washington, DC 20015, USA}
\affiliation{Centre for Astrophysics Research, University of Hertfordshire, College Lane, AL10 9AB, Hatfield, UK}
\author{Maksym Lisogorskyi}
\affiliation{Centre for Astrophysics Research, University of Hertfordshire, College Lane, AL10 9AB, Hatfield, UK}
\author{Hugh R. A. Jones}
\affiliation{Centre for Astrophysics Research, University of Hertfordshire, College Lane, AL10 9AB, Hatfield, UK}
\author{Sergei M. Kopeikin}
\affiliation{Department of Physics \& Astronomy, University of Missouri, Columbia, Missouri 65211, USA}
\author{R. Paul Butler}
\affiliation{Department of Terrestrial Magnetism, Carnegie Institution of Washington, Washington, DC 20015, USA}
\author{Guillem Anglada-Escud\'e}
\affiliation{School of Physics and Astronomy, Queen Mary University of London, 327 Mile End Road, E1 4NS, London, UK}
\author{Alan P. Boss}
\affiliation{Department of Terrestrial Magnetism, Carnegie Institution of Washington, Washington, DC 20015, USA}

\correspondingauthor{Fabo Feng}
\email{ffeng@carnegiescience.edu}
\begin{abstract}
The ability to make independent detections of the signatures of
  exoplanets with complementary telescopes and instruments brings a
  new potential for robust identification of exoplanets and precision
  characterization. We introduce PEXO, a package for Precise
  EXOplanetology to facilitate the efficient modeling of timing,
  astrometry, and radial velocity data, which will benefit not
    only exoplanet science but also various astrophysical studies in general. PEXO is general enough to account for binary motion and stellar reflex
motions induced by planetary companions and is precise enough to treat
various relativistic effects both in the solar system and in the
target system. We also model the post-Newtonian barycentric motion for
future tests of general relativity in extrasolar systems. We benchmark
PEXO with the pulsar timing package TEMPO2 and find that PEXO produces
numerically similar results with timing precision of about 1\,ns,
space-based astrometry to a precision of 1$\mu$as, and radial velocity
of 1\,$\mu$m/s and improves on TEMPO2 for decade-long timing data of
nearby targets, due to its consideration of third-order terms of Roemer
delay. PEXO is able to avoid the bias introduced by decoupling
  the target system and the solar system and to account for the
  atmospheric effects which set a practical limit for ground-based
  radial velocities close to 1 cm/s. Considering the various caveats
  in barycentric correction and ancillary data required to realize
  cm/s modeling, we recommend the preservation of original
  observational data. The PEXO modeling package is available at GitHub (\url{https://github.com/phillippro/pexo}). 
\end{abstract}
%%%%%%%%%%%%%%%%%%%%%%%%%%%%%%%%%%%%%%%%%%%%%%%%%%%%%%%%%%%%%%%%%%%%%%%%%%%%

\section{Introduction} \label{sec:introduction}
The first decades of exoplanet science have enabled detection and some
characterization of exoplanets with a much wider range of properties
than anticipated. In turn, this has prompted a reinvention of the
formation history of the solar system. However, so far we barely have the capability to be sensitive to the planetary systems, like our own solar system, around nearby stars. New high-precision
facilities such as {\it TESS} \citep{ricker14}, {\it Gaia} \citep{brown18},
ESPRESSO \citep{hernandez17}, and the upcoming {\it James Webb Space Telescope } \citep{beichman14} bring us to a golden age of exoplanet science
where a comprehensive survey of nearby planets becomes feasible
and the discoveries of nearby Earth-like planets around Sun-like stars
(or ``Earth twins'') become possible. This in turn leads to the ability
to detect biosignatures and begin habitability studies and to test planet formation theories. The high-precision and overlapping
constraints afforded by these new instruments might also suffice for
tests of relativity theory \citep{jordan08,mignard10} analogous to that achieved for pulsar timing \citep{hulse75,weisberg05,wex14}. For example, short-period binaries on eccentric orbits would show strong variation of gravitational Doppler shift. 

Five primary methods are used to detect exoplanets: radial
velocity, transit, astrometry, microlensing, and direct imaging. We
classify these methods into four categories according to the
dimension of the data used in them. Modern instruments produce timing,
photometric, spectroscopic, and astrometric data. The radial velocity
method uses the timing and spectroscopic data; the transit method uses the timing and photometry data;
the astrometry method uses the timing and astrometry data; the
microlensing method uses the timing and photometry or
spectroscopic data; direct imaging typically uses all four types of
data. Thus, a general model for exoplanet detection would require a
precise modeling of timing, photometry, astrometry, and
spectroscopy. Considering that radial velocity and transits are the main working
methods for exoplanet detection and the astrometry method will probably
be used to find thousands of exoplanets by {\it Gaia} \citep{perryman14}, the
immediate need for general and combined analysis of precision
exoplanet data is a combined model of timing, radial velocity, and astrometry. In
the pulsar timing community, TEMPO2 (\citealt{edwards06}, here ``E06''; \citealt{hobbs06}) is currently the only known package used to test general
relativity (GR) and indirectly detect gravitational waves due to its
unprecedented timing precision at a level of a nanosecond
(ns). However, a similar high-precision package is not available for
independent pulsar timing analysis and for
the search for exoplanets despite various efforts being made to improve
the precision in some special cases \citep{eastman10,wright14}.

In this work, we introduce a new package called ``PEXO'' \footnote{PEXO is an acronym for {\bf P}recise {\bf EXO}planetology.} to model the
timing, radial velocity, and astrometry simultaneously and precisely in order to
detect and characterize small planets such as Earth twins and test GR. PEXO is able to model timing to a precision of about 1\,ns, radial velocity to a precision of 1\,$\mu$m/s, and space-based astrometry
to a precision of 1\,$\mu$as. PEXO models the motion of
  the target star around the target system barycenter (TSB) due to its
  companion (so-called ``reflex motion''), heliocentric motion of the
TSB, and the Earth's motion simultaneously to avoid any bias caused by
decoupling and separating these motions. PEXO can be used for combined
analysis of timing, radial velocity, and astrometry data and to determine the orbital parameters of potential companions, as well as refining the astrometric
parameters and the motion of the observing instrument with respect to the barycenter
of the solar system. PEXO is also able to model the relativistic
effects in the binary motion and thus is able to test GR in systems with multiple stars and companions. 

PEXO is developed in particular to address the following issues
  in previous exoplanet packages and studies:
  \begin{itemize}
    \item The decoupling of remote and local effects or the so-called
      ``barycentric correction'', though efficient for single stars
      hosting planets, is not appropriate for detection of low-mass planets around stars with massive companions. We will discuss this issue in section \ref{sec:decoupling}. 
    \item The exoplanet community might be more focused on exoplanet
      detection and characterization than tests of GR
      although the classical astrometric effects induced by small planets could be comparable to relativistic effects. In section \ref{sec:relativity_test}, we
      address this issue by proposing the companion-induced
      gravitational redshift as a unique way to test gravity
      theories. 
    \item The relativistic effects in extrasolar systems are not well
      modeled, leading to potential bias in transit timing
      variation and radial velocity detection of exoplanets. This
      issue will be addressed in section \ref{sec:ttv}.
    \item The current packages are not able to analyze multiple types
      of exoplanet data in a consistent way, due to a lack of
      simultaneous modeling of timing, radial velocity, photometry, and
      astrometry. We will briefly discuss this problem in section \ref{sec:comparison_table}. 
    \end{itemize}

This paper is structured as follows. In section \ref{sec:geometric},
we introduce the geometric and kinematic model of astrometry and
radial velocity. We then introduce the relativistic effects in timing, astrometry,
and radial velocity in section \ref{sec:relativistic}. We compare PEXO with TEMPO2
and other packages to examine the precision of PEXO in section
\ref{sec:comparison}. This is followed by assessments of the
significance of various relativistic effects on a few key nearby
objects using two example transit systems and $\alpha$ Centauri A and B in section \ref{sec:effects}. Finally we conclude in section \ref{sec:conclusion}.

\section{Geometric and kinematic model of astrometry and radial velocity}\label{sec:geometric}
\begin{figure}
  \centering 
  \includegraphics[scale=0.5]{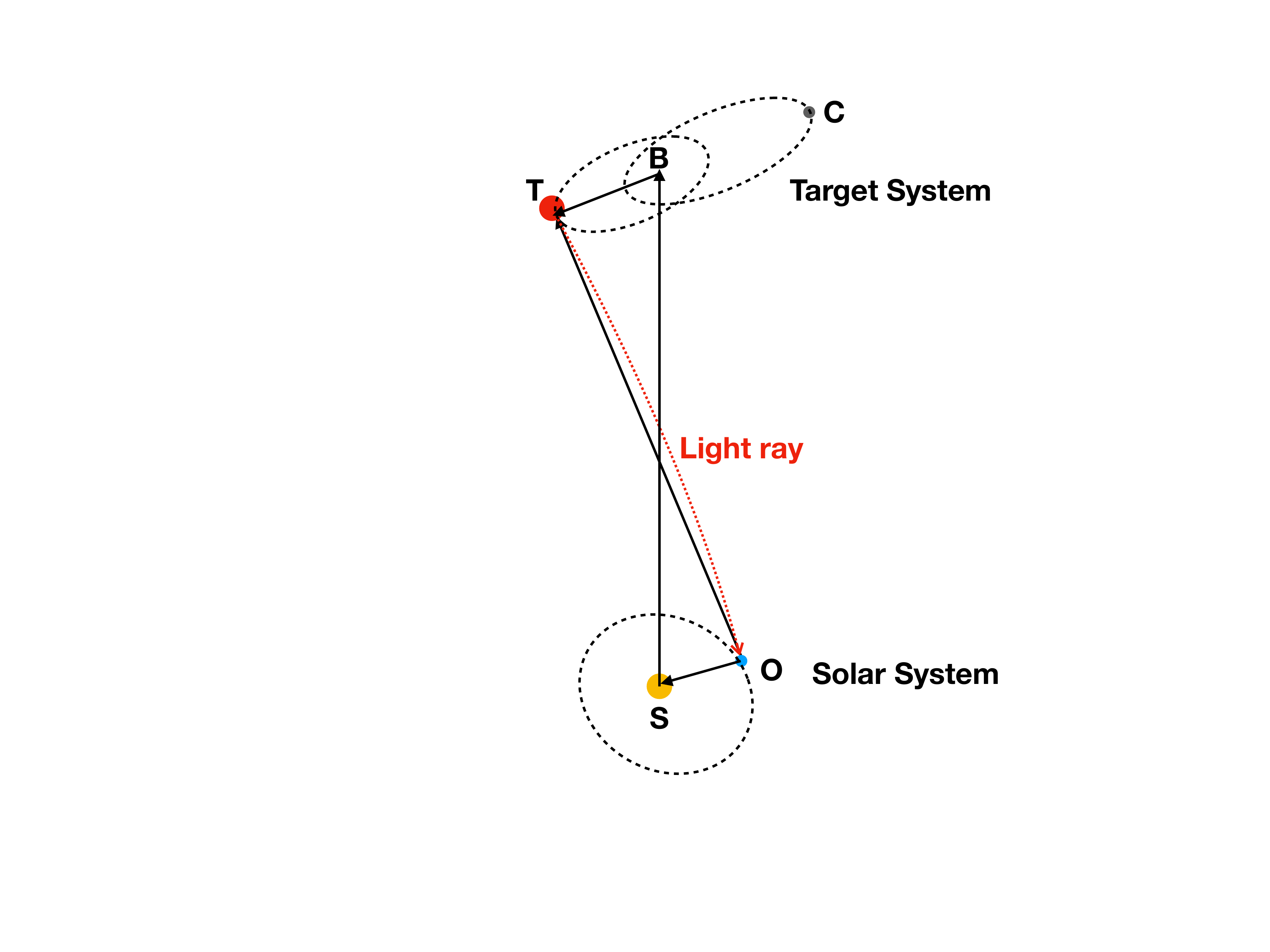}
  \caption{Illustration of the propagation of a photon vastly
    exaggerated in order to represent the underlying geometry of the
    model. A photon emitted from the target star (T) is delayed and
    deflected by other bodies in the target system with barycenter at
    B and deflected and delayed by solar system bodies with barycenter at
    S before arriving at the observation site (O). The companion in
    the target system is denoted by C. The vectors in the diagram are
    related according to ${\bf r}_{\rm OT}={\bf r}_{\rm OB}+{\bf
      r}_{\rm BT}$ and ${\bf r}_{\rm OB}={\bf r}_{\rm OS}+{\bf r}_{\rm
      SB}$. The time derivatives of these vectors are corresponding
    velocities, which are related in the same way. }
  \label{fig:light_ray}
\end{figure}

We follow the {\it Hipparcos} and {\it Gaia} team \citep{esa97,lindegren11} and use
vectors to model astrometry. In this section, we assume that the speed
of light is infinite and ignore the relativistic effects on the light
rays. In other words, we consider the kinematics and geometry of stars
and observers. We show the propagation of the light in Fig. \ref{fig:light_ray}. As we develop the model, the elements are described, though there are many of these, so we also provide a tabulation in the appendix.

\subsection{Astrometry Model}\label{sec:astrometry}
The observed position of the target star is
\begin{equation}
  {\bm r}_{\rm OT}={\bm r}_{\rm OS}+{\bm r}_{\rm SB}+{\bm r}_{\rm BT}~,
  \label{eqn:rOT}
  \end{equation}
where ${\bm r}_{\rm OS}$ is the position of the solar system
barycenter (SSB) with respect to the observer, ${\bm r}_{\rm SB}$ is
the position of the TSB with respect to the SSB, and ${\bm r}_{\rm
  BT}$ is the target star with respect to the TSB. We also define the
opposite of these vectors as ${\bm r}_{\rm TO}= -{\bm r}_{\rm OT}$,
${\bm r}_{\rm SO}= -{\bm r}_{\rm OS}$, ${\bm r}_{\rm BS}= -{\bm
  r}_{\rm SB}$, and ${\bm r}_{\rm TB}= -{\bm r}_{\rm BT}$. In the
following sections, the variable in bold is a vector, while the variable in normal font is a scalar or the mode of the corresponding vector. 

  %  %and the velocity of the target star relative to the TSB as ${\bm v}_{\rm BT}$, 
%  For a single star without perturbations from companions, its position in the Barycentric Celestial Reference System (BCRS; \citealt{rickman01}) time $t$ is 
  Denoting the velocity of the TSB relative to the SSB as ${\bm
    v}_{\rm SB}$ and assuming ${\bm v}_{\rm SB}$ to be constant, that
  is, ${\bm v}_{\rm SB}(t)={\bm v}_{\rm SB}(t_0)$, we find that \ref{eqn:rOT} becomes
    \begin{equation}
      {\bm r}_{\rm OT}(t)={\bm v}_{\rm SB}(t_0)(t-t_0) +{\bm r}_{\rm SB}(t_0)+{\bm r}_{\rm BT}(t) -{\bm r}_{\rm SO}(t)~,
      \label{eqn:rOT2}
    \end{equation}
where $t_0$ is a reference time. Here, ${\bm r}_{\rm BT}$ is determined by
the motion of the target star around the TSB (or reflex motion),
and ${\bm r}_{\rm SO}$ is determined by the ephemeris of the observer. 

\begin{figure}
    \centering
    \includegraphics[scale=0.5]{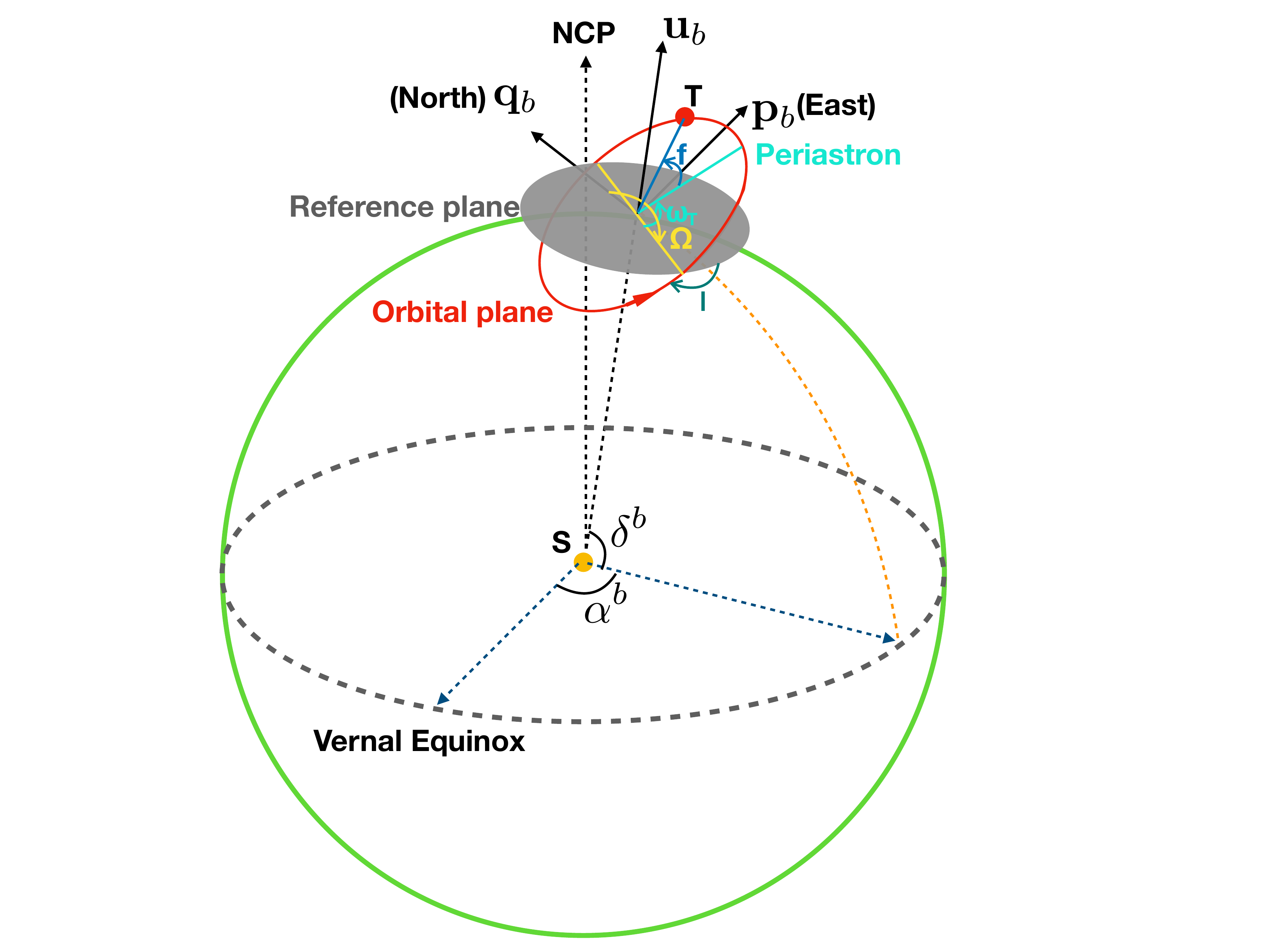}
    \caption{Illustration of the binary motion in the sky plane
      reference frame defined by the triad $[{\bm p}_b~
  {\bm q}_b~ {\bm u}_b]$. This coordinate system is
  fixed at the reference epoch and does not rotate as the TSB moves
  with respect to the Sun. In the orbital-plane coordinate system, the
  x axis ${\bm e}_x$ points to the periastron, the y axis ${\bm e}_y$
  is 90$^\circ$ in the direction of orbital motion from ${\bm
    e}_y$ in the orbital plane, and the z axis ${\bm e}_z$ is
  perpendicular to the orbital plane (parallel to the angular
  momentum). The three axes form a right-handed Cartesian coordinate
  system. Here, $\bm q_b$ points to the north while $\bm p_b$ points to the east from the barycentric observer's perspective. The reflex
  motion of the target star $T$ is determined by the five orbital
  elements: semi-major axis $a_{\rm T}$, eccentricity $e$, inclination $I$
  (the angle between -$\bm u_b$ and the angular momentum),
  longitude of ascending node $\Omega$ (counterclockwise angle
    from the north to the ascending node viewed from the observer),
  and argument of periastron $\omega_{\rm T}$. The true anomaly $f$ is
  needed to determine the location of the target star. For the orbit
  of a companion around the barycenter, the semi-major axis is
  $a_{\rm C}=\frac{m_{\rm T}}{m_{\rm C}}a_{\rm T}$ where $m_{\rm T}$ and $m_{\rm C}$ are respectively the masses of the target and
  the companion. The argument of periastron is $\omega_C=\omega_T+\pi$
  while the other orbital elements are the same as the reflex orbit of
  the target star. We call the convention illustrated by this
    figure ``astrometric convention.'' This and other conventions of
    binary motion and the transformation between the orbital plane
    frame and the sky plane
    frame are explained in detail in Appendix
    \ref{sec:convention}. According to equation \ref{eqn:triad}, the
    sky plane system can be further transformed into the equatorial coordinate system defined by the Vernal Equinox and the north
  Celestial Pole (NCP). }
\label{fig:binary_orbit}
  \end{figure}

Considering that ${\bm v}_{\rm SB}(t_0)$ and ${\bm r}_{\rm SB}(t_0)$ are provided by astrometric observations, we replace them with astrometry, and equation \ref{eqn:rOT2} becomes
  \begin{equation}
    {\bm r}_{\rm OT}(t)= {\bm r}_{\rm SB}(t_0)+\frac{A}{\widetilde\omega^b}({\bm p}_b\mu_\alpha^b+{\bm q}_b\mu_\delta^b+{\bm u}_b\mu_r^b)(t-t_0)+{\bm r}_{\rm BT}(t) -{\bm r}_{\rm SO}(t)~,
    \label{eqn:rOT3}
\end{equation}
where the relevant notations are defined as follows: ${\bm u}_b$ is the Barycentric Celestial Reference System (BCRS; \citealt{rickman01}) coordinate direction to the TSB at the reference TCB epoch $t_0$; 
 $\alpha^b$ is the BCRS R.A. of the TSB at the reference epoch;
 $\delta^b$ is the BCRS decl. the TSB at the reference epoch;
 $\widetilde\omega^b$ is the annual parallax of the TSB at the reference epoch; 
 $\mu_\alpha^b$ is the proper motion in R.A. of the TSB at the reference epoch; 
 $\mu_\delta^b$ is the proper motion in decl. of the TSB at the reference epoch;
 ${\bm \mu}={\bm p}_b\mu_\alpha^b+{\bm q}_b\mu_\delta^b$ is the total proper motion of the TSB at the reference epoch;  
 $\mu_r^b=v_r^b \widetilde\omega^b/A$ is the so-called ``radial proper
 motion'' of the TSB at the reference epoch, where $v_r$ is the radial velocity of the TSB and $A$ is the astronomical unit;
the unit vectors of ${\bm p}_b$, ${\bm q}_b$, and ${\bm u}_b$ form a triad, which is
\begin{equation}
  \begin{bmatrix}
    {\bm p}_b& {\bm q}_b& {\bm u}_b
  \end{bmatrix}
  =
      \begin{bmatrix}
                -\sin\alpha^b & -\sin\delta^b\cos\alpha^b & \cos\delta^b\cos\alpha^b \\
                \cos\alpha^b &-\sin\delta^b\sin\alpha^b  & \cos\delta^b\sin\alpha^b \\
                0 & \cos\delta^b & \sin\delta^b
                      \end{bmatrix}
      ~.
          \label{eqn:triad}
        \end{equation}
Here, ${\bm p}_b$ and ${\bm q}_b$ are respectively the unit vectors in the
directions of increasing $\alpha$ and $\delta$ at the reference
epoch. The coordinate system determined by the triad $[{\bm p}_b~
  {\bm q}_b~ {\bm u}_b]$ is illustrated in
  Fig. \ref{fig:binary_orbit}.

Because astrometric observations measure the direction of a star on the sky, we estimate the observed direction of a star $\hat{\bm u}_o$ from equation \ref{eqn:rOT3}, following \cite{esa97} and \cite{lindegren11}, 
  \begin{equation}
    {\bm u}_{\rm OT}(t)=\langle{\bm u}_b+({\bm p}_b\mu_\alpha^b+{\bm q}_b\mu_\delta^b+{\bm u}_b\mu_r^b)(t-t_0)+\left[{\bm r}_{\rm BT}(t) -{\bm r}_{\rm SO}(t)\right]\widetilde\omega^b/A\rangle~, 
    \label{eqn:uOT}
  \end{equation}
  where the angular brackets denote vector normalization. Similarly, the BCRS direction of the TSB is 
  \begin{equation}
    {\bm u}_{\rm SB}(t)=\langle{\bm u}_b+({\bm p}_b\mu_\alpha^b+{\bm q}_b\mu_\delta^b+{\bm u}_b\mu_r^b)(t-t_0)\rangle~, 
    \label{eqn:uSB}
  \end{equation}
  and the BCRS direction of the target star is
  \begin{equation}
    {\bm u}_{\rm ST}(t)=\langle{\bm u}_b+({\bm p}_b\mu_\alpha^b+{\bm q}_b\mu_\delta^b+{\bm u}_b\mu_r^b)(t-t_0)+{\bm r}_{\rm BT}(t) \widetilde\omega^b/A\rangle~, 
    \label{eqn:uST}
  \end{equation}
The only difference between equation \ref{eqn:uOT} in this paper and the one in equation 4 of \cite{lindegren11} is the inclusion of stellar reflex motion. 
%In the above astrometry model, we do not account for the
%light-propagation time from the source to the solar system, stellar
%aberration, and gravitational deflection of the stellar light by the
%solar system. These effects only contribute at most to a sub-mas
%absolute astrometry change \citep{lindegren11} and thus the contribution of these relativistic effects to the change in relative astrometry is sub-$\mu$as and so negligible.

Although a robust model of astrometry is typically expressed with
vectors, one is typically interested in the variation of the sky
position of a star rather than its absolute position. To model the
astrometry relative to a reference epoch, we follow \cite{esa97} to
project the position of a component relative to the TSB onto the
offset coordinates ($\xi,\eta$), which are defined as rectangular
coordinates in the tangent plane at the reference point ${\bm r}_{\rm
  SB}(t_0)$, with $\xi$ and $\eta$ increasing in the directions of
${\bm p}_b$ and ${\bm q}_b$. The offset of the target with respect to
the TSB in the topocentric reference frame is
\begin{align}
  \sin{\xi(t)}&=\frac{\mu_\alpha^b(t-t_0)+{\bm p}_b\cdot \left[{\bm r}_{\rm BT}(t) -{\bm r}_{\rm SO}(t)\right]\widetilde\omega^b/A}{1+\mu_r^b(t-t_0)+{\bm u}_b\cdot \left[{\bm r}_{\rm BT}(t) -{\bm r}_{\rm SO}(t)\right]\widetilde\omega^b/A}~,\nonumber\\
  \sin\eta(t)&=\frac{\mu_\delta^b(t-t_0)+{\bm q}_b\cdot \left[{\bm r}_{\rm BT}(t) -{\bm r}_{\rm SO}(t)\right]\widetilde\omega^b/A}{1+\mu_r^b(t-t_0)+{\bm u}_b\cdot \left[{\bm r}_{\rm BT}(t) -{\bm r}_{\rm SO}(t)\right]\widetilde\omega^b/A}~.
           \label{eqn:offset}
  \end{align}
In the above equation, $\sin\xi(t)$ and $\sin\eta(t)$ differ from
$\xi(t)$ and $\eta(t)$ by about 0.2\,mas over 100 yr for the
case of $\alpha$ Centauri. The above formula differs from Equation
1.2.26 of \cite{esa97} in terms of the consideration of the stellar
reflex motion and the use of the sinusoidal function for offset
coordinates. Although the companion-induced offset is small, the
integration of this offset over time would strongly bias the predicted
position of a star. While equation \ref{eqn:offset} can
  provide high-precision geometric modeling in offset coordinates, we
need to model absolute astrometry to account for relativistic
effects. Thus we model the geometric astrometry in the equatorial
coordinate system using equation \ref{eqn:uOT}. We then consider
relativistic effects (section \ref{sec:relativistic}) to form the full
astrometry model. This model is formulated in the offset coordinates
according to equation \ref{eqn:offset} if the astrometric data are
given in the offset coordinate system. 

To compare equation \ref{eqn:offset} with the astrometry model in
previous studies, we expand the offsets to second-order in a Taylor
series. Because all terms in the equations are small quantities compared with 1, the second-order Taylor expansion is
\begin{align}
  \xi(t)&=\mu_\alpha^b(t-t_0)+{\bm p}_b\cdot {\bm R}-\mu_\alpha^b\mu_r^b(t-t_0)^2-[\mu_\alpha^b{\bm u}_b \cdot {\bm R}+\mu_r^b{\bm p}_b\cdot  {\bm R}](t-t_0)-({\bm p}_b \cdot {\bm R})({\bm u}_b \cdot {\bm R})+\mathcal{O}(\xi^3)~,\nonumber\\
\eta(t)&=\mu_\delta^b(t-t_0)+{\bm q}_b\cdot {\bm R}-\mu_\delta^b\mu_r^b(t-t_0)^2-[\mu_\delta^b{\bm u}_b \cdot {\bm R}+\mu_r^b{\bm q}_b\cdot  {\bm R}](t-t_0)-({\bm q}_b \cdot {\bm R})({\bm u}_b \cdot {\bm R})+\mathcal{O}(\eta^3)~,
           \label{eqn:2nd}
  \end{align}
  where ${\bm R}\equiv\left[{\bm r}_{\rm BT}(t) -{\bm r}_{\rm SO}(t)\right]\widetilde\omega^b/A$. We explain the terms in the above equations as follows:\\
  \begin{itemize}
  \item $\mu_\alpha^b(t-t_0)$ and $\mu_\delta^b(t-t_0)$ are linear
    displacements from the TSB due to proper motions and so are first-order terms.
  \item ${\bm p}_b\cdot {\bm R}$ and ${\bm q}_b\cdot {\bm R}$ are the
    parallax of a star if there is no companion around the star. If a star hosts companions, these terms reflect the combined effect of the motion of the observer and the reflex stellar motion on the displacement of the star with respect to the TSB. This is a first-order term. 
  \item $\mu_\alpha^b\mu_r^b(t-t_0)^2$ and
    $\mu_\delta^b\mu_r^b(t-t_0)^2$ are second-order terms related to
    the so-called ``perspective acceleration.'' Because this effect is
    proportional to the square of time, it becomes significant for
    long-baseline astrometry. For example, $\mu_\alpha$ will change by
    about 6\,mas/yr due to the perspective acceleration over 10 years
    of observations of $\alpha$ Centauri. 
    \item $[\mu_\alpha^b{\bm u}_b \cdot {\bm R}+\mu_r^b{\bm p}_b\cdot
      {\bm R}](t-t_0)$ and $[\mu_\delta^b{\bm u}_b \cdot {\bm
        R}+\mu_r^b{\bm q}_b\cdot  {\bm R}](t-t_0)$ are second-order
      terms and are linearly proportional to time. These terms are
      related to the coupling of the proper motion or radial velocity
      with the motion of the Earth and stellar reflex motion. Because they are linear functions of time, they only become important for
   the interpretation of observations taken over decades. For example,
   this term will contribute to 0.1\,mas offset over a decade of
   observation of $\alpha$ Centauri. 
 \item $({\bm p}_b \cdot {\bm R})({\bm u}_b \cdot {\bm R})$ and
      $({\bm q}_b \cdot {\bm R})({\bm u}_b \cdot {\bm R})$ are terms
      related to the coupling of Earth's motion and stellar reflex motion in different directions. This term does not significantly increase
      with time because the orbits of the observer and the stellar
      reflex motion are periodic and the corresponding semi-major axis
      does not change much over time. Thus this term only contributes
      to a $\mu$as displacement even for nearby stars such as $\alpha$ Centauri. 
\end{itemize}
Although equation \ref{eqn:2nd} only expands the offset to the second-order, there are two third-order terms that become important for decades-long astrometry observations: 
\begin{itemize}
\item $\xi(t)^3/6=\xi(t)-\sin\xi(t)$ and
  $\eta(t)^3/6=\eta(t)-\sin\eta(t)$ are the third-order terms for a
  Taylor expansion of a sinusoidal function in the vicinity of zero. This
  term can introduce a sub-mas offset for high proper motion stars. For
  example, this term is about 0.3\,mas for $\alpha$ Centauri for a
  100 yr observational baseline. It would be about 8.8\,mas for a
  century of observations of Barnard's star. 
\item $\mu_\alpha\mu_r^2(t-t_0)^3$ and $\mu_\delta\mu_r^2(t-t_0)^3$
  are related to the coupling of proper motion and radial motion of
  the TSB. The sums of these two terms are respectively about 1\,mas
  and 40\,mas for 100 yr observations of $\alpha$ Centauri and Barnard's star. 
\end{itemize}
Models that only account for the first-order terms are not
reliable for the detection or characterization of planets which induce
sub-mas reflex motion. For example, the maximum reflex offset of a
Sun-like star at 10\,pc is 0.50\,mas for a Jupiter-like planet,
0.27\,mas for a Saturn-like planet, 0.08\,mas for Uranus, and 0.16\,mas
for Neptune. Without including these higher-order terms, it would be
impossible to robustly detect them even if data from all the
individual {\it Gaia} epochs were to be available.

Reference stars are typically difficult to obtain due to the large
relative brightness of stars such as $\alpha$ Centauri A and B. Thus, relative astrometry is more reliable than absolute astrometry in terms of constraining the orbit of $\alpha$ Centauri \citep{pourbaix16, kervella16}. 
By removing the third-order terms from equation \ref{eqn:2nd}, we derive the relative offset of the secondary with respect to the primary, 
\begin{align}
  \Delta\xi(t)&={\bm p}_b\cdot \Delta{\bm R}(t)-[\mu_\alpha^b{\bm u}_b \cdot \Delta{\bm R}(t)+\mu_r^b{\bm p}_b\cdot  \Delta{\bm R}(t)](t-t_0)~,\nonumber\\
\Delta\eta(t)&={\bm q}_b\cdot \Delta{\bm R}(t)-[\mu_\alpha^b{\bm u}_b \cdot \Delta{\bm R}(t)+\mu_r^b{\bm q}_b\cdot  \Delta{\bm R}(t)](t-t_0)~,
           \label{eqn:relative}
  \end{align}
  where $\Delta{\bm R}(t)\equiv {\bm R_2}(t)-{\bm R_1}(t)=\Delta{\bm
    r}(t)\widetilde\omega^b/A$, and $\Delta{\bm r}\equiv{\bm r}_{\rm
    BT2}(t) -{\bm r}_{\rm BT1}(t)$ denotes the Keplerian motion of the
  secondary with respect to the primary. It is notable that the
  relative astrometry depends not only on the reflex motion but also
  on the astrometry and radial velocity of the TSB when considering
  secondary effects. This linear-time secondary effect could
  contribute to sub-mas offsets that are comparable with the signal
  caused by Jupiter-like planets around nearby stars. 
  %Since relative astrometry is typically measured in the reference frame centered on the primary, the coordinate system centered on the TSB and determined by the triad $[{\bm p}_b~ {\bm q}_b~ {\bm u}_b]$ should be transformed into the primary-centered system at the reference epoch. 

  \subsection{Radial Velocity Model}\label{sec:rv_model}
  The time derivative of ${\bm r}_{\rm OT}$ (equation \ref{eqn:rOT}) determines the observed velocity of a star:
  \begin{equation}
    {\bm v}_{\rm OT}={\bm v}_{\rm OS}+{\bm v}_{\rm SB}+{\bm v}_{\rm BT}~.
    \label{eqn:vOT}
    \end{equation}
The observed radial velocity is the projection of $ {\bm v}_{\rm OT}$ onto the observed direction of the star:
      \begin{equation}
        v_r(t)={\bm v}_{\rm OT}(t)\cdot {\bm u}_{\rm OT}(t)
        \label{eqn:vr_kepler}
  \end{equation}
The terms ${\bm v}_{\rm OT}(t)$ and ${\bm u}_{\rm OT}$ can be respectively
  calculated from Eqns. \ref{eqn:uOT} and \ref{eqn:vOT} given the
  astrometry and radial velocity of a star and its reflex motion as
  well as the Jet Propulsion Laboratory (JPL) ephemeris such as DE430 \citep{folker14} and the rotation of the Earth. Thus, the above vectorized formula is the most robust nonrelativistic modeling of radial velocity. However, to compare with other radial velocity models, we need to approximate this model through Taylor expansions. We expand ${\bm u}_{\rm OT}(t)$ to the first-order Taylor series
\begin{align}
 {\bm u}_{\rm OT}(t)&={\bm u}_b+\Delta {\bm u_t}(t)+\mathcal{O}(|\Delta {\bm u}_t(t)|^2)~,
  \label{eqn:uobs2}
\end{align}
where 
\begin{equation}
 \Delta {\bm u_t}(t)= ({\bm p}_b\mu_\alpha^b+{\bm q}_b\mu_\delta^b)(t-t_0)+{\bm R_t}(t)
\end{equation}
is the tangential component of the change of ${\bm u}_{\rm OT}$, and ${\bm R_t}(t)$ is the tangential component of ${\bm R}(t)$. Then the radial velocity becomes
  \begin{equation}
 v_r(t)= {\bm v}_{\rm OT}(t)\cdot{\bm u}_b+ {\bm v}_{\rm OT}(t)\cdot\Delta {\bm u_t}(t). 
  \end{equation}
  In the above equation, the first term is the classical radial velocity model,
  which does not account for the influence of the perspective
  variation on the radial velocity. This perspective change is approximated by the second term, which is related to the tangential reflex motion and the tangential motion of the observer perpendicular to ${\bm u}_b$. Because the radial velocities are typically measured with respect to a reference epoch $t_0$, we derive the variation of radial velocity, which is
    \begin{equation}
      \Delta v_r(t)={\bm v}_{\rm OT}(t) {\bm u}_{\rm OT}(t)-{\bm
        v}_{\rm OT}(t_0) {\bm u}_{\rm OT}(t_0)
      \label{eqn:dRV}
    \end{equation}
The above geometric model of radial velocity does not account for relativistic
effects, which will be discussed in section \ref{sec:relativistic_rv}. 

\subsection{Stellar Reflex Motion}\label{sec:reflex}
We calculate the stellar reflex motion in the coordinate system formed
by the triad $[{\bm p}_b~ {\bm q}_b~ {\bm u}_b]$. Because this
coordinate system is determined at the reference epoch, the orbital
parameters are also calculated with respect to the reference
epoch. This is different from the rotation reference frame, where the
orbital elements may vary with time due to the adoption of a
noninertial frame \citep{kopeikin96}.

In the orbital-plane coordinate system shown in
Fig. \ref{fig:binary_orbit}, the position of the target star is
denoted by ${\bm r}_{\rm orb}=(x,y,0)^{\rm T}$. The orbital-plane
coordinate system is transformed into the sky plane coordinate system through a sequence of rotations such that
  the position of the target star with respect to the TSB is 

  \begin{equation}
    \begin{bmatrix} 
 x_{\rm BT} \\
 y_{\rm BT}\\
 z_{\rm BT}
\end{bmatrix}
=
\begin{bmatrix}
  B'& G'&\cos\Omega\sin I\\
  A'& F'&-\sin\Omega\sin I\\
  C'& H' &-\cos I
\end{bmatrix}
    \begin{bmatrix} 
 x \\
 y\\
0
\end{bmatrix} ~,
\end{equation}
where
\begin{align*}
  A'&=\cos\Omega \cos\omega_{\rm T} - \sin\Omega \sin\omega_{\rm T} \cos{I}\\
  B'&=\sin\Omega \cos\omega_{\rm T} + \cos\Omega \sin\omega_{\rm T} \cos{I}\\
  F'&=-\cos\Omega \sin\omega_{\rm T}-\sin\Omega \cos\omega_{\rm T} \cos{I}\\
  G'&=-\sin\Omega \sin\omega_{\rm T} + \cos\Omega \cos\omega_{\rm T} \cos{I}\\
  C'&=\sin\omega_{\rm T} \sin{I}\\
  H'&=\cos\omega_{\rm T} \sin{I}~,
\end{align*}
are the elements of the rotation matrix and a scaled
version of the so-called ``Thiele Innes constants''. In the above
equations, $\Omega$ is the longitude of ascending node, $\omega_T$ is
the argument of periastron for the orbit of the target star around the
TSB, and $I$ is the inclination. The argument of periastron for the
barycentric motion of the companion is $\omega_C=\omega_T+\pi$. Since
this convention of binary motion is consistent with the triad used for
the astrometry model, we call it the ``astrometric convention'' which is
described in details in Appendix \ref{sec:conv3}. 

The Keplerian motion of the target star with respect to the TSB is
\begin{align}
  {\bm r}_{\rm BT}(t)&=[B'x(t)+G'y(t)]{\bm p}_b + [A'x(t)+F'y(t)]{\bm q}_b + [C'x(t)+H'y(t)]{\bm u}_b~,\nonumber\\
  {\bm v}_{\rm BT}(t)&=[B'\dot{x}(t)+G'\dot{y}(t)]{\bm p}_b + [A'\dot{x}(t)+F'\dot{y}(t)]{\bm q}_b + [C'\dot{x}(t)+H'\dot{y}(t)]{\bm u}_b~,
\label{eqn:rvBT}
\end{align}
and the Keplerian motion in the orbital-plane (see Fig. \ref{fig:binary_orbit}) is
\begin{eqnarray}
  x(t)&=&a_{\rm T}\left[\cos E(t) -e\right] \nonumber\\
  y(t)&=&a_{\rm T}\left[\sqrt{1-e^2}\sin E(t)\right] \nonumber\\
  \dot{x}(t)&=&-a_{\rm T}\frac{n\sin E(t)}{1-e\cos E(t)}\\
  \dot{y}(t)&=&a_{\rm T}\sqrt{1-e^2}\frac{n\cos E(t)}{1-e\cos E(t)},\nonumber
  \label{eqn:orbital_plane}
\end{eqnarray}
where $a_{\rm T}=\frac{m_{\rm C}}{m_{\rm C}+m_{\rm T}}a$ is the semi-major axis of the
target star with respect to the TSB, $a$ is the semi-major
axis of the target star with respect to its companion, $E(t)$ is the
eccentric anomaly which is determined by solving the Kepler's equation
$M(t)=E(t)-e\sin E(t)$ for a given time $t$, $e$ is the
eccentricity; $n\equiv 2\pi/P$ is the mean orbital motion; and $P$ is
the orbital period. By transforming the Keplerian motion from the
orbital-plane reference frame into the sky plane reference frame using equations \ref{eqn:rvBT}, we derive ${\bm r}_{\rm BT}$ and ${\bm v}_{\rm BT}$ to model astrometry and radial velocity fully using equations \ref{eqn:uOT} and \ref{eqn:vr_kepler}. 
  
\section{Relativistic and high-order geometric effects}\label{sec:relativistic}
Relativistic effects can be important for nearby or binary systems
such as $\alpha$ Centauri. For example, assuming an orbital period of
80\,yr, a semi-major axis of 17.6\,au, and an inclination of 79$^\circ$
for $\alpha$ Centauri A and B, the arrival time of light from A or B is
changed by 6.3\,hr due to binary motion over 40\,yr. Although this is small compared with the
binary orbital period, it would introduce timing noise and thus bias the detection of
potential exoplanets in this system.

For the convenience of calculation of relativistic terms, we
define various times following E06. The proper emission time $\tau_e$ of a photon is derived from the proper observed arrival time $\tau_o$ by including various delays as follows:
\begin{equation}
  \tau_e=\tau_o-\Delta_{\rm S}-\Delta_{\rm is}-\Delta_{\rm T}
  \label{eqn:te}
\end{equation}
where $\Delta_{\rm S}$ is the time delay due to effects in the solar
system, $\Delta_{\rm is}$ is related to the travel time of a photon in
the interstellar medium, and $\Delta_{\rm T}$ is the delay related to the target system. We introduce the coordinate time of light arrival at the SSB,
$t_{a}^{\rm SSB}=\tau_o-\Delta_{\rm S}$, and the coordinate time of
light arrival at the TSB, $t_{a}^{\rm TSB}=t_a^{\rm SSB}-\Delta_{\rm
  is}$. The proper emission time and the arrival time at TSB are related
by $\tau_e=t_a^{\rm TSB}-\Delta_{\rm T}$. We also define the
coordinate reference time $t_{\rm pos}$ as the epoch when the position
or astrometry of the target star is measured. At the reference epoch,
$t_{a}^{\rm SSB}=t_{a}^{\rm TSB}=t_{\rm pos}$. We will introduce post-Newtonian and GR models of binary motion in section \ref{sec:PN}, and
describe various relativistic terms in the models of timing (section
\ref{sec:timing}), astrometry (section \ref{sec:relativistic_astrometry}) and radial velocity (section \ref{sec:relativistic_rv}). 

\subsection{Post-Newtonian Stellar Reflex Motion}\label{sec:PN}
In section \ref{sec:reflex}, we derive the formula for the classical
Keplerian motion in the reference frame formed by the triad $[{\bm
  p}_b~ {\bm q}_b~ {\bm u}_b]$. Here we model the post-Newtonian
Keplerian (PPK) motion in terms of proper emission time $\tau_e$ according to previous works (\citealt{damour86}, \citealt{taylor89}, and E06):
\begin{equation}
 {\bm r}_{\rm BT}=[{\bm p}_b~ {\bm q}_b~ {\bm u}_b] 
\begin{bmatrix}
  \sin\Omega & -\cos\Omega & 0 \\
  \cos\Omega & \sin\Omega & 0 \\
  0 & 0 & 1 
\end{bmatrix}
\begin{bmatrix}
 1 & 0 & 0 \\
  0 & -\cos I& -\sin I \\
  0 & \sin I & -\cos I 
\end{bmatrix}
               \begin{bmatrix}
                 r_{\rm reflex} \cos \theta\\
                 r_{\rm reflex} \sin \theta\\
                 0
               \end{bmatrix}~,
               \label{eqn:rBT}
\end{equation}
where 
\begin{align}
  r_{\rm reflex}&=a_r(1-e_r\cos U)~,\nonumber\\
  M&=n(\tau_e-\tau_p)=U-e_\theta\sin U~,\nonumber\\
  \theta&=\omega+A_{e_\theta}(U)~,\nonumber\\
  A_{e_\theta}(U)&=2\tan^{-1}\left[\left(\frac{1+e_\theta}{1-e_\theta}\right)^2\tan{\frac{U}{2}}\right]~,\nonumber\\
  e_r&=e(1+\delta_r)~,\nonumber\\
  e_\theta&=e(1+\delta_\theta)~,\\
  n&=\frac{2\pi}{P_0}+\frac{\pi \dot{P}(\tau_e-\tau_p)}{P_0^2}~,\nonumber\\
  \omega &=\omega_0+kA_e(U)~,\nonumber\\
  k&=\frac{\dot{\omega}}{n}~,\nonumber\\
  e&=e_0+\dot{e}(\tau_e-\tau_p)~,\nonumber\\
  x_a&=x_{a0}+\dot{x_a}(\tau_e-\tau_p)\nonumber~,
  \label{eqn:x}
\end{align}
where $\omega_0$, $e_0$, $P_0$ are the Keplerian parameters at the reference
epoch, $\tau_p$ is the proper time of periastron, $U$ is the relativistic
eccentric anomaly, $a_r$ is the semi-major axis of the primary with
respect to the barycenter of the target system, $\delta_r$ and
$\delta_\theta$ are PPK terms used to define
eccentricities, and $x_a\equiv a\sin{I}/c$ is the light travel time across the
projected semi-major axis. Because this model was first proposed by
\cite{damour86}, we call it ``DD'', following \cite{taylor89} and
E06. Because the x axis of the orbital-plane is in the direction of
  the ascending node rather than periastron as in the astrometric
  convention, we call this coordinate system framework
 the ``precession-compatible convention,'' which is described in detail
  in Appendix \ref{sec:conv3}.

Considering GR, we define $m_{\rm tot}=m_{\rm C}+m_{\rm T}$ and give the following relativistic terms after E06's eqs. 71 and 80-88, 
\begin{align}
  \dot{\omega}^{\rm GR}&=3T_\odot^{2/3}n^{5/3}\frac{m^{2/3}_{\rm tot}}{1-e^2}~,\nonumber\\
  g^{\rm GR}&=T_\odot^{2/3}n^{-1/3}e\frac{m_{\rm C}(m_{\rm T}+2m_{\rm C})}{m_{\rm tot}^{4/3}}~,\nonumber\\
  r_s^{\rm GR}&=T_\odot m_{\rm C}~,\nonumber\\
  s_s^{\rm GR}&=\sin I=T_\odot^{-1/3}n^{2/3}x_a\frac{m_{\rm tot}^{2/3}}{m_{\rm C}}~,\\
  \delta_r^{\rm GR}&=T_\odot^{2/3}n^{2/3}\frac{3m_{\rm T}^2+6m_{\rm T}m_{\rm C}+2m_{\rm C}^2}{m_{\rm tot}^{4/3}}~,\nonumber\\
 \delta_\theta^{\rm GR}&=T_\odot^{2/3}n^{2/3}\frac{\frac{7}{2}m_{\rm T}^2+6m_{\rm T}m_{\rm C}+2m_{\rm C}^2}{m_{\rm tot}^{4/3}}~,\nonumber\\
 \dot{P}^{\rm GR} &=-\frac{192\pi}{5} T_\odot^{5/3} n^{5/3}\frac{m_{\rm T}m_{\rm C}}{m_{\rm tot}^{1/3}}\frac{1+73e^2/24 + 37e^4/96}{(1-e^2)^{7/2}}~,\nonumber
\label{eqn:grPar}
\end{align}
where ``GR'' denotes GR, $c$ is the speed of light,
$T_\odot=Gm_\odot/c^3$ is half the light travel time across the solar
Schwarzschild radius, $m_\odot$ is the Solar mass, $G$ is the
gravitational constant, $g$ is the timing model parameter, $r_s^{\rm GR}$
and $s_s^{\rm GR}$ are parameters for the Shapiro delay in the target
system assuming GR. We call this model ``DDGR'' following the syntax of TEMPO2. 

In a combined model of radial velocity and astrometry, the free classical orbital
parameters are $\{ P_0 ,e_0, \omega_0, I, \Omega, \tau_p\}$ as well as
masses $\{m_{\rm C}, m_{\rm T}\}$. For general post-Newtonian theories, the
additional fittable parameters are $\{\dot{\omega}, g, \dot{P}, s_s,
r_s, \dot{x}_a, \dot{e}\}$, where $s_s$ and $r_s$ are respectively the
shape and range parameters of Shapiro delay. For a classical Keplerian
orbit, $\dot{\omega},\dot{P},\dot{x}_a, \dot{e}$ are all zero.

\subsection{Timing model}\label{sec:timing}
We transform the proper light arrival time at the observatory
$\tau_o$ to the
barycentric coordinate time ($t_a^{\rm SSB}$) by calculating the
``tropospheric delay'' ($\Delta_{\rm tropo}$), ``Roemer delay'' ($\Delta_{\rm rS}$), ``Shapiro delay'' ($\Delta_{\rm sS}$), and ``Einstein delay'' ($\Delta_{\rm eS}$). Then we transform
$t_a^{\rm SSB}$ to the light arrival coordinate time at the TSB
($t_a^{\rm TSB}$) by calculating the vacuum propagation time of the
light traveling from SSB to TSB ($\Delta_{\rm vp}$) as well as the
Einstein delay ($\Delta_{\rm ei}$) due to the relative motion between TSB
and SSB. Finally, we derive the proper emission time $\tau_e$ from
$t_a^{\rm TSB}$ by calculating the Roemer delay ($\Delta_{\rm rT}$),
Shapiro delay ($\Delta_{\rm sT}$), and Einstein delay ($\Delta_{\rm eT}$) in the target system. 

The purpose of modeling the light emission time is to calculate the
mean anomaly of the stellar reflex orbit precisely given an
observation proper time $\tau_o$. The formulae in the following sections are similar to the formulae given in E06 for pulsar timing but are adapted and implemented to be more suitable for exoplanets.

\subsubsection{Tropospheric Delay}\label{sec:tropo}
  The time delay in the solar system is
\begin{equation}
  \Delta_{\rm S}\simeq \Delta_{\rm tropo}+\Delta_{\rm eS}+\Delta_{\rm
    rS}+\Delta_{\rm pS}+\Delta_{\rm sS}~,
  \label{eqn:delay_solar}
\end{equation}
where $\Delta_{\rm pS}$ is a second-order Roemer delay, called
``parallax delay'' (E06).

An incident light ray is refracted by the Earth's atmosphere and is delayed by \citep{nilsson13}
  \begin{equation}
    \Delta_{\rm tropo}=c\int_{\mathcal L}[n(l)-1]dl+(t_{\mathcal L}- t_{\mathcal G})~,
    \label{eqn:Dtropo1}
  \end{equation}
  where $\mathcal L$ is the light ray path in the atmosphere, $n$
  is the refractive index, and $t_{\mathcal L}$ and  $t_{\mathcal G}$ are respectively
  the vacuum light propagation times for deflected and straight light
  rays. Because the atmosphere model typically consists of the
  hydrostatic and wet components, the tropospheric delay is typically
  split into two parts:
  \begin{align}
    \Delta_{\rm tropo}&=\frac{10^{-6}}{c}\int_{\mathcal
                        L}N_{\rm hydro}(l)dl+\frac{10^{-6}}{c}\int_{\mathcal L}N_{\rm wet}(l)dl+(t_{\mathcal L}- t_{\mathcal G})\\
    &=\Delta_{\rm hydro}+\Delta_{\rm wet}+(t_{\mathcal L}- t_{\mathcal
 G})~,
      \label{eqn:Dtropo2}
  \end{align}
where $N_{\rm hydro}$ and $N_{\rm wet}$ are respectively the
hydrostatic and wet refractivity. The refractivity is related to
refractive index by $N=10^{-6}(n-1)$. Each of these two components is a
product of a zenith delay and a mapping function
. The geometric delay term $t_{\mathcal L}-
t_{\mathcal G}$ is typically included in the mapping function of the
hydrostatic component \citep{nilsson13}. Hence, the tropospheric delay becomes
\begin{equation}
  \Delta_{\rm tropo}=\Delta_{\rm hydro}m_h(\Theta)+\Delta_{\rm
    wz}m_w(\Theta)~,
  \label{eqn:Dtropo3}
\end{equation}
where $\Theta$ is the observed elevation angle of the source, and
$\Delta_{\rm hydro}$ is\footnote{Note that the routine {\small tropo.C}
  in the TEMPO2 package contains an error. The cosine function in the
  denominator should be $\cos(2\phi_{\rm O})$ rather than
  $\cos{\phi_{\rm O}}$. However, this error may not significantly influence the TEMPO2 precision because $\cos(2\phi_{\rm O})$ is multiplied by 0.00266.}
\begin{equation}
  \Delta_{\rm hydro}=\frac{0.02268 (\frac{p_{\rm O}}{\rm kPa})}{(\frac{c}{{\rm
        m\,s}^{-1}})[1-0.00266\cos(2\phi_{\rm O})-2.8\times
    10^{-7}(\frac{h_{\rm O}}{\rm m})]}~,
\end{equation}
where $\phi_{\rm O}$ is the latitude of the observatory, $p_{\rm O}$ is the air
pressure, and $h_{\rm O}$ is the telescope altitude. The zenith hydrostatic
delay is typically a few nanoseconds. On the other hand, the wet zenith delay is not
well modeled and is highly variable. However, it is about one order of
magnitude smaller than the hydrostatic delay and thus is only
important for high-precision applications. Following E06, we adopt the
Niell mapping function (NMF; \citealt{niell96}) to calculate $m_h$ and
$m_w$ in Equation \ref{eqn:Dtropo3}. Like E06, we only consider the wet
component if the zenith wet delay is given by the observatory. Similar
to the wet component, the refraction caused by the ionosphere is
highly variable and cannot be separated from the dispersion in
interstellar and interplanetary medium. Thus, we consider them as
time-correlated noise, which can be modeled using a red noise model
such as the moving-average model \citep{feng17a}. 

\subsubsection{Time Delay in the solar system}\label{sec:delay_solar}
The Einstein delay is caused by the gravitational effect on the time measurement in different reference systems. According to E06 and \cite{irwin99}, the Einstein delay is 
\begin{equation}
  \Delta_{\rm
    eS}=\frac{1}{c^2}\int_{t_0}^t\bigg(U_\oplus+\frac{v_\oplus^2}{2}+\Delta
  L_C^{\rm (PN)}+\Delta L_C^{\rm (A)}\bigg){\rm d} t+\frac{{\bm
      s}\cdot{\bm \dot{\bm r}_\oplus}+W_0\tau_o}{c^2}~,
  \label{eqn:einstein}
\end{equation}
where $U_\oplus$ is the gravitational potential of all solar system
objects apart from the Earth, $v_\oplus$ is the barycentric velocity
of the geocenter, $W_0$ is approximately the gravitational and spin
potential of the Earth at the geoid, and $\Delta L_C^{\rm (PN)}$ and $\Delta L_C^{\rm (A)}$
respectively characterize the post-Newtonian effects and asteroidal
effects. The integral is to transform the Geocentric Coordinate Time
(TCG) into the barycentric time (TCB) at the geocenter. In the above
equation, the last term corresponds to the time difference between the
observer and the geocenter and transforms the terrestrial time (TT)
to TCG. The rate of TT with respect to TCG at the geocenter is
$L_G=W_0/c^2=6.969290134\times 10^{-10}$. The term $\Delta_{\rm rot}={\bm s}\cdot{\bm \dot{r}_\oplus}/c^2$ induces a
periodic delay with an amplitude of about 2\,$\mu$s. We model the
Earth's rotation using equation 26 in E06. 

Instead of calculating the integral in equation \ref{eqn:einstein}
directly, we use the time ephemeris of the Earth in JPL DE430 to
transform TT at the geocenter to Barycentric Dynamical Time (TDB) and
use $L_B=1.550519768\times10^{-8}$ to transform TDB into TCB according to ${\rm TCB}={\rm TDB}/(1-L_B)$. Because the rotation-induced delay ${\bm s}\cdot{\bm
  \dot{r}_\oplus}/c^2$ is not accounted for in the transformation from
TT to TDB by the JPL ephemeris, we add it in the transformation and
determine TDB in an iterative way as follows:

\begin{enumerate}
\item Transform Coordinate Universal Time (UTC) to the International
  Atomic Time (TAI; {\it Temps Atomique} in French) using the
  SOFA\footnote{\url{http://www.iausofa.org/}} routine {\small iauUtctai}. 
\item Transform TAI to TT(BIPMXY). ``BIPM'' denotes the International
  Bureau of Weights and Measures, and XY represents the year when the
  BIPM realization of TT is published. TT(BIPMXY)=TAI+32.184\,s+$\delta
  t$, where $\delta T$ is a the difference between the BIPMXY and TAI
  realizations of TT \citep{petit04} and can be downloaded from
  \url{https://www.bipm.org/jsp/en/TimeFtp.jsp?TypePub=ttbipm}. In
  this work, we use the TT(BIPM17) realization by default. The BIPM file is
    automatically updated to the latest version by PEXO.  
\item Determine TT-TDB as a function of TT at the geocenter using the latest JPL time ephemeris (e.g., DE430t). 
\item For a ground-based observer, determine the observer's geocentric
  position and velocity using the Earth rotation model recommended by
  IAU2006 resolutions \citep{capitaine06,wallace06}. For space
  telescopes, their ephemerides are determined using the JPL HORIZONS
  system. We have implemented an automated downloading of JPL ephemerides in PEXO.
\item Calculate $\Delta_{\rm rot}$ in the TDB coordinate system based
  on step 4 and add it onto TT-TDB. Note that {$\Delta_{\rm rot}$ is
    calculated using TT and thus needs to be scaled with $d$TT/$d$TDB
    although this scaling is a negligible secondary effect and only
    contributes at most 1\,ps (1\,ps=picosecond=$1\times 10^{-12}$\,s)}. 
\item Transform TDB to TCB using ${\rm TCB}={\rm TDB}/(1-L_B)$. 
\end{enumerate}
In summary, the transformation chain of various time standards is UTC$\rightarrow$TAI$\rightarrow$TT$\rightarrow$TCG$\rightarrow$TDB$\rightarrow$TCB. 

To derive the barycentric time, we need to account for the difference in the light travel time to the observer and to the SSB. This is the so-called Roemer delay, which is
\begin{equation}
  \Delta_{\rm rS}=-\frac{{\bm r}_{\rm SO}\cdot{\bm u}_{\rm SB}}{c}~,
  \label{eqn:dRS}
\end{equation}
where ${\bm r}_{\rm SO}={\bm r}_\oplus+{\bm s}$ is the sum of the BCRS
position of the geocenter ${\bm r}_\oplus$ and the position of the
observatory with respect to the geocenter ${\bm s}$. For space
telescopes, ${\bm r}_{\rm SO}$ can be obtained from the ephemeris of
the telescope from JPL HORIZONS. However, the Roemer delay assumes that the fiducial observer at the SSB receives plane waves from the light source. To account for the curvature of the wave, we include the second-order ``Roemer delay'':
\begin{equation}
  \Delta_{pS}=\frac{|{\bm r}_{\rm SO}\times{\bm u}_{\rm SB}|^2}{2cr_{\rm SB}}~.
  \label{eqn:dpS}
\end{equation}
This so-called ``parallax delay'' is included in the Roemer delay in
some other studies (e.g., \citealt{lindegren03} and
\citealt{eastman10}). For example, the parallax delay for $\alpha$
Centauri is about 0.7\,ms. Note that this parallax delay is equal
  to the one in equation (8) of (\citealt{eastman10}; hereafter E10), who use ${\bm u}_{\rm OT}$ rather than ${\bm u}_{\rm SB}$ as the reference unit vector, leading to an
  opposite sign of parallax delay. However, equations \ref{eqn:dRS} and
  \ref{eqn:dpS} do not account for higher-order astrometric effects, as mentioned
  in section \ref{sec:astrometry}. To improve the precision of PEXO
  for solar system objects, we calculate the total Roemer delay (including parallax delay) using
\begin{equation}
  \Delta_{\rm rS}=\frac{r_{\rm OT}-r_{\rm BT}}{c}~.
  \label{eqn:dRS1}
\end{equation}
Because the third-order astrometric terms contribute sub-mas position
offsets over decades, we expect tens of nanoseconds bias introduced by using
equations \ref{eqn:dRS} and \ref{eqn:dpS}. Such a bias is inversely
proportional to the heliocentric distance and increases with time, as
we will see in section \ref{sec:compare_timing}. Because this bias is
cumulative, the estimation of $<1$\,ns for third-order delays in E10 is not representative for long-term timing observations. 

A photon is deflected by the gravitational field of the solar system, leading to the so-called ``Shapiro delay'' \citep{shapiro64}, which is
\begin{equation}
  \Delta_{\rm sS}=(1+\gamma)\sum_i\frac{Gm_i}{c^3}\left\{\ln\left(\frac{2r_{\rm ST}}{A}\right)-\ln\left[\frac{r_{\rm SO}(1-\cos\psi_i)}{A}\right]\right\}~,
  \label{eqn:DSO}
\end{equation}
where $A=1$\,au, and $\psi_i$ is the coordinate angle distance between the
center of the body $i$ and the target star from the perspective of the
observer. The angle between the Sun and the target star dominates the
Shapiro delay and is determined by $\cos\psi_i=\frac{{\bm r}_{\rm
    OT}{\bm r}_{\rm OS}}{r_{\rm OT}r_{\rm OS}}$. The Shapiro delay
formulated in equation \ref{eqn:DSO} differs from equation (5) of \cite{eastman10}, who ignore the terms related to $r_{\rm ST}$ and $r_{\rm SO}$. However, $r_{\rm SO}$ is not constant for an observer on an eccentric orbit. Although this change may not be important for current exoplanet research, it is crucial for high-precision pulsar timing and thus is included in the model of TEMPO2 by E06. 

In summary, the barycentric Julian date (BJD) in the TCB standard
(${\rm BJD_{\rm TCB}}$) is determined by the corresponding Julian date
 (${\rm JD_{\rm TCB}}$) through ${\rm BJD_{TCB}}={\rm
  JD_{TCB}}-\Delta_{\rm rS}-\Delta_{\rm pS}-\Delta_{\rm sS}$,
  where ${\rm BJD}_{\rm TCB}$ and ${\rm JD}_{\rm TCB}$ are BJD and JD in the TCB time standards, respectively. BJD can only be determined
precisely if the precise location of the observed target is known at a
given epoch. However, this is impossible even with {\it Gaia} astrometry
because the astrometric solution is based on the assumption of a single
star. Thus, BJD is known {\it a posteriori} rather than {\it a priori}
by simultaneously modeling the motions of the Earth, the barycenter of the
target system and the stellar reflex motion. We will discuss the
  influence of decoupling the Solar and target systems on timing in
  section \ref{sec:decoupling}. Although PEXO does not separate the
  solar system dynamics and the target system dynamics in its timing
  model, we provide $\rm BJD_{\rm TDB}$ and BJD$_{\rm TCB}$ for users
  who do not require a timing precision of $<$0.02\,s. The upper
  limit of this precision corresponds to the timing bias amplitude for
  $\alpha$ Centauri A due to the decoupling of $\alpha$ Centauri and the solar system over one decade (see section \ref{sec:comparison_table}). For users who need high-precision timing model, PEXO provides a combined modeling of all motions and various times as optional outputs.

PEXO generates quantities compatible with both TDB and TCB time
standards. TCB is used as the time standard for Gaia \citep{brown18}
while TDB is used by TESS \citep{ricker14}. Since TDB is a time
standard compatible with JPL ephemeris and has a time increasing rate
very similar to that of TT and TAI, it is frequently used in the
exoplanet community. TCB by definition is not a relativistic time
standard and is not sensitive to relativistic effects in the solar
system although its realization may depend on the relativistic simulation of the solar system. Both TDB and TCB systems have particular advantages, we provide the ability to introduce data from both time standards for example for the combined analysis of the data from Gaia and TESS. The critical matter is the consistent transformation when using data sets with different time standards. PEXO is designed to provide for this. We refer the readers to \cite{klioner10} and \cite{iers10} for detailed discussion of different time standards. 

\subsubsection{Interstellar Time Delay}\label{sec:delay_interstellar}
Ignoring the interaction between a photon with the interstellar medium, the arrival time at the SSB is delayed with respect to the TSB by 
\begin{equation}
\Delta_{\rm is}\simeq \Delta_{\rm vp} + \Delta_{\rm ei}
\end{equation} 
where $\Delta_{\rm vp} =|{\bm v}_{\rm SB}(t_a^{\rm SSB}-t_{\rm
  pos})+{\bm r}_{\rm SB}(t_{\rm pos})|/c$. Because the vacuum
propagation of light at the reference time ($|{\bm r}_{\rm
  SB}(t_{\rm pos})|/c$) is a constant, we only model the relative vacuum
propagation delay, $\Delta_{\rm vp} =|{\bm v}_{\rm SB}(t_a^{\rm SSB}-t_{\rm
  pos})+{\bm r}_{\rm SB}(t_{\rm pos})|/c-|{\bm r}_{\rm SB}(t_{\rm
  pos})|/c$. The Einstein delay due to the relative motion
between TSB and SSB is
\begin{equation}
  \Delta_{\rm ei}=\frac{v_{\rm SB}^2}{2c^2}(t_a^{\rm SSB}-t_{\rm pos}-\Delta_{\rm vp})~.
\end{equation}
  
\subsubsection{Time Delay in the Target System}\label{sec:delay_target}
Similar to the time delay in the solar system, the delay in the target system is
\begin{equation}
\Delta_{\rm T}\simeq \Delta_{\rm rT}+\Delta_{\rm pT}+\Delta_{\rm
  eT}+\Delta_{\rm sT}~.
\label{eqn:dT}
\end{equation}
According to E06, the Roemer delay is
  \begin{equation}
    % \Delta_{\rm rT}=\frac{{\bm r_{\rm reflex}}(t)\cdot{\bm u}_{\rm B}(t)}{c}
        \Delta_{\rm rT}=\frac{{\bm r_{\rm BT}}\cdot {\bm u}_b}{c}+\frac{1}{cr_{\rm SB}}\left({\bm \mu}\cdot {\bm r}_{\rm BT,\perp}-{\bm r}_{\rm SO, \perp}\cdot {\bm r}_{\rm BT, \perp}+\frac{|{\bm r}_{\rm BT, \perp}|^2}{2}\right)~,
      \label{eqn:drT}
    \end{equation}
    where ${\bm r}_{\rm SO, \perp}={\bm u}_b\times({\bm r}_{\rm
      SO}\times{\bm u}_b)$ and ${\bm r}_{\rm BT, \perp}={\bm
      u}_b\times({\bm r}_{\rm BT}\times {\bm u}_b)$. In the above
    equation, the first term is related to the Roemer delay that is
    due to the motion of TSB, while the other terms are named
    ``Kopeikin terms'' related to the orbital variation of the target
    system that is due to the changing perspective caused by the proper motion of TSB \citep{kopeikin96}. In a rotation reference frame perpendicular to the line of sight, these terms can ``change'' the orbital elements of the target system. However, such an apparent change disappears if the orbit is defined at the reference epoch in a fixed reference frame, as in equation \ref{eqn:rOT3}. 

Instead of using the reference unit vector ${\bm u}_b$, we use the
time-varying vector ${\bm u}_{\rm SB}$ to calculate the combined
Roemer and parallax delay as
  \begin{equation}
        \Delta_{\rm rT}+\Delta_{\rm pT}=\frac{{\bm r_{\rm BT}}\cdot {\bm u}_{\rm SB}}{c}-\frac{|{\bm r}_{\rm BT}\times {\bm u}_{\rm SB}|^2}{2cr_{\rm SB}}~.
      \label{eqn:drT-pT}
    \end{equation}
    This delay is similar to its counterpart in the solar system, as expressed in equations \ref{eqn:dRS} and \ref{eqn:dpS}. 

 According to \cite{blandford76} and \cite{damour86}, the Einstein delay in the target system is 
\begin{equation}
  \Delta_{\rm eT}=g U~,
    \label{eqn:deT}
  \end{equation}
  where $g$ is the timing model parameter. 

According to \cite{damour86}, the Shapiro delay for the target system is
\begin{equation}
    \Delta_{\rm sT}=-2r_s\log\left\{1-e\cos{U}-s_s\left[\sin{\omega}(\cos{U}-e) +(1-e^2)^{1/2}\cos{\omega}\sin{U}\right]\right\}~,
  \label{eqn:DST}
\end{equation}
where all of the variables are given in section \ref{sec:PN}. Although
higher-order Shapiro delay terms are available \citep{kopeikin99},
they are insignificant because the first-order term is of the order of $(v_{\rm BT}/c)^3$. 

\subsection{Astrometry Model}\label{sec:relativistic_astrometry}
The direction of a light ray observed by an observer is deflected by
the gravitational field between the source and the frame
transformation between the observer and the target. Thus we aim to
find the observed direction of the target star by tracing the
direction of a photon forward from the emission time to the arrival
time at the observatory. To avoid confusion with
the geometric modeling of the observed direction of the source derived
in section \ref{sec:astrometry}, we use $\bm l$ to denote the
direction of a light ray at a given time. 

\subsubsection{Stellar aberration}\label{sec:abberration}
According to special relativity, the Lorentz transformation from a static reference frame to a moving reference frame would introduce a change in the direction of the target star. This effect is called ``stellar aberration''. After equation (7) of \cite{klioner03},
 the direction of the observed light ray is
\begin{align}
  \hat{\bm u}_o&=\langle-{\bm l}_o+c^{-1}{\bm l}_o\times({\bm v}_{\rm SO}\times {\bm l}_o)+c^{-2}[({\bm l}_o\cdot {\bm v}_{\rm SO}) {\bm l}_o\times({\bm v}_{\rm SO}\times {\bm l}_o)+\frac{1}{2}{\bm v}_{\rm SO}\times({\bm l}_o\times {\bm v}_{\rm SO})]\nonumber\\
&+c^{-3}\left\{\left[({\bm l}_o\cdot {\bm v}_{\rm SO})^2+(1+\gamma)w(r_{\rm SO})\right] {\bm l}_o\times({\bm v}_{\rm SO}\times {\bm l}_o)
+\frac{1}{2}({\bm l}_o\cdot {\bm v}_{\rm SO}) {\bm v}_{\rm SO}\times({\bm l}_o\times {\bm v}_{\rm SO})\right\}+\mathcal{O}(c^{-4})\rangle~,
\label{eqn:uo}
  \end{align}
where ${\bm l}_o$ is the light ray direction when it is observed, the
absolute value of potential $w(r_{\rm SO})$ is approximated by a spherically symmetric Sun by
  \begin{equation}
    w(r_{\rm SO})\approx Gm_\odot/r_{\rm SO}
  \end{equation}
and $\gamma$ is a dimensionless parameter in the Parameterized
  post-Newtonian formalism (PPN; \citealt{nordtvedt72}). It is equal
  to 1 if GR is true. It could be fitted to astrometry
  data in the case of weak-field relativity tests although a fully
  post-Newtonian formalization of the timing, astrometry, and radial
  velocity models are required to test GR
  consistently. For strong-field relativity tests, only the PPK
  parameters (see Equation \ref{eqn:x}) are fitted. For the difference
  between PPN and PPK parameters, we recommend \cite{taylor92} for
  more details. Due to gravitational lensing, ${\bm l}_o\neq -{\bm u}_{\rm OT}$.  

\subsubsection{Atmospheric refraction}\label{sec:refraction}
As mentioned in section \ref{sec:tropo}, a light ray is refracted when
it propagates in the Earth's atmosphere. This effect is one of the
main factors that limits the precision of ground-based astrometry
\citep{gubler98,mangum15}. We use the routine {\it slaRefro} in
{\small
  SLALIB}\footnote{\url{http://star-www.rl.ac.uk/star/docs/sun67.htx/sun67.html}}
to calculate the refraction, 
\begin{equation}
  \mathcal{R}=\int_{1}^{n_{\rm O}}\frac{\tan{Z}}{n}dn~,
  \label{eqn:R1}
\end{equation}
where $n_{\rm O}$ is the refractive index at the telescope and $Z$ is the refracted
zenith angle. The observed zenith angle $Z_o$ is the sum of the
incident zenith angle above the atmosphere $Z_i$ and the refraction:
\begin{equation}
  Z_o=Z_i+\mathcal{R}~.
\end{equation}
As $\tan{Z}$ diverges when $Z$ approaches 90$^\circ$ (see equation
\ref{eqn:R1}), \cite{auer00} reformulate the integrant as a function of zenith
angle, and the refraction becomes
\begin{equation}
  \mathcal{R}=-\int_{0}^{Z_o}\frac{rdn/dr}{n+rdn/dr}dZ~,
\end{equation}
where $r$ is the distance from the geocenter. Because refraction is
wavelength dependent, the effective temperature or wavelength of a
star should be known in order to calculate the refraction. By adopting the
atmospheric model developed by \cite{rueger02} and using the {\it
  slaRefro} routine adapted from the {\it AREF} routine given by
\cite{hohenkerk85}, we can calculate refraction $R$ to a precision of
about 1\,arcsec \citep{mangum15} and the differential refraction $\Delta
R$ to a precision of 10\,$\mu$as \citep{gubler98}. However, in order to achieve such relative astrometric precision for a typical binary, those authors find that the effective
temperature of stars should be measured to a precision of 100\,K,
absolute zenith angle to a precision of 36\,arcsec, relative zenith
angle to a precision of 30\,mas, air temperature at the observatory to
a precision of 0.6\,K, air pressure to a precision of 160\,Pa,
and relative humidity to a precision of 10\%. Because the refraction is
calculated using the observed zenith in {\it slaRefro}, we set
$Z_o=Z_i$ and repeat the calculation of $R$ until it converges. Because the refraction occurs in the plane formed by the zenith and the
incident light ray and is perpendicular to the incident light ray, the refraction vector is
\begin{equation}
  \bm{\mathcal{R}}=\frac{{\bm u}_Z-({\bm u}_Z\cdot {\bm u}_{\rm
      OT}){\bm u}_{\rm OT}}{\sin{Z}}\mathcal{R}~,
  \label{eqn:refro}
\end{equation}
where ${\bm u}_Z$ is the unit vector in the zenith direction. Then the light ray direction when it is observed
is
\begin{equation}
  {\bm l}_o={\bm l}_i-\bm{\mathcal{R}}~,
\end{equation}
where ${\bm l}_i$ is the
approximately $-{\bm u}_{\rm OT}$. Such an assumption would at most
induce third order effects. 

\subsubsection{Gravitational light deflection}\label{sec:deflection}
For a target system outside of the solar system (with heliocentric
distance $>10^5$\,au), the emitted light from the target star would be
deflected by the gravitational fields of companions. This effect is
also called gravitational lensing and will also contribute to the
Shapiro delay, as discussed in section \ref{sec:timing}. After equation
(70) of \cite{klioner03}, we convert the light ray direction at the emission time ${\bm l}_e$ into the direction after leaving the target system as
\begin{equation}
  % {\bm l}_l =  {\bm l}_e-\sum_{\rm C}\frac{(1+\gamma) GM_C }{c^2r_{\rm
%  OC}}\cot{\frac{\psi_C}{2}}~,
  {\bm l}_l =  {\bm l}_e-\sum_A\frac{(1+\gamma) Gm_A}{c^2}\frac{{\bm r}_{\rm
      OT}\times({\bm r}_{\rm TA}\times{\bm r}_{\rm OA})}{ |{\bm r}_{\rm OT}||{\bm r}_{\rm OA}|(|{\bm r}_{\rm TA}||{\bm r}_{\rm OA}|+{\bm r}_{\rm OA}{\bm r}_{\rm TA})}~,
\label{eqn:ll}
\end{equation}
where A denotes the body in the target system  parameter and
$\gamma=1$ if GR is assumed. We ignore the gravitational
deflection of light that is due to the nonspherical gravitational potential
of lenses because it only contributes 1$\mu$as when the light source is very close to the lens (see Table 1 of \cite{klioner03} for details). 

Assuming vacuum propagation of the light ray between the target and
the solar system, the direction of the incident light beyond the atmosphere is
\begin{equation}
  {\bm l}_i =  {\bm l}_l-\sum_{\rm L}\frac{(1+\gamma) Gm_{\rm L} {\bm
      d}_{\rm L}}{c^2d_{\rm L}^2}(1+\cos{\psi_{\rm L}})~,
\label{eqn:lo}
\end{equation}
  where $\cos{\psi_{\rm L}} ={\bm u}_{\rm OT}\cdot{\bm r}_{\rm OL}/r_{\rm
    OL}$ is the angular distance between the light ray and lens L, and
  ${\bm d}_{\rm L}={\bm l}_e\times({\bm r}_{\rm OL} \times {\bm
    l}_e)$. For an observer at the geocenter, the light ray does not
  bend if one assumes the gravitational field of the Earth is spherically
  symmetric. According to \cite{klioner03}, the main light deflection
  is caused by the Sun and the Earth, while the Moon and other planets
  are only important if the light ray passes them closely. 
  
In summary, the emitted light ray direction is derived from the
geometric observed direction using ${\bm l}_e=-{\bm u}_{\rm OT}$ with
${\bm u}_{\rm OT}$ derived from equation \ref{eqn:uOT}. Here, ${\bm l}_i$ is
calculated using equations \ref{eqn:ll} and \ref{eqn:lo}. The
  incident light is further refracted by the atmosphere by
  $\bm{\mathcal{R}}$. The direction of the light ray at the telescope
  is $\bm{l}_o=\bm{l}_i-\bm{\mathcal{R}}$. Then $\hat{\bm u}_o$ is calculated using equation \ref{eqn:uo} to model the observed direction of star ${\bm u}_o$. 
    
\subsection{Radial Velocity Model}\label{sec:relativistic_rv}
In this section, we model the observed radial velocity related to the kinematics, geometry, and relativistic effects of the target star and the observer. 

\subsubsection{Einstein Doppler Shift}\label{sec:VG_shift}
In an inertial reference frame, the Schwarzschild solution to the Einstein field equations leads to the following exact ratio between the rate of proper time and the rate of coordinate time for a clock:
\begin{equation} 
\frac{d\tau}{dt}=\sqrt{1-\left(\frac{v^2}{c^2}+\frac{v_e^2}{c^2}+\frac{(v_{||}/c)^2(v_e/c)^2}{1-(v_e/c)^2}\right)}~,
\end{equation}
where $v_{||}$ is the radial velocity of the clock with respect to the
inertial frame, and
\begin{equation}
  v_e=\sqrt{\sum_i{\frac{2Gm_i}{r_i}}}
\end{equation}
is the escape velocity determined by the sum of the gravitational potential of
nearby bodies. Applying the above formula to the solar system and the
target system and ignoring $c^{-4}$ terms, we derive the increment
ratio of the proper observation time $\tau_o$ and the proper emission
time $\tau_e$ as
\begin{align}
  1+z&\equiv \frac{\lambda_o}{\lambda_e}=\frac{\nu_e}{\nu_o}=\frac{d\tau_o}{d\tau_e}=\frac{d\tau_o}{dt_o}\frac{dt_o}{dt_i}\frac{dt_i}{dt_e}\frac{dt_e}{d\tau_e}\nonumber\\
     & =(1-\frac{\Phi_{\rm S}}{c^2}-\frac{v_{\rm SO}^2}{2c^2})
       (1-\frac{\Phi_{\rm T}}{c^2}-\frac{v_{\rm ST}^2}{2c^2})^{-1} \frac{dt_o}{dt_i}\frac{dt_i}{dt_e}
       \label{eqn:VG_shift}
\end{align}
  where $\lambda_o$ and $\lambda_e$ are respectively the observed and
  emission wavelength, $\lambda_o$ and $\lambda_e$ are respectively
  the observed and emitted light frequency, $\Phi_{\rm
    S}=\sum\limits_i{\frac{Gm_i}{r_i}}$ is the absolute value of gravitational potential of the solar system at the observer's location
  while $\Phi_{\rm T}=\sum\limits_j{\frac{Gm_j}{r_j}}$ is the absolute
  value of gravitational potential of the target system when it emits
  the light, $dt_o/dt_i$ is determined by the atmospheric refraction, and $dt_i/dt_e$ is determined by Shapiro delay and vacuum propagation. We define
  \begin{equation}
    z_{\rm
      grS}\equiv\Phi_{\rm S}/c^2
                \label{eqn:zgrS}
  \end{equation}
  and
  \begin{equation}
    z_{\rm grT}\equiv\Phi_{\rm T}/c^2
            \label{eqn:zgrT}
  \end{equation}
  as gravitational Doppler shifts in the solar and target systems,
  respectively. We also define
  \begin{equation}
    z_{\rm srS}\equiv\frac{v_{\rm SO}^2}{2c^2}
        \label{eqn:zsrS}
  \end{equation}
  and
  \begin{equation}
    z_{\rm srT}\equiv\frac{v_{\rm ST}^2}{2c^2}
    \label{eqn:zsrT}
  \end{equation}
  as the Doppler shifts due to special
  relativity effects in the solar and target systems, respectively. Because
  \begin{equation}
    \frac{d\tau_o}{dt_o}=\frac{\rm dTT}{\rm dTCB}=1-z_{\rm grS}-z_{\rm srS}~,
    \label{eqn:zrS}
  \end{equation}
the relativistic effects on the Doppler shifts in the SS can be derived
from $\Delta_{\rm eS}$ later in section \ref{sec:timing}.

 For photons emitted from different places on the surface of a star,
 they experience different gravitational Doppler shifts especially if
 there is a massive companion close to the target star. For example,
 the velocity variation corresponding to the gravitational Doppler
 shift caused by a Sun-like star located about 1\,au from the target
 star is 3\,m/s. Assuming that the radius of the target star is
 comparable with the solar radius, which is about $1/215$\,au, the
 differential Doppler shift would lead to about 1\,cm/s of radial velocity variation. Such a differential Doppler shift should be accounted for together with the rotation-induced differential Doppler shift in the case of exoplanet detection in close binary systems. 

\subsubsection{Kinematic, lensing, and tropospheric Doppler shift}\label{sec:special_shift}
As discussed in section \ref{sec:timing}, the emission
coordinate time $t_e$ is delayed from the coordinate arrival time
  of nonrefracted light ray $t_i$ by
\begin{equation}
  t_i-t_e=\Delta_{\rm geo}+\Delta_{\rm sS}+\Delta_{\rm sT}~,
\end{equation}
where
\begin{equation}
  \Delta_{\rm geo}=\frac{r_{\rm OT}}{c}
  \label{eqn:Dgeo}
\end{equation}
is the vacuum propagation time from the target to the observer.
The differential of the above delay gives
\begin{equation}
  \frac{{\rm d}t_e}{{\rm d}t_i}=\frac{1+z_{\rm kS}-z_{\rm lS}}{1+z_{\rm kT}-z_{\rm lT}}~,\\
\end{equation}
where $z_{\rm lS}$ and $z_{\rm lT}$ are respectively the lensing Doppler shift corresponding to the Shapiro delay in the solar system and in the target system. Thus the ratio of time rate is

\begin{equation}  
  z_{\rm kS}=\frac{{\bm u}_{\rm OT}\cdot{\bm v}_{\rm SO}}{c},
\end{equation}
and
\begin{equation}  
  z_{\rm kT}=\frac{{\bm u}_{\rm OT}\cdot{\bm v}_{\rm ST}}{c},
\end{equation}
are respectively the kinematic Doppler shift in the solar and target
systems. We calculate the relativistic effects by adopting the
direction from the observer to the target as in \cite{kopeikin99} and
\cite{lindegren03}.

In \cite{lindegren03}, the gravitational deflection and the Shapiro
delay of the light are not thoroughly treated because of their negligible
effects. For example, \cite{lindegren03} dropped the Shapiro delay
term because it contributes at most 0.3\,m/s radial velocity
precision. Although this upper limit is determined from the extreme
situation when the light ray grazes the solar limb, the lensing effect
for stars with a large angular distance from the Sun can still be important for achieving 1\,cm/s radial velocity precision and we
consider this further below. Based on a more rigorous treatment of the
Shapiro effect in
equations (169), (173), and (238) of \cite{kopeikin99}, the lensing Doppler shift in the solar system is
\begin{equation}
  z_{\rm lS}= \left(\frac{\delta \nu}{\nu_{\rm o}}
  \right)_S=\sum_{\rm L}\frac{1}{c}({\bm v}_{\rm SL}-\frac{r_{\rm
      LT}}{r_{\rm OT}}{{\bm v}_{\rm SO}}-\frac{r_{\rm OL}}{r_{\rm
      OT}}\cdot{{\bm v}_{\rm ST}}) \cdot{{\bm \alpha}({{\bm
        \lambda}_{\rm L}})}
\label{eqn:zlS1}
\end{equation}
where ${\bm \lambda}_{\rm L}$ is the impact parameter of the
unperturbed path of photons with respect to lens L, and
\begin{equation}
  {\bm \alpha}({{\bm \lambda}_{\rm L}})=2(1+\gamma)\frac{G
    m_{\rm L}}{c^2\lambda_{\rm L}^2}{\bm \lambda}_{\rm L}~,
  \label{eqn:alphaL}
\end{equation}
where $m_{\rm L} $ is the mass of lens L. 
Ignoring the lensing effects of planets, assuming a static Sun with
respect to the SSB, and considering $r_{\rm OA}\ll r_{\rm AT}$ and
$r_{\rm OT}\simeq r_{\rm OA}+ r_{\rm AT}$, we find that Equation \ref{eqn:zlS1} becomes
\begin{equation}
  z_{\rm lS}= \left( \frac{\delta \nu}{\nu_{\rm o}} \right)_S=-\frac{{\bm v_{\rm SO}} \cdot{\bm \alpha}({\bm \lambda_S})}{c}~,
  %\simeq -{\bm v_o} \cdot{\bm \alpha}({\bm \lambda_\odot})~,
\label{eqn:zlS}
\end{equation}
where ${\bm \lambda}_S={\bm u}_{\rm OT}\times({\bm r}_{\rm OS} \times{\bm
  u}_{\rm OT})$. Because the lensing effect is proportional to $c^{-3}$,
the above assumptions would at most have fourth-order effect. Equation
\ref{eqn:zlS} is the lensing formula used in most literature. Similarly, the lensing Doppler shift in the target system is approximately
\begin{equation}
  z_{\rm lT}= \left( \frac{\delta \nu}{\nu_{\rm o}} \right)_{\rm T}=-\frac{{\bm v_{\rm CT}} \cdot{\bm \alpha}({\bm \lambda_{\rm C}})}{c}~,
  %\simeq -{\bm v_o} \cdot{\bm \alpha}({\bm \lambda_S})~,
\label{eqn:zlT}
\end{equation}
where ${\bm v_{\rm SC}}$ and ${\bm v_{\rm ST}}$ are respectively the
velocity of the companion and target star with respect to the SSB and
${\bm \lambda}_{\rm C}={\bm u}_{\rm OT}\times({\bm r}_{\rm OC}
\times{\bm u}_{\rm OT})$. The Sun is the main gravitational {\rm lens} in the solar system which
induces a gravitational shift of about $\frac{1{\rm
    AU}}{\lambda_S}$\,mm/s assuming the observer's tangential velocity
of 30\,km/s. For an impact parameter comparable with the solar radius,
the shift would be about 0.3\,m/s. The angle between the target source
and the light ray from the perspective of the observer $\psi$ should
be less than 7$^\circ$ based on equations \ref{eqn:zlS} and \ref{eqn:zlT} in order to induce
$>1$\,cm/s line shift. If the target system is like the solar system,
this effect leads to $>1$\,cm/s doppler shift in edge-on systems. Although the
lensing effect is typically ignored in current exoplanet packages such
as {\small EXOFAST} \citep{eastman13}, it could become significant in
the search for small planetary signals whose amplitude is comparable with the
lensing effect. 

 Atmospheric refraction not only causes timing delay and deflects light
 rays but also leads to Doppler shift. The Doppler shift induced by
 tropospheric refraction is
 \begin{equation}
   z_{\rm tropo}=\frac{dt_o}{dt_i}-1=\frac{d\Delta_{\rm
       tropo}}{dt_i}\approx (\Delta_{\rm hydro}m_h'(\Theta)+\Delta_{\rm
     wz}m_w'(\Theta))\frac{d\Theta}{dt_i}~,
   \label{eqn:ztropo}
 \end{equation}
where $m_h'=\frac{{\rm d}m_h}{{\rm d}\Theta}$ and $m_w'=\frac{{\rm
     d}m_w}{{\rm d}\Theta}$. 
The differential tropospheric delay is derived numerically using {\it
  slaRefro}. The rotation of the Earth leads to a continuous change of
the elevation and thus changes the mapping functions $m_h$ and
$m_w$. This effect would induce diurnal radial velocity variation of a
few mm/s for elevation angles of lower than 30$^\circ$ if only
hydrostatic delay was considered. For elevation angles less than
10$^\circ$, the refraction could induce up to a few m/s radial velocity
variation due to the exponential variation of refraction near the horizon
(see P4 of Fig. \ref{fig:acrel}). 

By combining all Doppler effects, the Doppler shift is
\begin{equation}
  \frac{v_r^{\rm obs}}{c}\equiv z=\frac{1-z_{\rm grS}-z_{\rm
      srS}}{1-z_{\rm grT}-z_{\rm srT}}\frac{1+z_{\rm kT}-z_{\rm
      lT}}{1+z_{\rm kS}-z_{\rm lS}-z_{\rm tropo}}-1~.
  \label{eqn:vr_obs}
\end{equation}
Unlike \cite{wright14} and \cite{butkevich14}, we do not explicitly add
a term related to the light travel effect. Rather, we calculate the
quantities at the corresponding retarded time for a given light ray. We
calculate the emitted frequency at the proper emission time according
to the time transformation described in section \ref{sec:timing}. In
equation \ref{eqn:vr_obs}, the special and general relativistic Doppler shifts
($z_{\rm srS}$, $z_{\rm grS}$, $z_{\rm srT}$, and $z_{\rm grT}$) are
proportional to $c^{-2}$, and lensing effects ($z_{\rm lS}$ and $z_{\rm
  lT}$) lead to $\mathcal{O}(c^{-3})$ Doppler shift. The kinematic
Doppler shifts ($z_{\rm kS}$ and $z_{\rm kT}$) are proportional to
$c^{-1}$ and are thus significant radial velocity
variations. In the case of detection of small planets like the
Earth, $z_{\rm kT}$ corresponds to $<$1\,m/s radial velocity
variation, as large as the radial velocity effects caused by some
relativistic effects. Thus, a comprehensive modeling of these effects
is essential for reliable detection of Earth-like planets.

\subsection{Caveats in the decoupling of the Solar and target systems}\label{sec:decoupling}
The so-called ``barycentric correction'' is typically used to
transform the measured radial velocity into the BCRS radial
velocity. However, it is only possible if we can separate the local effects and the remote effects caused by the target system. Specifically, the total Doppler shift is split into local and remote Doppler shifts:
\begin{align}
  z&=(1+z_{\rm T})(1+z_{\rm S})-1~,\\
  1+z_{\rm S}&=\frac{1-z_{\rm grS}-z_{\rm srS}}{1+z_{\rm kS}-z_{\rm lS}-z_{\rm tropo}}~,\\
1+z_{\rm T}&=\frac{1+z_{\rm kT}-z_{\rm lT}}{1-z_{\rm grT}-z_{\rm srT}}~.
  \label{eqn:zb}
\end{align}
Although most terms in $z_{\rm S}$ can be precisely determined, ${\bm
  u}_{\rm OT}$ is typically not known {\it a priori}. For single
stars, the error in proper motion may bias the barycentric correction
for decades-long radial velocity data. For example, a 10\,mas/yr uncertainty in
proper motion would lead to 1.5\,cm/s uncertainty in barycentric
correction over one year.

For stars with massive planet companions, the catalog astrometry of
the barycenter is biased by the typical assumption of a single
star in the data reduction. Because the stellar reflex motion is coupled
with the Earth's motion (see equation \ref{eqn:uOT}), a barycentric
correction for binaries would not only lead to a spurious trend but
also introduce false periodic signals in the corrected radial velocity
data. For $\alpha$ Centauri, these false signals lead to sub-m/s radial velocity
variation, hindering the detection of Earth-like planets in this
system. Thus, precise barycentric correction is only possible if the
stellar reflex orbit is accurately determined {\it a priori}. However,
this is rarely the case even in the {\it Gaia} era because the five-parameter astrometry solution assumes no companions around a target star.
Even if companions are considered in astrometry modeling, potential
uncertainty and bias are expected because of a lack of precise modeling of
instrumental bias, stellar activity, and other noise terms.

Two types of biases are caused by barycentric correction:
\begin{itemize}
\item A trend bias is caused by using the barycentric
  velocity of the target system as the velocity of the target star
  without considering the stellar reflex motion. This assumption would
  bias the astrometric solution and thus induce long-term trend in the
radial velocity data. Thus this bias is related to the velocity of the stellar
reflex motion and is important for long-term observations.

\item A periodic bias is caused by ignoring the position offset of the target
star with respect to the barycenter of the target system. Because the
stellar reflex motion is periodic, this assumption would cause
periodic variation of the visual direction of the target star, leading
to periodic variation of radial velocity. It is important for observations with
baselines longer than the period of stellar reflex motion.
\end{itemize}

Because the Earth's motion is coupled with the barycentric and binary motions, the
annual and diurnal Earth motions are manifested in both biases. To estimate the trend bias, we calculate the average reflex motion
of the target star as
\begin{equation}
  \bar{v}_{\rm reflex}=\frac{m_{\rm C}}{m_{\rm C}+m_{\rm T}}\sqrt{\frac{G(m_{\rm C}+m_{\rm T})}{a}}~.
\end{equation}
The corresponding proper motion bias caused by ignoring this reflex
motion is
\begin{equation}
  \delta\mu=\bar{v}_{\rm reflex}\frac{\widetilde\omega^b}{A}~.
\end{equation}
This proper motion offset leads to a positional bias of
\begin{equation}
\delta u=\delta\mu \delta t
\end{equation}
over a time span of $\delta t$. 
Assuming that the characteristic radial velocity caused by the motions of the
target star and the Earth is $v_{\rm tot}$=50\,km/s, we estimate the radial
velocity bias related to $\delta u$ as
\begin{equation}
  \delta v_r^{\rm trend}=v_{\rm tot} \delta
  u=1.52\left(\frac{m_{\rm C}}{M_\odot}\right)\left(\frac{m_{\rm C}+m_{\rm T}}{M_\odot}\right)^{-1/2}\left(\frac{a}{{\rm
        au}}\right)^{-1/2}\left(\frac{\widetilde\omega^b}{{\rm
        mas}}\right)\left(\frac{\delta t}{{\rm yr}}\right)~ {\rm mm}~{\rm s}^{-1}~.
    \label{eqn:vr_trend}
\end{equation}
The corresponding acceleration of the trend bias is
\begin{equation}
\delta \dot{v}_r^{\rm trend}=\frac{\delta {v}_r^{\rm trend}}{\delta t}=1.52\left(\frac{m_{\rm C}}{M_\odot}\right)\left(\frac{m_{\rm C}+m_{\rm T}}{M_\odot}\right)^{-1/2}\left(\frac{a}{{\rm
        au}}\right)^{-1/2}\left(\frac{\widetilde\omega^b}{{\rm
        mas}}\right)~ {\rm mm}~{\rm s}^{-1}~{\rm yr}^{-1}~.
\end{equation}
The periodic bias is determined by the semi-major axis of the
stellar reflex motion and is
\begin{equation}
    \delta v_r^{\rm period}=v_{\rm tot}a\frac{m_{\rm C}}{m_{\rm C}+m_{\rm T}}\frac{\widetilde\omega^b}{A}=0.24\left(\frac{a}{{\rm au}}\right)\left(\frac{\widetilde\omega^b}{{\rm mas}}\right)~ {\rm mm}~{\rm s}^{-1}~.
    \label{eqn:vr_period}
  \end{equation}

Considering that the barycentric correction is also frequently used in
astrometry and timing, we calculate the time delay biases
corresponding to the trend and periodic radial velocity biases, which
are
\begin{equation}
   \delta \Delta^{\rm trend}=\frac{\delta v_r^{\rm period}}{v_{\rm tot}}\frac{A}{c}=15.17\left(\frac{m_{\rm C}}{M_\odot}\right)\left(\frac{m_{\rm C}+m_{\rm T}}{M_\odot}\right)^{-1/2}\left(\frac{a}{{\rm
        au}}\right)^{-1/2}\left(\frac{\widetilde\omega^b}{{\rm
        mas}}\right)\left(\frac{\delta t}{{\rm yr}}\right)~\mu {\rm s}
    \label{eqn:timing_trend}
\end{equation}
and
\begin{equation}
   \delta \Delta^{\rm periodic}=\frac{\delta v_r^{\rm period}}{v_{\rm tot}}\frac{A}{c}=2.40\left(\frac{a}{{\rm au}}\right)\left(\frac{\widetilde\omega^b}{{\rm mas}}\right)~ \mu{\rm s}~.
   \label{eqn:timing_period}
 \end{equation}
In the above equations, we consider the light travel time from the Earth
to the Sun (about 1\,au) as the characteristic Roemer delay.

Similarly, the astrometric biases corresponding to the trend and
periodic radial velocity biases are
\begin{equation}
   \delta u^{\rm trend}=\frac{\delta v_r^{\rm trend}}{v_{\rm tot}}=6.27\left(\frac{m_{\rm C}}{M_\odot}\right)\left(\frac{m_{\rm C}+m_{\rm T}}{M_\odot}\right)^{-1/2}\left(\frac{a}{{\rm
        au}}\right)^{-1/2}\left(\frac{\widetilde\omega^b}{{\rm
        mas}}\right)\left(\frac{\delta t}{{\rm yr}}\right)~{\rm mas}~.
   \label{eqn:u_trend}
\end{equation}
and
\begin{equation}
 \delta u^{\rm periodic}=\frac{\delta v_r^{\rm period}}{v_{\rm tot}}=0.99\left(\frac{a}{{\rm au}}\right)\left(\frac{\widetilde\omega^b}{1~{\rm mas}}\right)~
 {\rm mas}~.
 \label{eqn:u_period}
\end{equation}
Therefore, a radial velocity bias of 1\,mm/s corresponds to a timing
bias of 10\,$\mu$s and an astrometric bias of about 4\,mas.

\begin{figure}
  \centering  
  \includegraphics[scale=0.5]{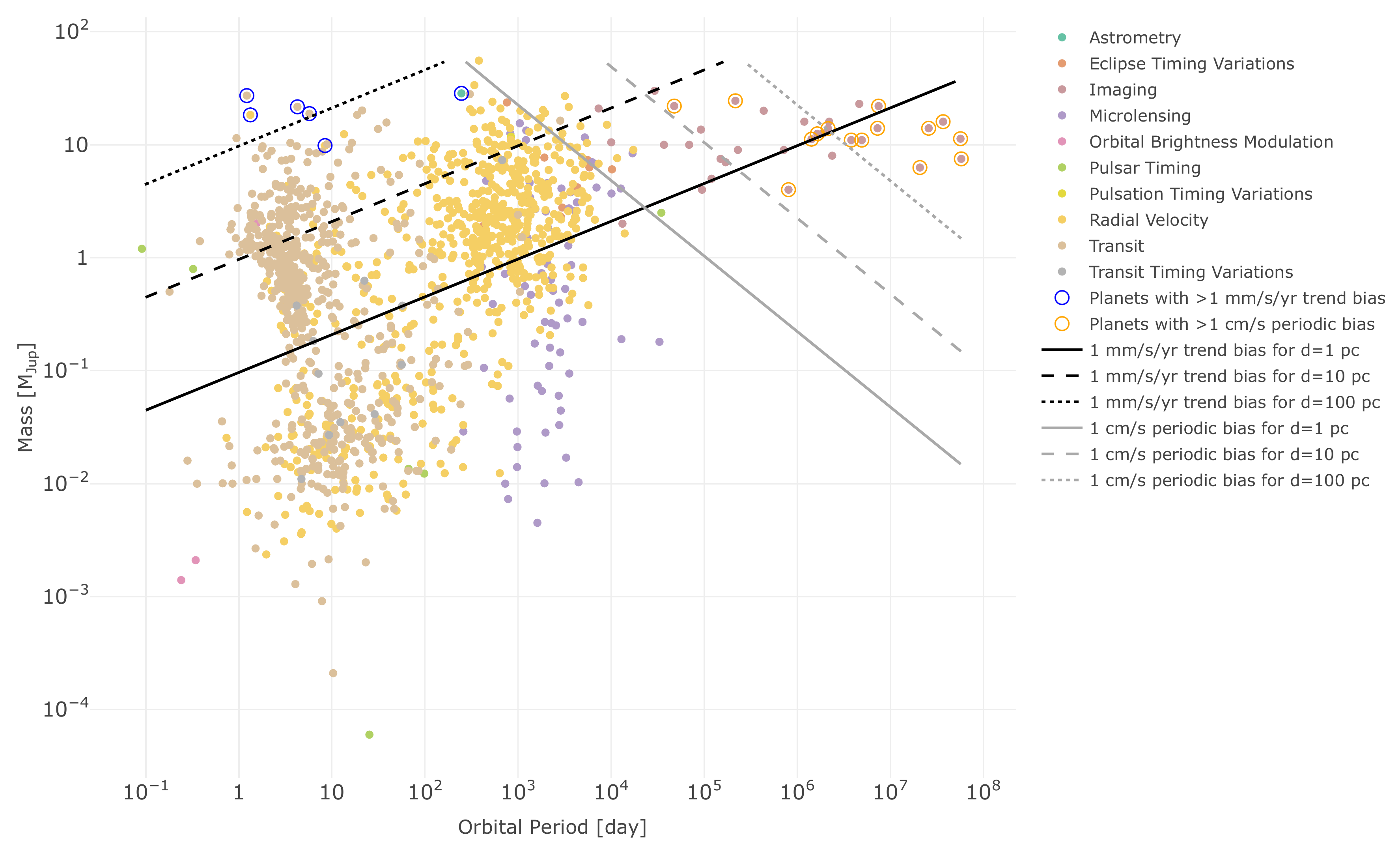}
  \caption{The plot shows exoplanets downloaded from the NASA
    Exoplanet Archive and color coded for different detection methods
    as a function of orbital period and planetary mass. The lines on
    the plot are to indicate potential radial velocity biases caused
    by barycentric corrections or more generally by decoupling local
    and remote effects with the assumption that plotted exoplanets are
    hosted by single stars. The black lines show the boundaries for
    the 1\,mm/s/yr trend bias while the gray lines show the boundaries for
    the 1\,cm/s periodic bias. The top left part of the phase space is
    particularly susceptible to large trend bias while the top right
    part of the phase space is particularly susceptible  to large
    periodic bias. The solid, dashed, and dotted lines show the biases
    for stars with distances $d=$1, 10, and 100\,pc, respectively. The
    planets denoted by open circles are influenced by at least
    1\,mm/s/yr trend bias or 1\,cm/s periodic bias because of decoupling. The orbital periods for directly imaged planets are derived from their semi-major axes by assuming a face-on circular orbit.}
  \label{fig:barycorr_bias}
\end{figure}

To investigate the influence of the barycentric correction or more
generally the decoupling of local and remote effects on the detection of exoplanets, we calculate the boundary of companion mass and orbital
period corresponding to a trend bias with an acceleration of 1\,mm/s/yr and a
periodic bias of 1\,cm/s for stars with a heliocentric distance of 1, 10, and
100\,pc. We show these boundaries together with the currently known
planets in Fig. \ref{fig:barycorr_bias}. In this figure, we add
circles around plotted points to denote planets that are subject to
trend bias with an acceleration larger than 1\,mm/s/yr or periodic
bias larger than 1\,cm/s. An acceleration of
1\,mm/s/yr corresponds to a 1\,cm/s trend bias for observations over one
decade. There are only six planets strongly influenced by trend
bias. Five of them are transit planets, while one of them is detected
through astrometry. They are all massive planets with relatively short
orbital periods. The boundaries for trend biases also suggest that
short-period massive planets such as hot Jupiters would induce large
stellar reflex motions and thus bias the initial proper motions,
leading to a spurious radial velocity trend. On the other hand, the
periodic bias is manifested in stars with long-period and massive companions, most of which are detected through direct imaging.

Nevertheless, our estimation of the bias for a given system
is only a lower limit because the bias is determined by the largest
companion in a system, which might not be detected. Therefore, the
boundaries shown in Fig. \ref{fig:barycorr_bias} are a better guide
for the estimation of decoupling bias because it is calculated for the most
massive companion in a system. 

Because the transit timing is not sensitive to $<0.02$\,s bias
(corresponding to decoupling bias for $\alpha$ Centauri A), a decoupling
in timing modeling is efficient and reliable for most transit
systems. However, the transit system is sensitive to remote effects
such as the transit timing variation (TTV) caused by binary
motions. Relative astrometry is sensitive to the astrometry and radial velocity of
the TSB (see Equation \ref{eqn:relative}), which could be biased
through decoupling. Absolute astrometry is
sensitive to decoupling, and this is why the high-precision astrometry
software GREM \citep{klioner03} is used to model all motions
simultaneously for {\it Gaia} astrometry although its timing model is biased by decoupling effects. 

For the radial velocity method, decoupling is unlikely to achieve
1\,cm/s radial velocity precision because even distant close binaries
($d>1$\,kpc) show 1\,cm/s trend bias for decade-long observations. On
the other hand, wide binaries (with orbital periods longer than one
decade) show strong periodic bias, which can be approximated as a trend
for observations with a baseline far shorter than the orbital
period. Because nearly half of the solar-type stars are binaries (e.g.,
\citealt{sana11,moe18}), a combined modeling of the target and local systems is essential to achieve 1\,cm/s precision
over decade-long observations. As illustrated in
Fig. \ref{fig:barycorr_bias}, for nearby stellar systems, planets
with hot and cold Jupiter companions are sensitive to trend and
periodic biases, respectively. Because most TESS targets are close to
the Sun, the radial velocity follow up for hot Jupiters detected by
TESS may need to consider the trend bias. Specifically, decoupling
could introduce $\sim$0.1\,m/s bias in ten years of radial velocity
measurements of a nearby star ($<$10 pc) with hot or cold Jupiters. It
could introduce $\sim$1\,m/s bias over one year for a nearby star hosting stellar-mass companions.

Considering the above difficulties, a separation of local and remote
radial velocity effects through decoupling is unlikely to achieve 1\,cm/s precision for decades-long radial velocity data especially for
nearby stars with massive companions (e.g., with a mass $>1~M_{\rm Jup}$). Because astrometry data is essential for a reliable
decoupling, a combined modeling of radial velocity and astrometry is
the proper way to avoid bias induced by decoupling. Another more efficient approach is to use astrometry offsets or jitter terms to model potential bias and fit
these offsets together with the radial velocity model parameters to
the radial velocity data ``corrected'' for barycentric effects. 

\subsection{Significance of Relativistic Effects in Extrasolar
  Systems}\label{sec:relativity_test}
\begin{figure}
  \centering  
  \includegraphics[scale=0.5]{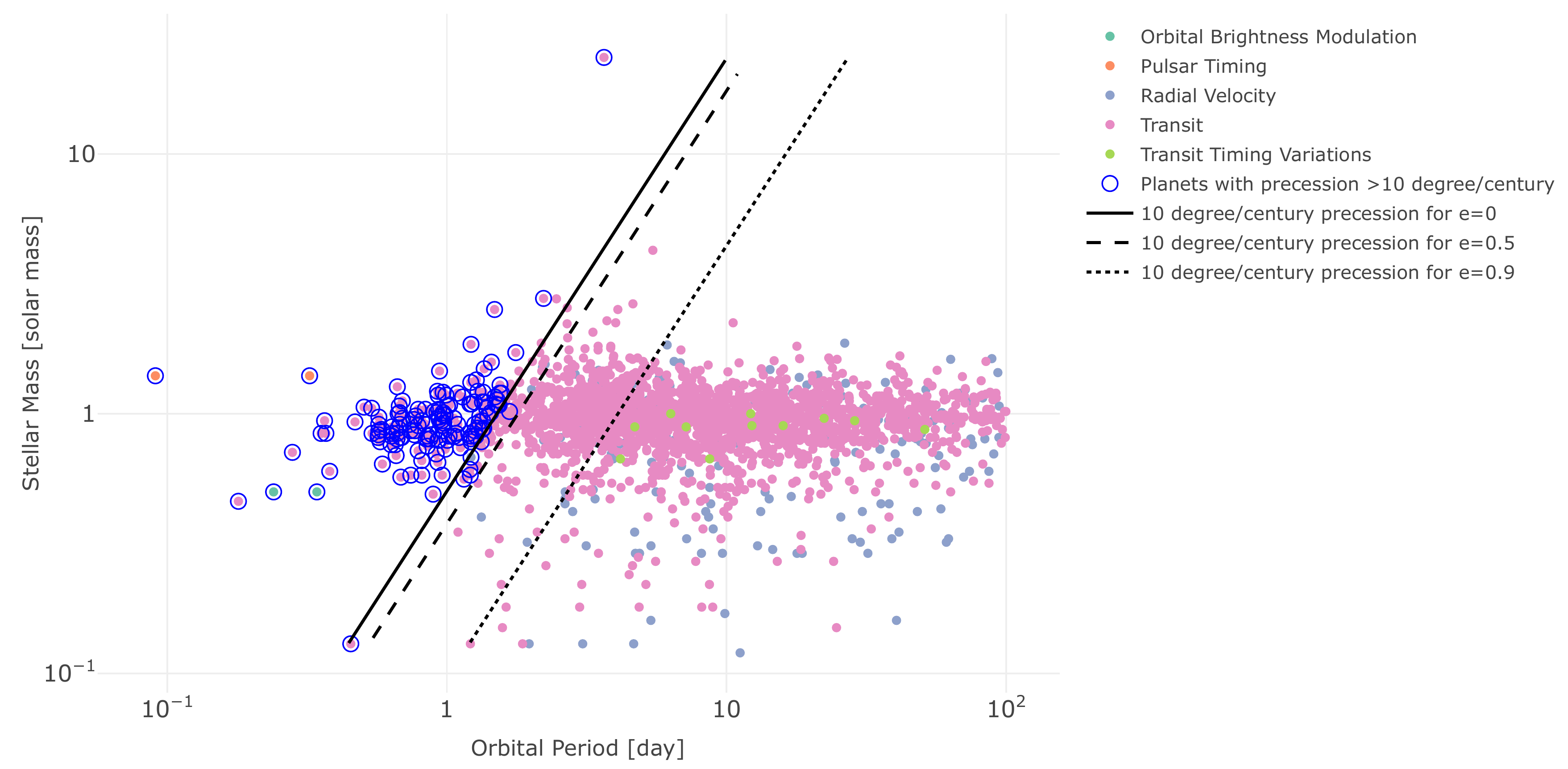}
  \caption{In order to illustrate the detectability of
    relativistic effects in currently known planetary systems, we
    select exoplanets with stellar mass larger than 0.1\,$M_\odot$ and
    orbital period less than 100 days from the known exoplanets
    selected for Fig. \ref{fig:barycorr_bias}. All of the exoplanets to
    the left of the solid, dashed, and dotted lines (for $e=$0, 0.5,
    and 0.9) are the exoplanets with a precession larger than 10$^\circ$ per century and are highlighted with open circles.}
  \label{fig:precession}
\end{figure}

In this section, we investigate the sensitivity of currently confirmed
exoplanets to relativistic effects. The main relativistic effect in
extrasolar systems is the precession of the longitude of
periastron. According to \cite{misner73} and \cite{jordan08}, it is
\begin{equation}
  \dot{\omega}_{\rm
    GR}=\frac{3Gm_{\rm T}}{ac^2(1-e^2)}n=\frac{7.78}{1-e^2}\left(\frac{m_{\rm T}}{1~M_\odot}\right)\left(\frac{a}{0.05~{\rm
      au}}\right)^{-1}\left(\frac{P}{1~{\rm
      day}}\right)^{-1}~^\circ/{\rm century}~,
\label{eqn:omega.dot}
\end{equation}
where $n\equiv [G(m_{\rm T}+m_{\rm C})/a^3]^{1/2}$ is the Keplerian mean
motion and $P$ is the orbital period. Assuming $m_{\rm C}\ll m_{\rm T}$, we derive the period-mass
boundaries for $e=$0,0.5, and 0.9 for $\dot{\omega}_{\rm GR}=$10$^\circ$ per
century or 1$^\circ$ per decade and show them in the period-mass
distribution of currently known planets in Fig. \ref{fig:precession}. There are about 144 transit planets with strong relativistic precession, although the
planetary perturbation and tidal deformations may also contribute at a
level comparable to the relativistic precession \citep{jordan08}. However, these nonrelativistic
effects only become important when the planet is very close to the
star. According to \cite{jordan08}, planetary orbits with semi-major axis
larger than 0.05\,au are suitable for relativity tests. The precession is
detectable in the variation of primary transit duration
\citep{miralda02} and in the changes of longitude of periastron in
radial velocity data \citep{jordan08}. Although such effects will
probably be detected in the near future, the current radial velocity
and transit timing data are not likely to be precise enough to put strong
constraints on various post-Newtonian theories and to test GR in particular. 

Unlike star-planet systems, binaries have relatively stronger
gravitational fields and thus are more suitable for relativity
testing. Considering that timing data is limited to one-dimensional
information, we investigate the feasibility of using astrometry and
radial velocity data to test relativity. The former has been
studied previously (e.g., \citealt{kopeikin99b,klioner03,kopeikin07}). We will focus on the latter by
assessing the significance of the Doppler shift induced by special
($z_{\rm srT}$) and general ($z_{\rm grT}$) relativity. Because these
two Doppler shifts are proportional to $c^{-2}$, they dominate the
relativistic Doppler shifts in the target system, compared with the
lensing Doppler shifts, which are proportional to
$c^{-3}$. Because the constant relativistic Doppler shift is not
detectable in radial velocity data, we only estimate the variation of
relativistic Doppler shifts. According to Equation \ref{eqn:zsrT} the
amplitude of the variation of $z_{\rm srT}$ is
\begin{equation}
\delta z_{\rm srT}= \frac{\delta v_{\rm ST}^2}{2c^2}=\frac{{\bm v}_{\rm ST, max}^2-{\bm v}_{\rm ST, min}^2}{2c^2}~,
\end{equation}
where $\bm v_{\rm ST,max}$ and $\bm v_{\rm ST,min}$ are respectively the
maximum and minimum $\bm v_{\rm ST}$. The variation of $({\bm
  v}_{\rm SB}+{\bm v}_{\rm BT})^2=v_{\rm SB}^2+v_{\rm BT}^2+2{\bm
  v}_{\rm SB}\cdot{\bm v}_{\rm BT}$ depends on the angle between ${\bm v}_{\rm SB}$ and ${\bm v}_{\rm BT}$ and thus depends on the inclination and angular parameters of the binary
orbit. To simplify the problem, we explore the range of $\delta z_{\rm
  srT}$ for a given eccentricity $e$ and barycentric velocity $v_{\rm
  SB}$. The minimum $\delta z_{\rm srT}$ is
\begin{equation}
\delta z_{\rm srT.min}=\frac{\delta{\bm v}^2_{\rm BT}}{2c^2}=\frac{v_{\rm BT, max}^2-v_{\rm BT,min}^2}{2c^2}=\frac{2e}{c^2(1-e^2)}\left[\frac{2\pi G}{P}\right]^{2/3}(m_{\rm T}+m_{\rm C})^{-4/3}m_{\rm C}^2~.
\end{equation}
Then the minimum amplitude of radial velocity variation induced by special relativity is
\begin{equation}
\delta v_{\rm srT,min} = c\delta z_{\rm
  srT}=5.92\frac{e}{1-e^2}\left(\frac{P}{{\rm
      year}}\right)^{-2/3}\left(\frac{m_{\rm T}+m_{\rm C}}{M_\odot}\right)^{-4/3}\left(\frac{m_{\rm C}}{M_\odot}\right)^2~{\rm m/s}~.
\label{eqn:dvsrTmin}
\end{equation}

The maximum $\delta z_{\rm srT}$ is
\begin{eqnarray}
\delta z_{\rm srT,max}&=&\frac{(v_{\rm SB}+v_{\rm BT,max})^2-(v_{\rm
    SB}-v_{\rm BT,min})^2}{2c^2}=\frac{v_{\rm SB}(v_{\rm
    BT,max}-v_{\rm BT,min})}{c^2}+\delta z_{\rm srT,min}\\
&=&v_{\rm SB}\frac{2e}{c^2\sqrt{1-e^2}}\left[\frac{2\pi
    G}{P}\right]^{1/3}(m_{\rm T}+m_{\rm C})^{-2/3}m_{\rm C}+\delta z_{\rm srT,min}~.
\end{eqnarray}
%Assuming a typical barycentric velocity $v_{\rm SB}=50$\,km/s \citep{feng14}
The maximum amplitude of radial velocity variation induced by special relativity is
\begin{equation}
\delta v_{\rm srT,max} = c\delta z_{\rm
  srT,max}=\delta v_{\rm srT,min}+\delta v_{\rm srT,couple}~,
\label{eqn:dvsrTmax}
\end{equation}
where
\begin{equation}
  \delta v_{\rm srT,couple}=9.94\frac{e}{\sqrt{1-e^2}}\left(\frac{P}{{\rm
      year}}\right)^{-1/3}\left(\frac{m_{\rm T}+m_{\rm C}}{M_\odot}\right)^{-2/3}\left(\frac{m_{\rm C}}{M_\odot}\right)\left(\frac{v_{\rm SB}}{50~{\rm km/s}}\right)~{\rm m/s}
\label{eqn:dvsrTcouple}
\end{equation}
is the relativistic radial velocity related to the coupling of the
heliocentric motion of the TSB and the binary motion. For $\alpha$
Centauri A, $\delta v_{\rm srT,min}\approx 0.08$\,m/s and $\delta
v_{\rm srT,max}\approx 0.61$\,m/s over half of the binary orbital period.

According to Equations \ref{eqn:zgrT} and \ref{eqn:orbital_plane}, the amplitude of the variation of
gravitational Doppler shift for a binary is
\begin{equation}
\delta z_{\rm grT}=\Phi_T/c^2=\frac{Gm_{\rm C}}{r_{\rm CT,min}}-\frac{Gm_{\rm C}}{r_{\rm CT,max}}=\frac{2m_{\rm C}}{c^2}\frac{e}{1-e^2}\left[\frac{4\pi^2G^2}{(m_{\rm C}+m_{\rm T})P^2}\right]^{1/3}~.
\end{equation}
Hence the corresponding amplitude of radial velocity variation is
\begin{equation}
\delta v_{\rm grT} =5.92\frac{e}{1-e^2}\left(\frac{P}{{\rm
      year}}\right)^{-2/3}\left(\frac{m_{\rm T}+m_{\rm C}}{M_\odot}\right)^{-1/3}\left(\frac{m_{\rm C}}{M_\odot}\right)~{\rm m/s}.
\label{eqn:dvgrT}
\end{equation}
For $\alpha$ Centauri A, $\delta v_{\rm grT}\approx 0.16$\,m/s over
half of the binary orbital period.

\begin{figure}
  \centering  
  \includegraphics[scale=0.6]{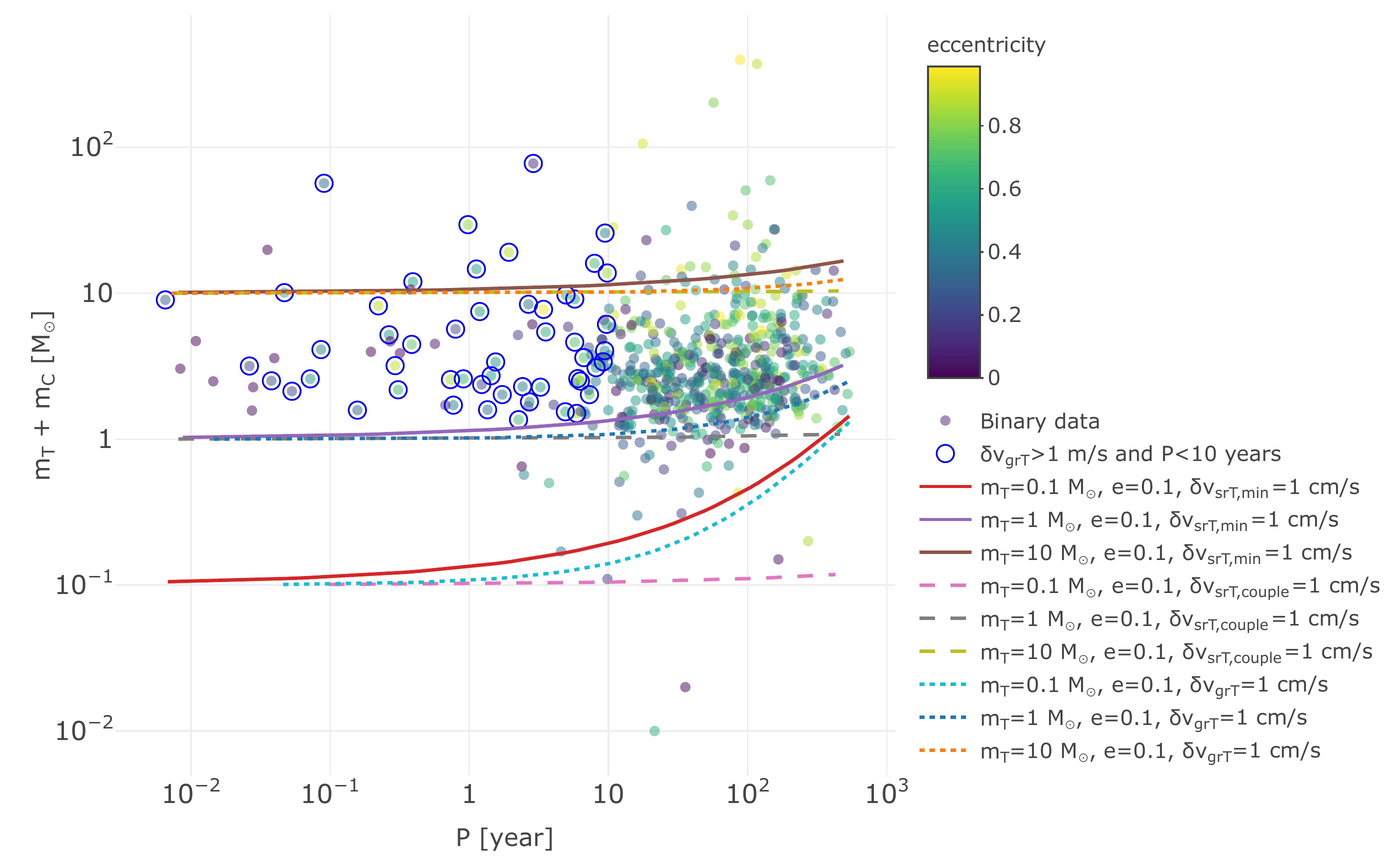}
  \caption{We show the observed binary masses and orbital periods
    from \cite{malkov12} along with period sensitivity for the
    relativistic radial velocity variation. The three groups of lines
    are for different values of the mass of target star $m_{\rm T}$
    (0.1, 1, and 10\,$M_\odot$). The solid upper solid lines are for
    minimum amplitude of radial velocity induced by special
    relativity (equation \ref{eqn:dvsrTmin}). The dashed lines are for
   the relativistic radial velocity related to the coupling of the
    heliocentric motion of the TSB and the binary motion (equation
    \ref{eqn:dvsrTcouple}). The dotted lines are the amplitude of radial
    velocity variation due to the gravitational Doppler shift of a
    binary (equation \ref{eqn:dvgrT}). On the basis that primaries and secondaries have the same
    mass, the 52 binaries with gravitational radial velocity $\delta
    v_{\rm grT} >$ 1 m/s and orbital period $P<10$\,yr are denoted
    by blue circles. We note that in all our predictions we assume GR to be true and use $e=0.1$ although a wide range of eccentricity values are observed, as denoted by the colored eccentricity legend.  }
  \label{fig:binary}
\end{figure}

To investigate the sensitivity of binary orbits to relativistic
effects, we show the relativistic radial velocity variation as a
function of binary mass and orbital period. We show a sample of 652
binaries with dynamical masses derived by \cite{malkov12} in
Fig. \ref{fig:binary}. According to equations
\ref{eqn:dvsrTmin}, \ref{eqn:dvsrTmax}, \ref{eqn:dvsrTcouple}, and
\ref{eqn:dvgrT}, the relativistic radial velocity variation is not
sensitive to eccentricity if the binary orbit is not circular
(e.g. $e>0.1$). Because only 8\% binaries in the binary sample have
$e<0.1$, we adopt $e=0.1$ to more easily calculate the relativistic radial velocity variation. Fig. \ref{fig:binary}
illustrates that the relativistic radial velocity is relatively sensitive to the mass of the secondary $m_{\rm C}$ for a low-mass primary compared with a high-mass
primary. Thus the optimal targets for detecting relativistic effects
are the low-mass companions of massive primaries. While many binaries
show a relativistic radial velocity variation of a few cm/s over one
orbital period, the detection of such a variation is at most marginal
and thus is not suitable for relativity tests. To select the optimal targets
for relativity tests, we note that there are 52 binaries with $v_{\rm grT}>$1\,m/s
and orbital period $P<10$\,yr. Based on a PPN formulation of the
gravitational redshift (e.g., \citealt{misner73,kopeikin99b,gravity18}), these
binaries can be used to constrain the strong principle of equivalence
to a relative precision of 1\% if a few cm/s radial velocity
precision can be achieved by high-precision spectrographs. Because the
gravitational redshift caused by a binary companion has a period
that differs from the orbital periods of potential planets around the
target star, the gravitational redshift variation can be detected
without considering planetary perturbations, although a combined modeling may reduce the residual and improve the significance
of detection. 

\begin{figure}
  \centering  
  \includegraphics[scale=0.7]{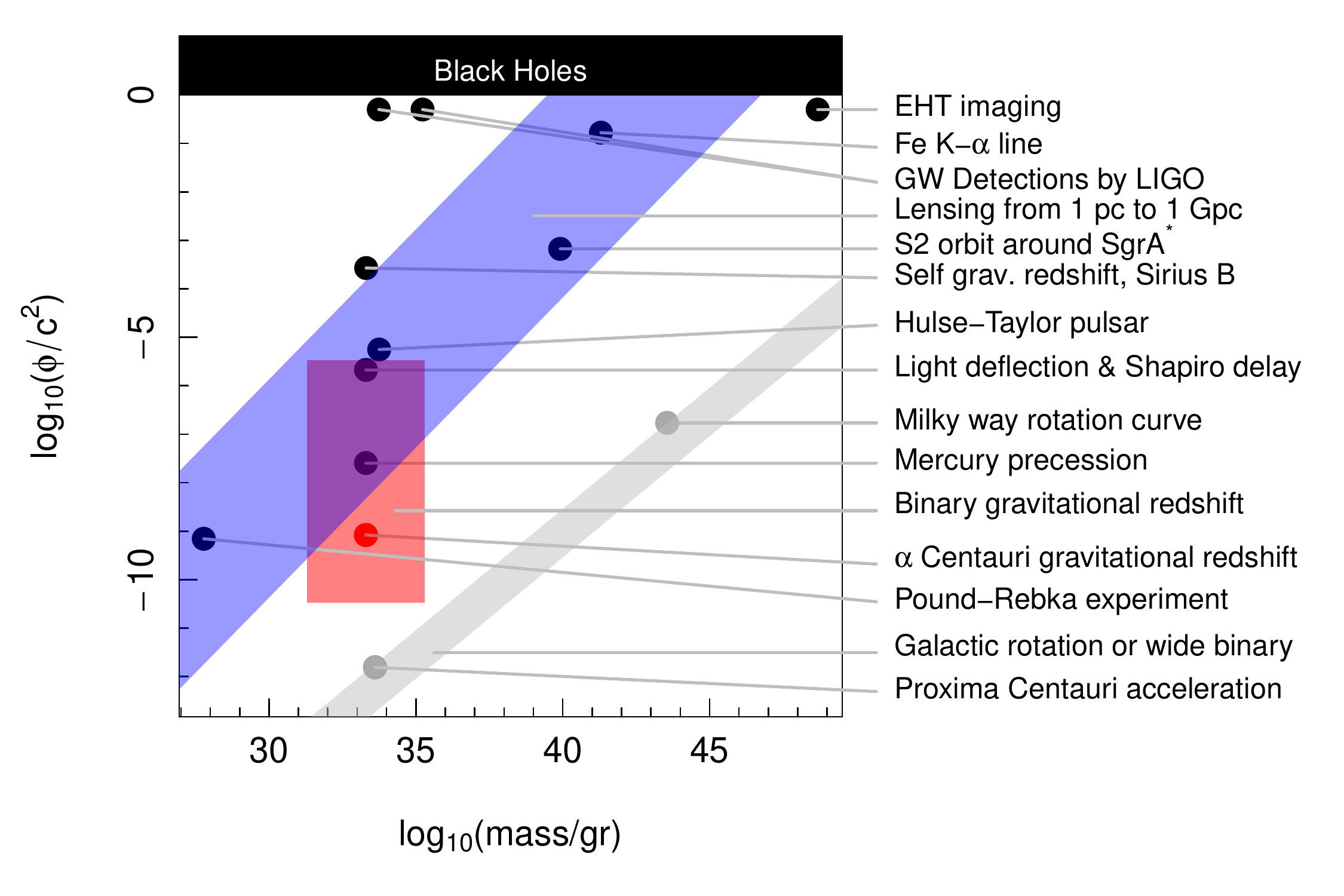}
   \caption{Relativity tests as a function of logarithmic mass and dimensionless gravitational potential, inspired
     by \cite{psaltis04,gravity18}. The blue region represents the gravitational lensing
 effect at the Einstein radius for a reduced distance $d=\frac{d_{\rm L}d_{\rm S}}{d_{\rm LS}}$ from 1\,pc to 1\,Gpc, where
 $d_{\rm L}$ and $d_{\rm S}$ are respectively the distances to the
 lens and to the source, and $d_{\rm LS}$ is the distance from the
 lens to the source. The red region represents the
 binary test with mass from 0.1 to 150\,$M_\odot$ and gravitational
 radial velocity variation from 1\,cm/s to 1\,km/s. The gray region
 represents the galactic rotation where the width of the gray region is determined by an
acceleration ranging from $10^{-12}$ to $10^{-10}$\,m/s$^2$
\citep{lelli17}. The black dots show well-established tests (from top
to bottom): the imaging of the M87 black hole horizon by \cite{eht1}, the
relativistic broadening of Fe K$\alpha$ lines
\citep{tanaka95,fabian00}, the LIGO/Virgo detection of gravitational
waves \citep{gw150914,gw170817}, the S2 orbit around the Galactic-centre massive black hole \citep{gravity18}, the self-gravitational
redshift of Sirius B \citep{greenstein71,barstow05}, the Hulse-Taylor
pulsar \citep{taylor82}, the light deflection and Shapiro delay in the
solar system (e.g., \citealt{shapiro64}), the precession of Mercury
\citep{einstein16}, and the \cite{pound59} experiment. The red dot
denotes the gravitational redshift in the $\alpha$ Centauri AB binary system. The gray dots
 represent tests of dark matter and MOND theories and thus also
 provide tests of GR in a nonrelativistic
 regime. They are galactic rotation curves represented by the Milky
 Way and the wide binary acceleration represented by $\alpha$ and Proxima Centauri
 \citep{banik18}. 
 }
  \label{fig:relativity_test}
\end{figure}

In summary, the gravitational redshift variation in binary systems
can provide a new method to test GR. To demonstrate the
uniqueness of this method, we show the mass and dimensionless
gravitational potential for various relativity tests in
Fig. \ref{fig:relativity_test}. Although current efforts are
focused on strong-field tests of GR, few tests have
been done in the weak-field regime. It is in the extreme weak-field
regime where dark matter needs to be invoked to explain phenomena such
as galactic rotation and gravitational lensing. However, in the extremely
weak gravitational field, relativistic effects become weak as well,
and thus it is not clear whether the weak-field anomaly is due to the
breakdown of the classical or relativistic predictions of GR if the null detection of dark matter over the past two decades
\citep{cosine100} indicates alternative gravity theories. To this end,
the binary test of relativity provides a unique way to probe the weak-field and stellar-mass regime in order to test GR and
alternative theories such as the modified Newtonian dynamics (MOND; \citealt{milgrom83}). 

\section{Comparison between PEXO and TEMPO2}\label{sec:comparison}
To estimate the precision of PEXO, we compare PEXO with TEMPO2, which
is able to model timing to a precision of $\sim$1\,ns. Because radial velocity is
simply the time derivative of various delay terms, we also use TEMPO2
to estimate the radial velocity model precision of PEXO. However, TEMPO2 does not
model astrometry precisely. Because our astrometry model is similar to the model used by GREM \citep{klioner03} which is able to achieve
$\mu$as precision, we expect a similar precision for
PEXO. Considering that our radial velocity and astrometry models are consistent with each
other, we also expect 1\,$\mu$as precision for the astrometry model if
the radial velocity modeling precision is 1\,cm/s for most stars over decades\footnote{The velocity precision is $\delta v=1$\,cm/s\,$=2.109\times
  10^{-6}$\,au/yr. The corresponding astrometry precision $\delta u=v\delta
  t/d=2.109\times 10^{-6}$\,au/yr\,$ \times 10$\,yr$/10$\,pc\,$=$\,2.109\,$\mu$as, where $d=10$\,pc is the distance of the target star and $\delta t=10$\,yr is the time span.}. 

\subsection{Timing}{\label{sec:compare_timing}}
We use $\tau$ Ceti as an example to compare the timing model of PEXO with the one in TEMPO2 and the one
introduced by \cite{eastman10}. The position of $\tau$ Ceti is
characterized by $\alpha=-15^\circ.93955572$
(ICRF), $\delta^\circ=26.02136459$ (ICRF),
$\widetilde\omega=273.96$\,mas, $\mu_\alpha=-1721.05$\,mas/yr,
$\mu_\delta=854.16$\,mas/yr, and radial velocity $v_r=-16.68$\,km/s
\citep{brown18}. We use the online applet developed by
E10 to calculate the ${\rm BJD_{TDB}}$ from JD$_{\rm UTC}$. Because
this online applet does not propagate the orbit of the target star, we
set $\mu_\alpha$, $\mu_\delta$, and $v_r$ to be zero in order to
compare with PEXO as well as TEMPO2. We use the GPS position of CTIO
determined by \cite{mamajek12} as an example observatory geocentric
coordinate. We calculate $\rm BJD_{TDB}$ for $\rm JD_{UTC}$ over
  a 10,000 day time span in a step of 10 days and use the ephemeris
of JPL DE405 \citep{standish98} to determine the motions of the Earth and
observatory\footnote{Because E10 uses DE405 by default, we use DE405
  in PEXO for comparison, though DE430 is used in other cases.}. We use the 2001 version (hereafter FB01) of the analytical method developed by
  \cite{fairhead90} and recommended by \cite{mccarthy04} to calculate TDB-TT for PEXO. The 1990 version (hereafter FB90) is used for TEMPO2 because it is the only available version in TEMPO2. We also use the 2000B model of the Earth rotation \citep{capitaine03,mccarthy03} for TEMPO2 and PEXO.

\begin{figure}
  \centering  
  \includegraphics[scale=0.38]{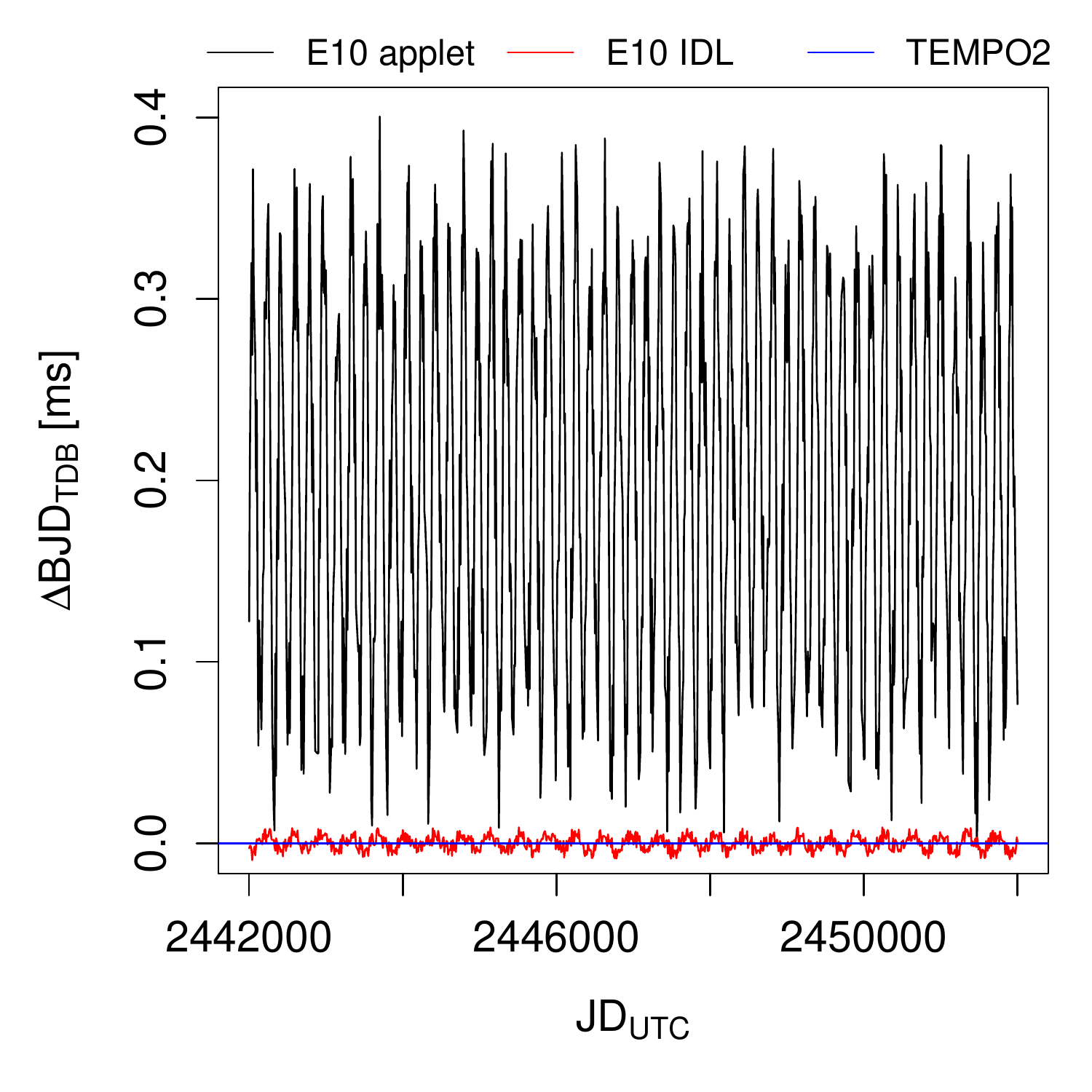}
  \includegraphics[scale=0.38]{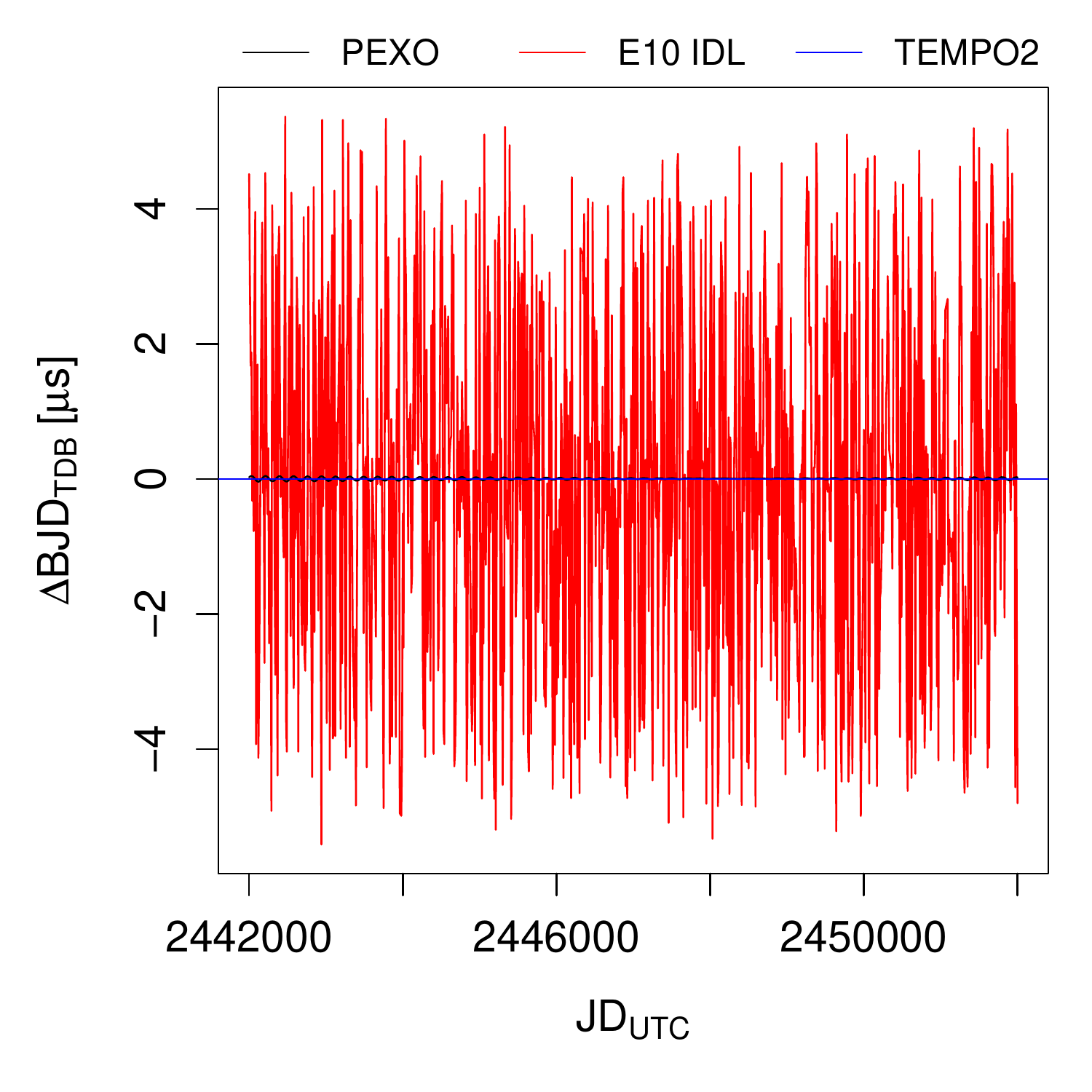}
  \includegraphics[scale=0.38]{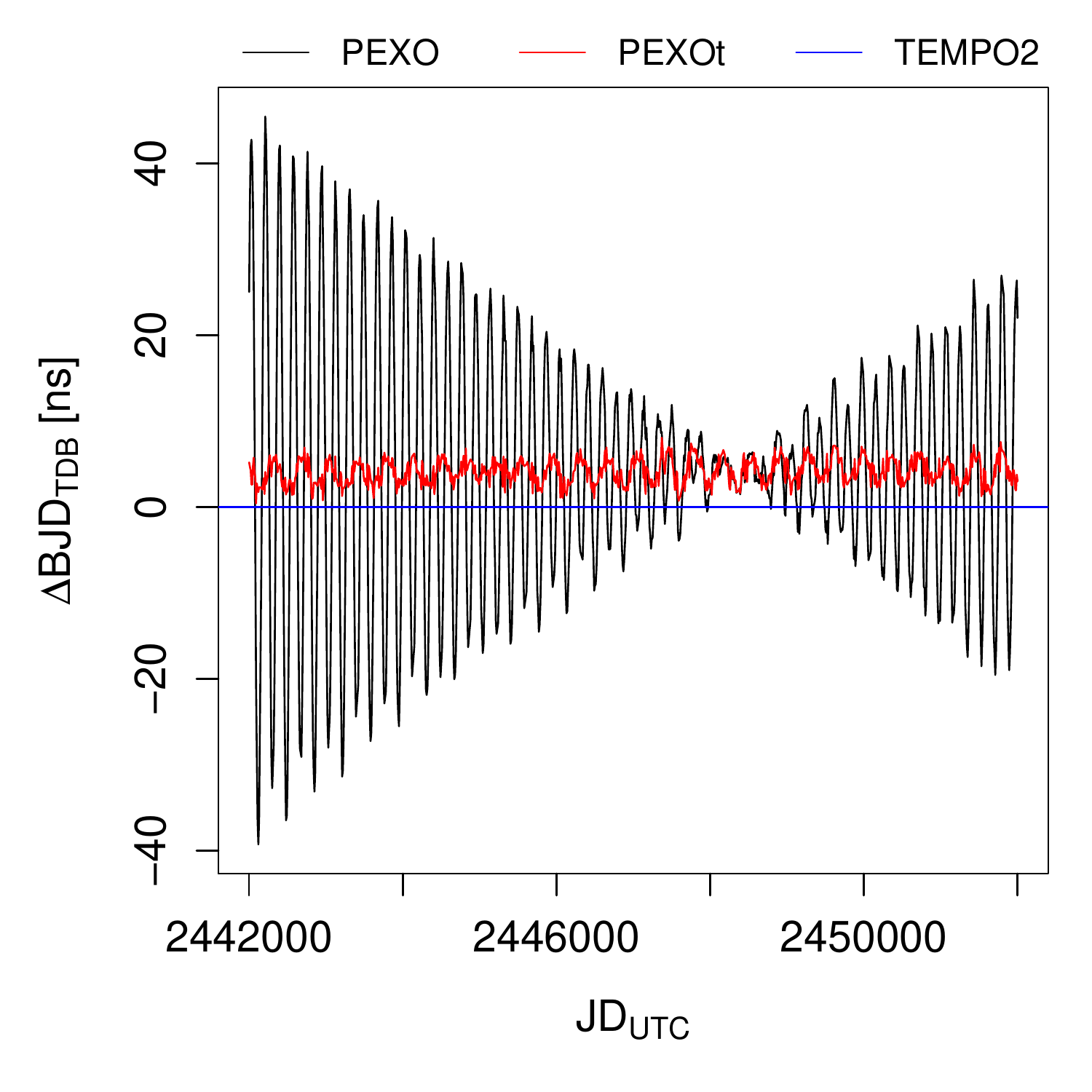}
  \caption{Comparison of $\rm BJD_{TDB}$ calculated by E10, PEXO,
     and TEMPO2. Left panel: $\rm BJD_{TDB}$ modeled by the online
     applet (black) and IDL version (blue) of E10 after subtraction by
     the TEMPO2 values. Considering that the applet does not propagate
     the coordinates of targets, zero proper motion of $\tau$ Ceti is
     assumed. Middle panel: $\rm BJD_{TDB}$ modeled by E10 and PEXO
     with respect to the TEMPO2 values. In this comparison, proper
     motion effects are considered. Right panel: $\rm BJD_{TDB}$
     modeled by PEXO with and without third-and-higher-order Roemer
     delay terms. Because the latter is the approach adopted by TEMPO2,
     we call it ``PEXOt''. Proper motion effects are considered in
     this comparison.} 
  \label{fig:bjd}
\end{figure}

We show the difference in $\rm BJD_{TDB}$ between the online applet
and IDL versions of E10 and TEMPO2 in the left panel of
Fig. \ref{fig:bjd}. The IDL version gives a few $\mu$s timing
  precision, while the applet gives sub-ms precision due to
  its use of double precision to store the unreduced JD. However, the original IDL version
of E10 has an error in the calculation of ``parallax delay''
(equation \ref{eqn:dpS}). In the E10 paper, they correctly add a positive
sign in the parallax delay shown in equation 8 by using ${\bm u}_{\rm
  OT}$ as the reference direction. In the E10 IDL code {\small utc2bjd.pro}, the input R.A. and
decl. are barycentric, and thus the reference unit vector is ${\bm
 u}_{\rm SB}$. However, E10, calculates the total Roemer delay by
adding the parallax delay onto rather than subtracting it from the
first-order Roemer delay. The latter is used to calculate the correct
Roemer delay shown in the left and middle panels of Fig. \ref{fig:bjd}.

To compare PEXO, E10, and TEMPO2 on the same footing, we include all
astrometric parameters for $\tau$ Ceti and use $u_{\rm OT}$ calculated
by PEXO in the IDL version of E10. We compare the three packages in
the middle and right panels of Figure \ref{fig:bjd}. We see that the E10 timing
precision is about 4\,$\mu$s, while the PEXO difference from TEMPO2 is less
than 50\,ns. In the right panel, we compare PEXO with its degraded
version (``PEXOt''), which does not include Roemer delay terms higher
than two orders (see section \ref{sec:astrometry}). The degraded
version differs from TEMPO2 by less than 8\,ns. There is an offset
of $\sim$4\,ns due to the different computation methods of
TDB-TT. Instead of using the FB90 method to derive TDB-TT like TEMPO2,
we use the FB01 method, which is updated and more accurate \citep{petit10}. We also
use DE430t to derive TDB-TT and find a similar offset, suggesting a bias
of a few nanoseconds in the FB90 method. The annual variation in $\Delta {\rm
  BJD}_{\rm TDB}$ suggests that the bias caused by FB90 depends on the
barycentric distance of the geocenter. The minimum difference between PEXO
and TEMPO2 occurs at the reference {\it Hipparcos} epoch. Therefore, third-order geometric terms shown in section \ref{sec:astrometry} are
necessary for a timing model with a precision of $\sim$1\,ns. The failure
to consider this in TEMPO2 might bias its modeling of decade-long pulsar timing data. 

\begin{figure}
  \centering  
  \includegraphics[scale=0.5]{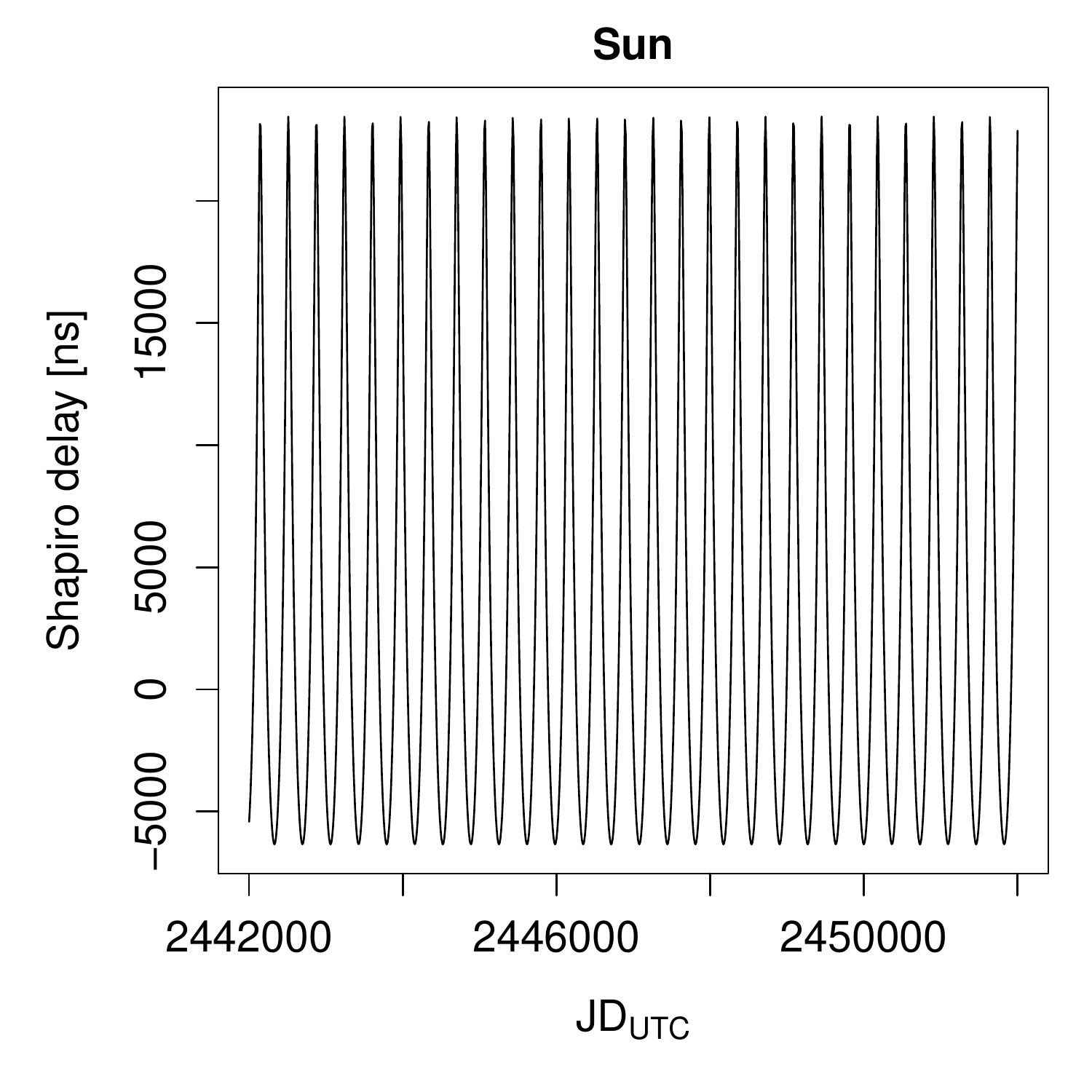}
  \includegraphics[scale=0.5]{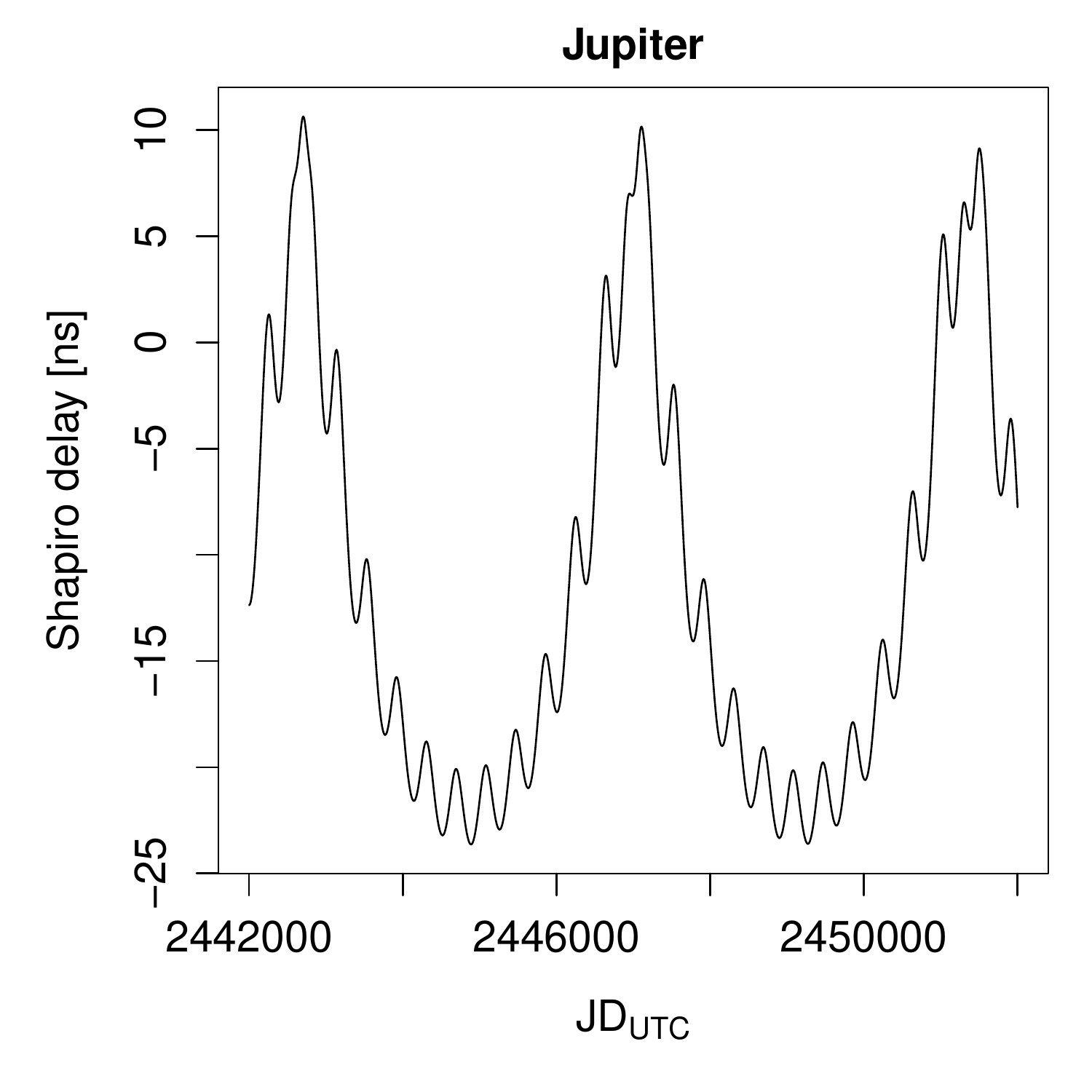}
  \includegraphics[scale=0.5]{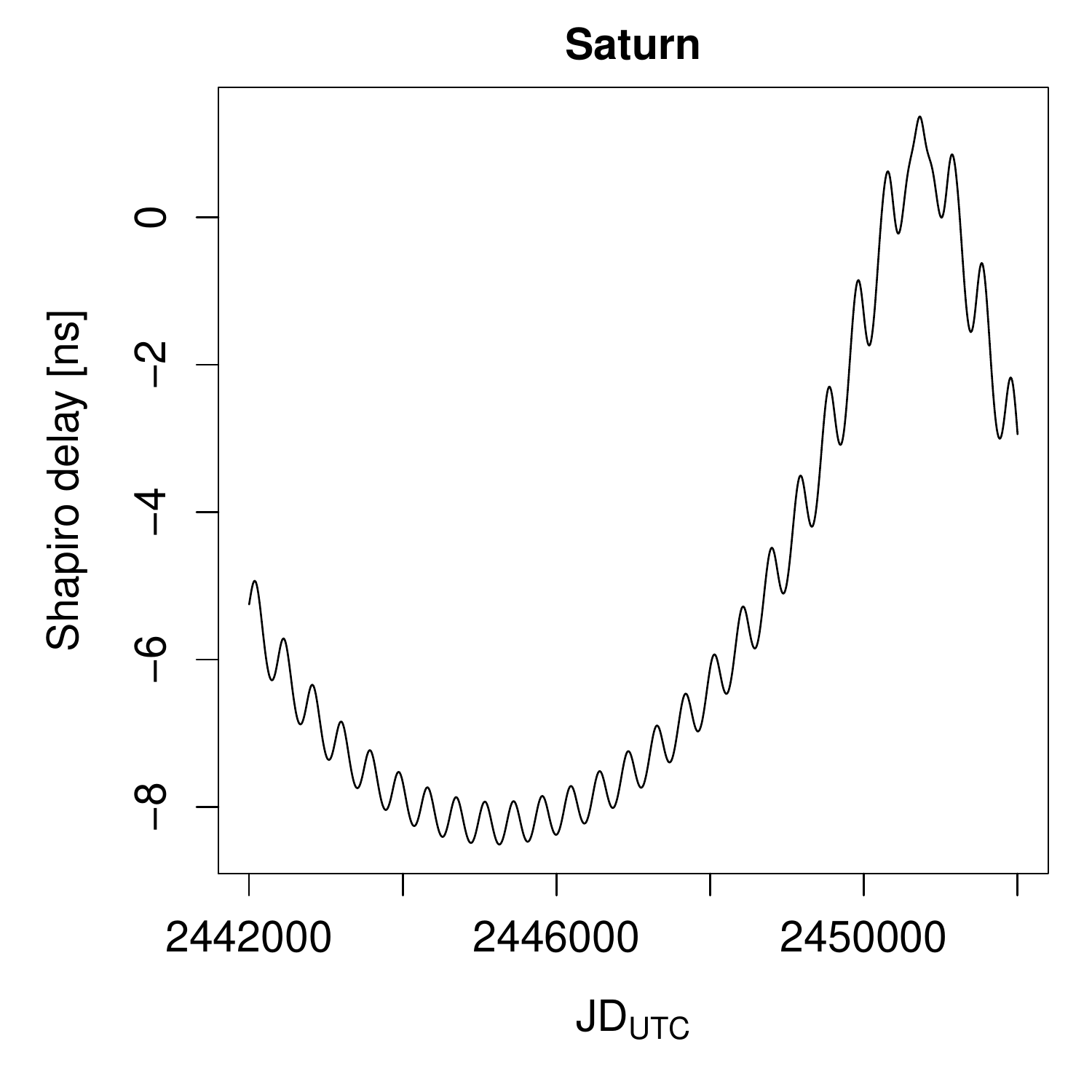}
  \includegraphics[scale=0.5]{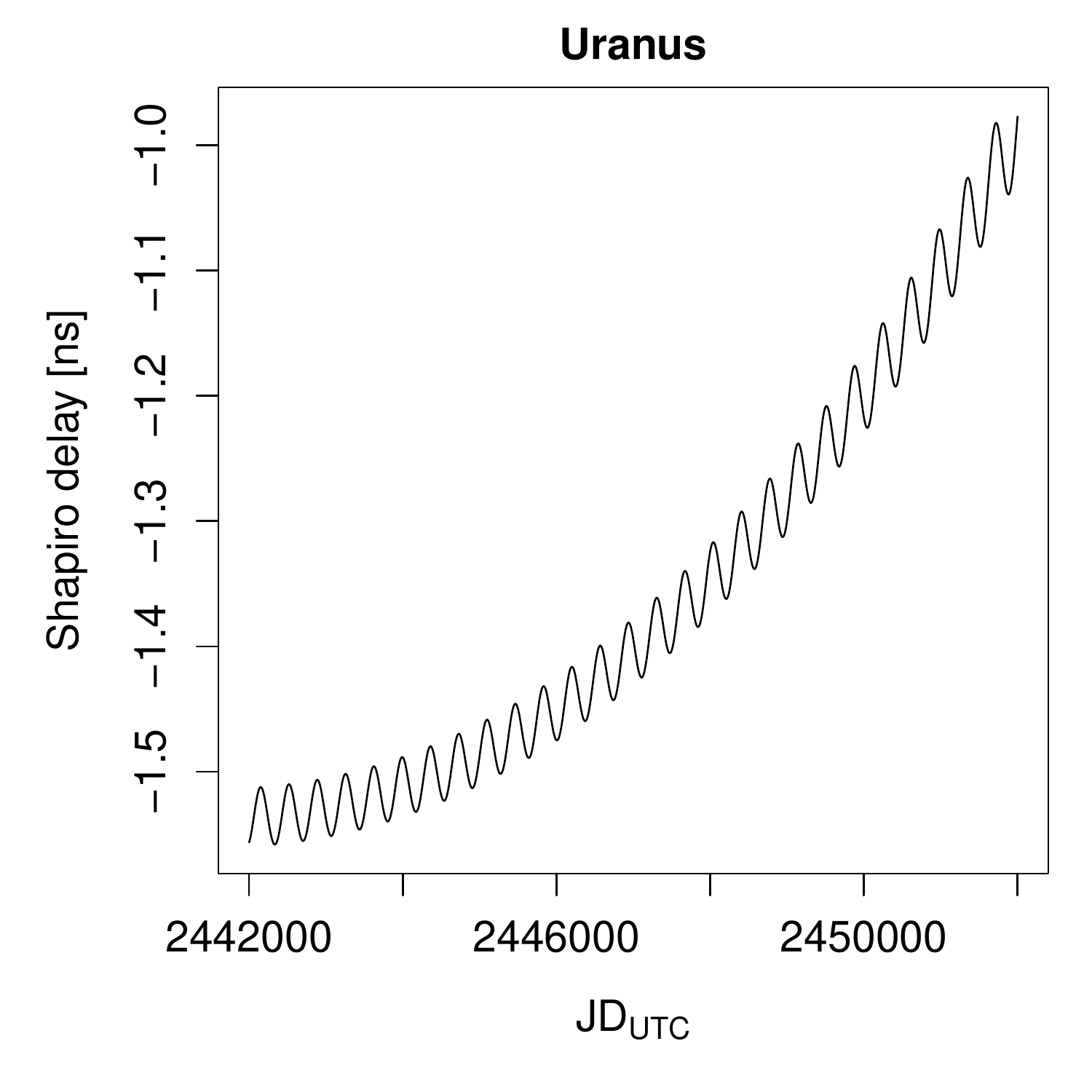}
   \caption{Shapiro delay induced by the Sun, Jupiter, Saturn, and
   Uranus.}
  \label{fig:shapiro}
\end{figure}

In the Fig. \ref{fig:bjd} comparison of the degraded PEXO (i.e. PEXOt) and TEMPO2, we only account for the Shapiro delay due to the Sun because we find a related bug in TEMPO2. The term $1-\cos(\psi)$ in equation \ref{eqn:DSO} is implemented as
  $1+\cos(\psi)$ in the TEMPO2 routine {\small shapiro\_delay.C}. This
  will lead to considerable bias in Shapiro delay caused by the solar
  system planets. We show the Shapiro delays induced by the Sun, Jupiter,
  Saturn, and Uranus in Fig. \ref{fig:shapiro}. The Sun is the dominant
  source of Shapiro delay. Jupiter contributes about 30\,ns to the
  total Shapiro delay and thus is the second important source. Saturn
  and Uranus contribute about 10 and 1.5\,ns, respectively. The other
  solar system planets only induce less than 1\,ns Shapiro
  delay. Therefore, the Shapiro delays due to the Sun, Jupiter, Saturn,
  and Uranus are essential components in the model for $\sim$1\,ns
  timing. The Shapiro delay and lensing effects due to Jupiter and
  Saturn have been detected using very-long-baseline interferometry
  \citep{fomalont03,fomalont09}. In the timing, astrometry, and
    radial velocity models of PEXO, Shapiro or lensing effects of the
    Sun, Mercury, Venus, Earth, Moon, Mars, Jupiter, Saturn, Uranus,
    and Neptune are considered as standard. 

\begin{figure}
  \centering  
  \includegraphics[scale=1]{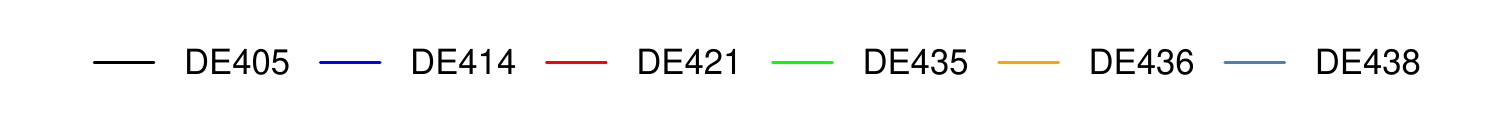}
  \includegraphics[scale=0.38]{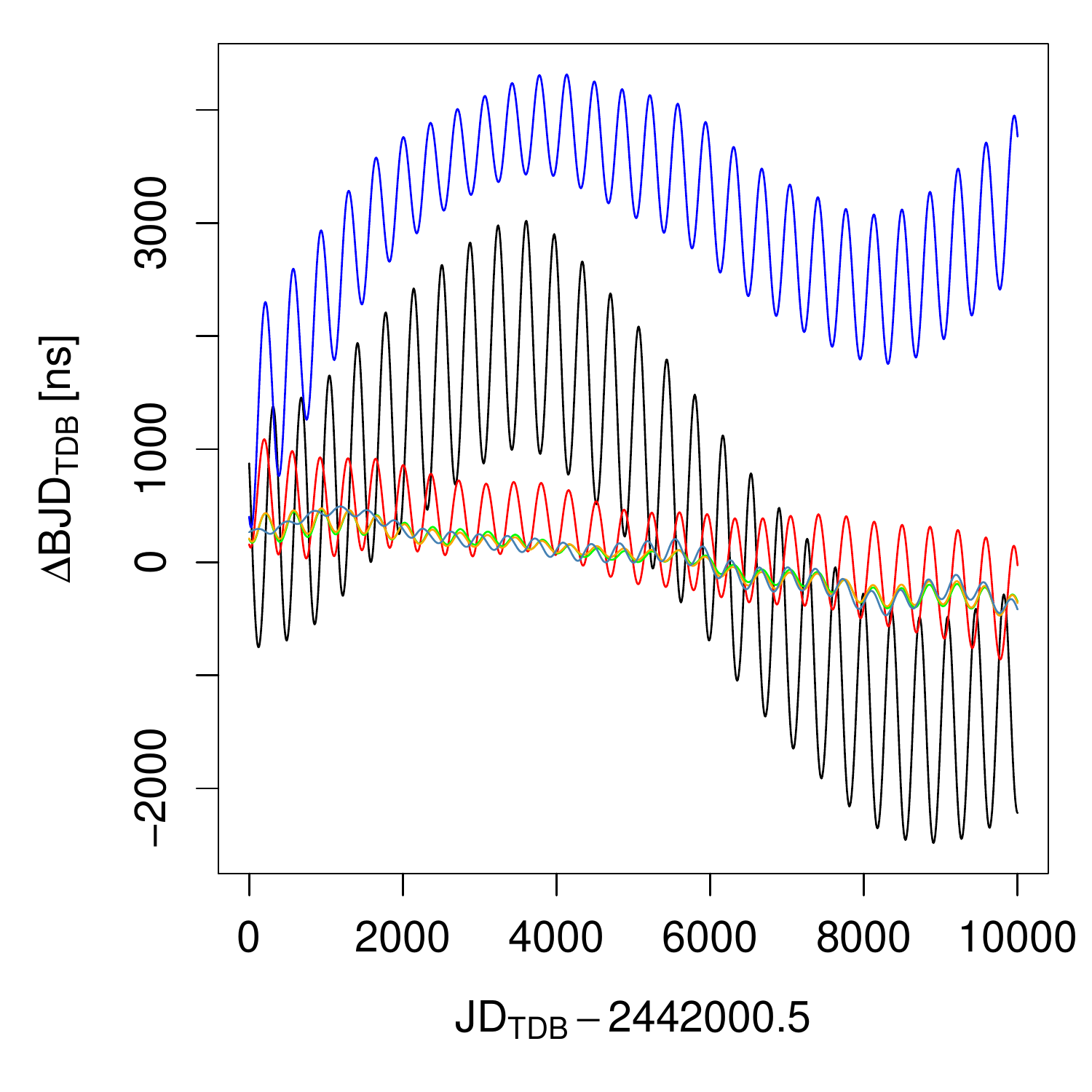}
  \includegraphics[scale=0.38]{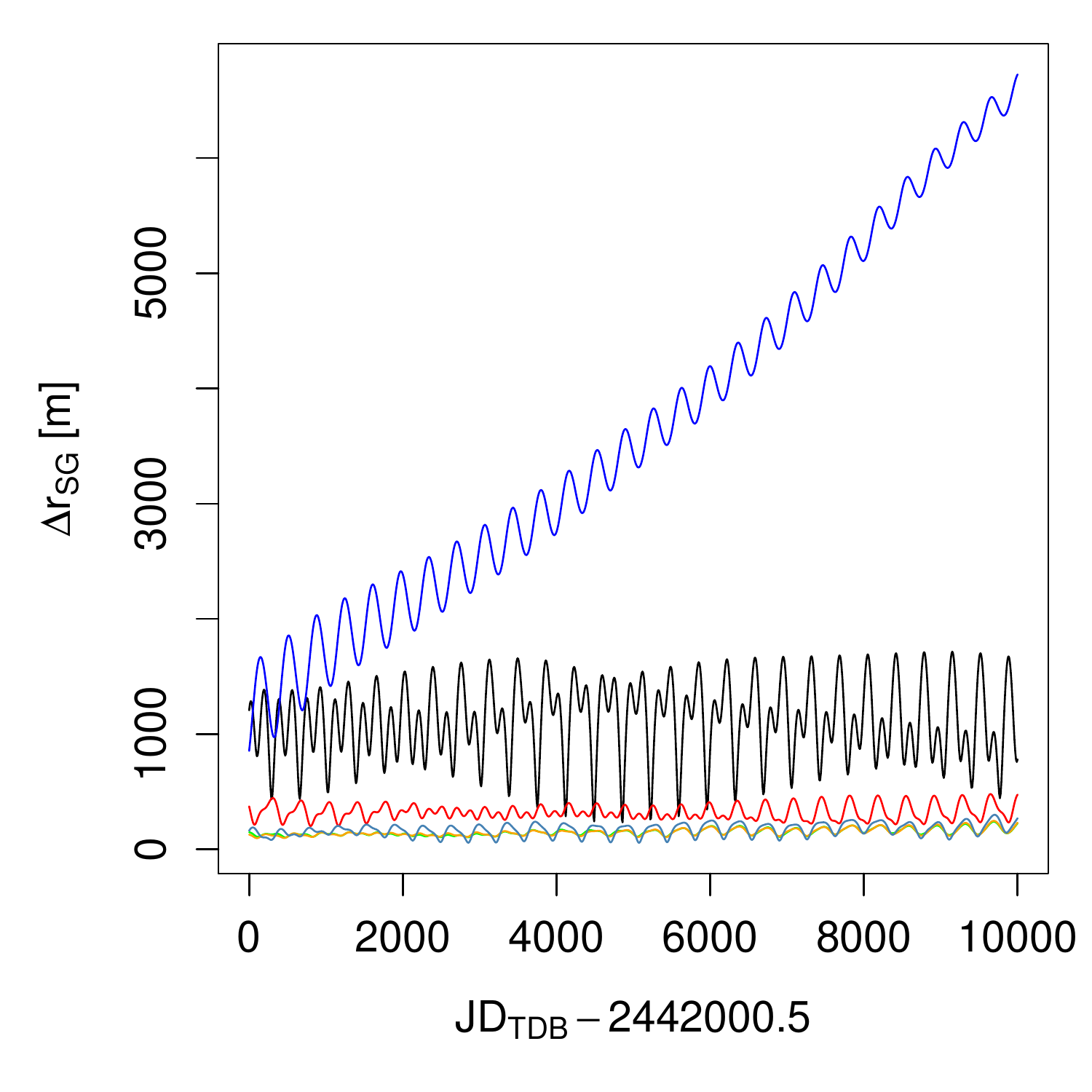}
  \includegraphics[scale=0.38]{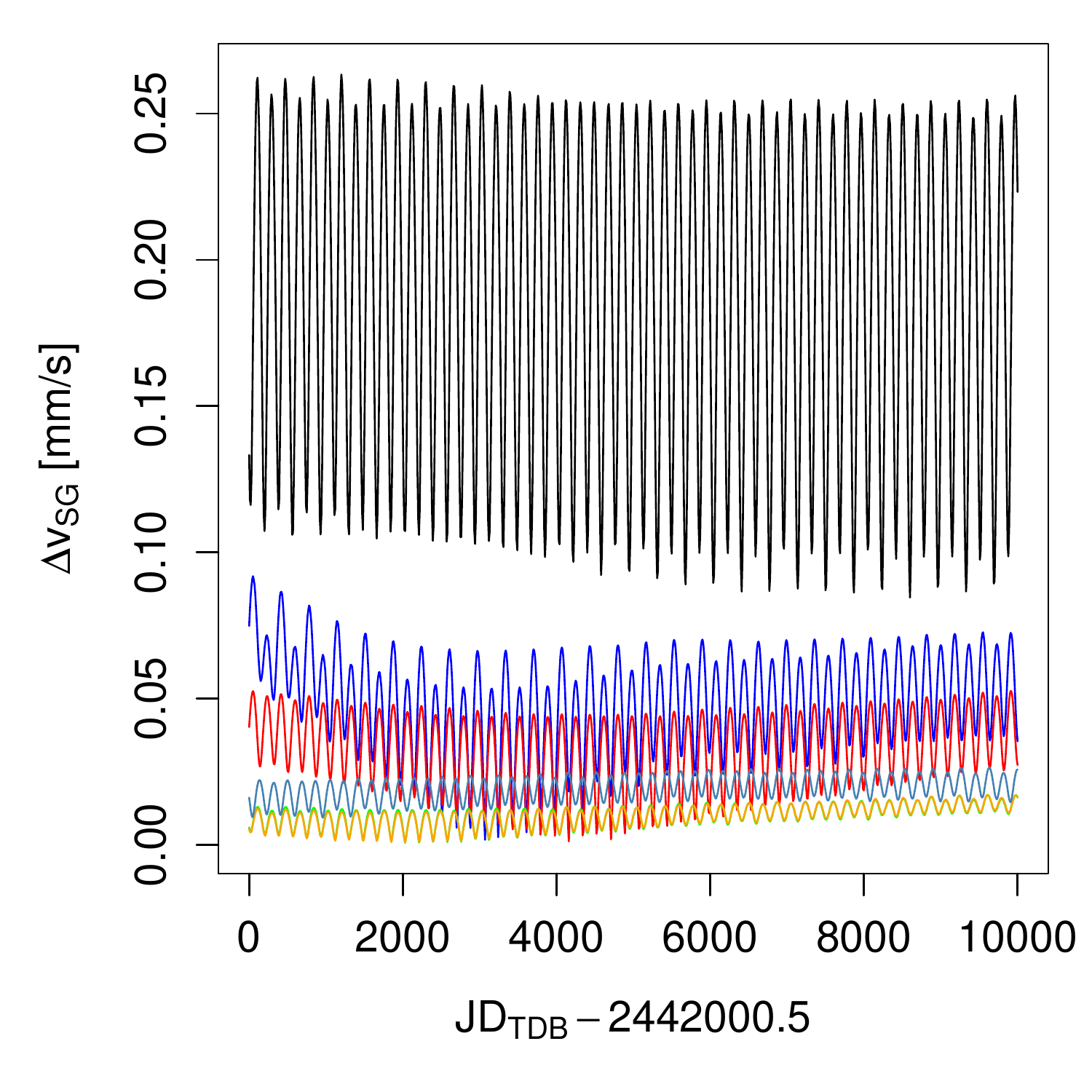}
   \caption{Difference in BJD$_{\rm TDB}$ (left), barycentric position
     (middle), and velocity (right) of the geocenter. }
  \label{fig:ephemeris}
\end{figure}

In the comparison shown in Fig. \ref{fig:bjd}, we modify PEXO to use a
relatively outdated ephemeris, DE405. To assess the significance of
ephemeris difference, in Fig. \ref{fig:ephemeris} we compare the JPL
ephemerides DE405 \citep{standish98}, DE414 \citep{standish06}, DE421
\citep{folkner08}, DE435, DE436 \citep{folkner16}, and DE438 with
DE430 \citep{folker14}. From left to right, the plots of
Fig. \ref{fig:ephemeris} indicate that DE405 and DE414 typically
differ from DE 430 and other recent ephemerides by more than 1000\,ns
in BJD$_{\rm TDB}$ ($\Delta {\rm BJD}_{\rm TDB}$), more than 1\,km in
barycentric position of the geocenter ($r_{\rm SG}$), and more than
0.05\,mm/s in barycentric velocity of the geocenter ($v_{\rm
  SG}$). DE436 and DE438 differ from DE430 by $\Delta {\rm BJD}_{\rm
  TDB}\sim 400$\,ns, $r_{\rm SG}\sim 200$\,m, $v_{\rm SG}\sim
0.02$\,mm/s. In contrast, DE435 and DE436 differ from each other by $\Delta {\rm BJD}_{\rm TDB}\sim 35$\,ns,
  $r_{\rm SG}\sim 20$\,m, $v_{\rm SG}\sim 0.0005$\,mm/s. These stated
  differences are only a guideline as they are based on average differences and constitute an annual variation superposed on the trend due to perspective change.

  The significant difference between DE405 and other ephemerides has been studied
  frequently (e.g., \citealt{viswanathan17,wang17}). The precision of
  an ephemeris is determined by the quality of the solar system model,
  as well as the amount of data available when the ephemeris was computed and fit. Thus we
  encourage a use of the most recent ephemeris if high-precision
  timing data are analyzed. For the ephemeris of DE430 and more recent ones,
  we expect a timing precision of about 100\,ns, a positional
  precision of about 100\,m for the geocenter, and a velocity precision of about
  0.01\,mm/s. Thus the timing precision of both PEXO and TEMPO2 is mainly limited by
  the solar system ephemeris. Potential signals in precise timing data
  should be analyzed with various ephemerides to confirm, as done by
  the team of the north
  American Nanohertz Observatory for Gravitational Waves (NANOGrav; \citealt{arzoumanian18}) to constrain the gravitational-wave background. 

\begin{figure}
  \centering  
  \includegraphics[scale=0.38]{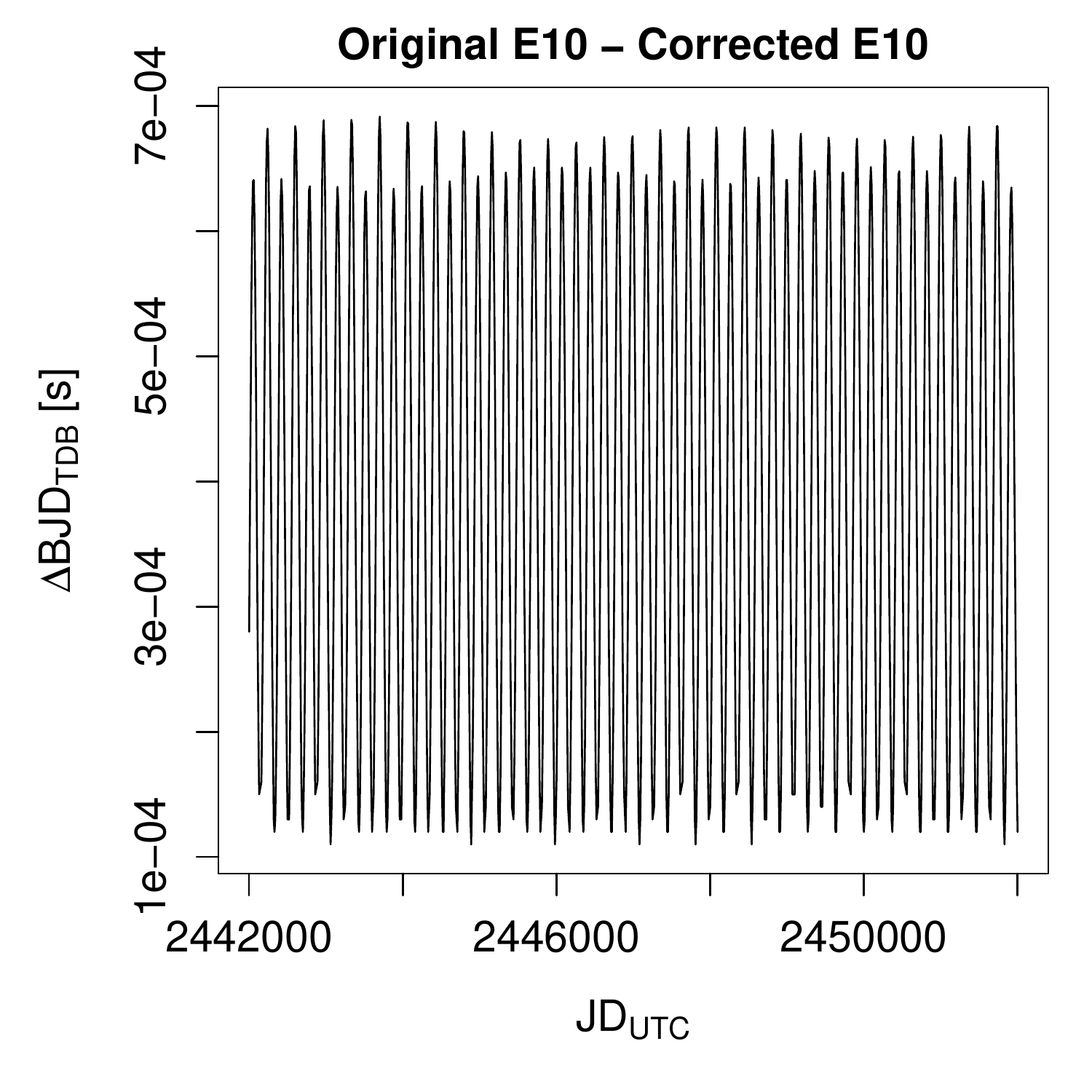}
  \includegraphics[scale=0.38]{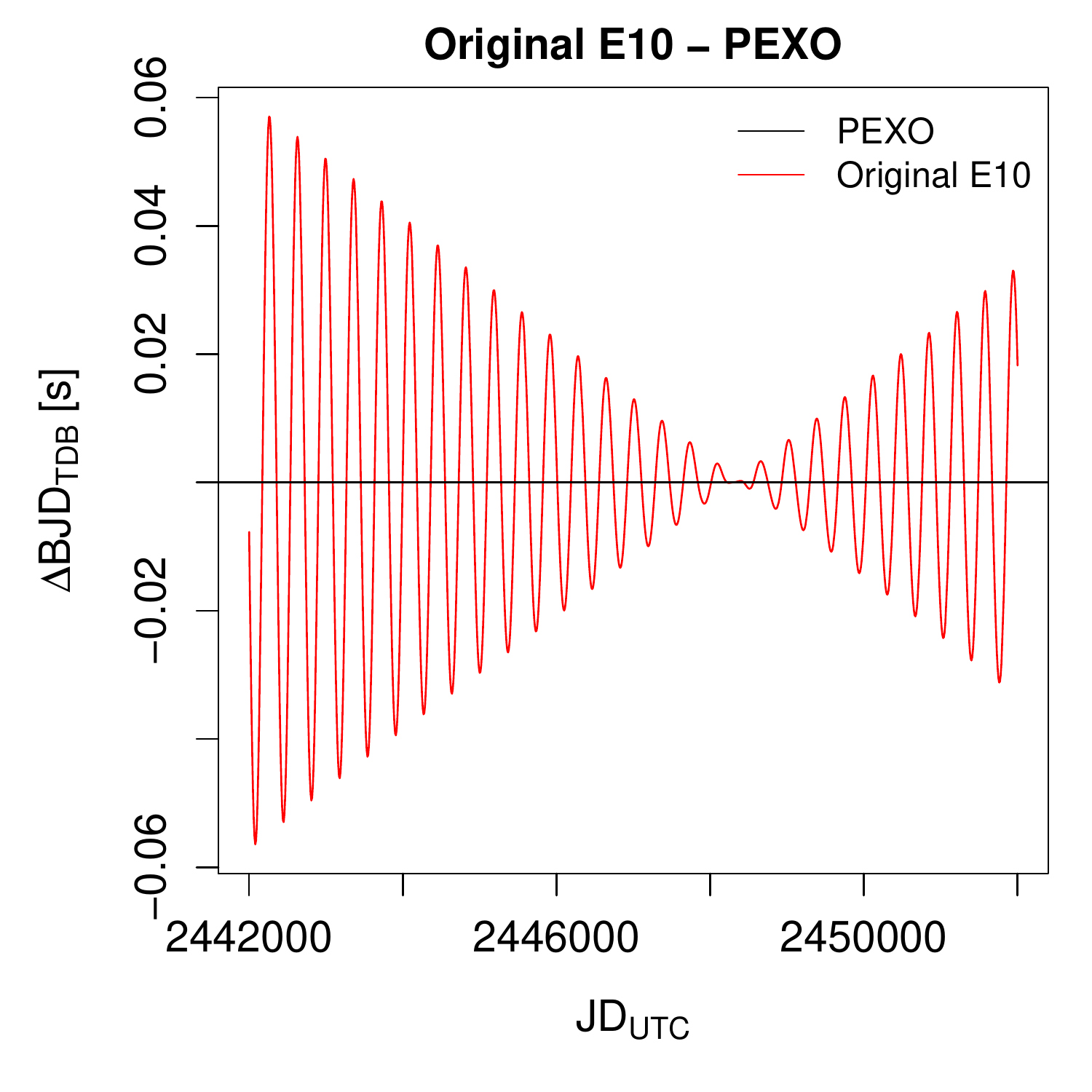}
    \includegraphics[scale=0.38]{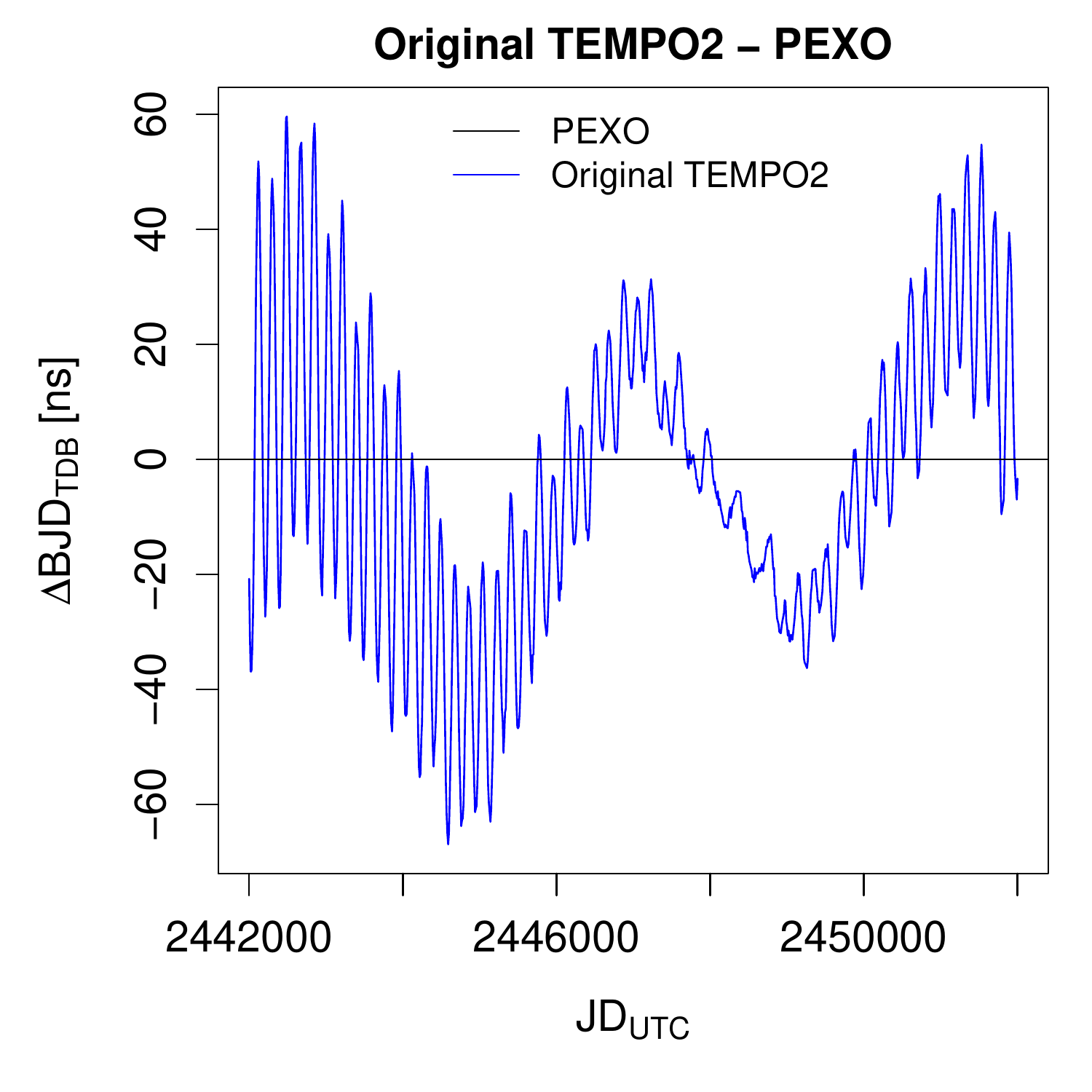}
   \caption{Comparison of the timing precision of E10 and TEMPO2 with
     PEXO for $\tau$ Ceti. }
  \label{fig:pexo}
\end{figure}

To explore precision limits of E10 and TEMPO2 relative to PEXO, we
apply them to the nearby star $\tau$ Ceti in
Fig. \ref{fig:pexo}. Considering the lack of third-order Roemer delays
and coding errors in TEMPO2, we consider PEXO as the package with the
highest precision and compare E10 and TEMPO2 to it in order to explore
the precision limit of these well-known packages. We use the Earth
rotation model recommended by IAU2006 resolutions \citep{capitaine06,wallace06} and the DE430 ephemeris of JPL as well
as DE430t to derive TT-TDB. First, we compare the original {\small
  utc2bjd.pro} routine without correcting the error of ``parallax delay'' with PEXO. We show the results in the left panel of
Fig. \ref{fig:pexo}. As seen from the left panel, the coding error in
E10 leads to about 0.1\,ms bias. The ignorance of proper motion leads
to about 0.06\,s timing bias for $\tau$ Ceti over the 30\,yr time
span (see middle panel of Fig. \ref{fig:pexo}). This timing bias will
lead to about 2\,mm/s radial velocity bias. This bias could be
significant for the analysis of data with high timing resolution, such
as fast radio bursts with millisecond resolution (e.g.,
\citealt{chime19}). Moreover, proper-motion-induced timing bias is comparable with relativistic precession for
some systems and thus needs to be modeled in order to detect
relativistic effects in timing data. As seen in the right panel of
Fig. \ref{fig:pexo}, TEMPO2 and PEXO with the same Earth rotation
model and with the DE430 ephemeris are similar at the level
of tens of nanoseconds. The original TEMPO2 (with coding error for planetary
Shapiro delays) deviates from PEXO due to a combined effects of third-order Roemer delays (see the right panel of Fig. \ref{fig:bjd}) and
planet Shapiro delays (see Fig. \ref{fig:shapiro}). For distant
pulsars, the third-order Roemer delays are not significant, although the
coding error in the calculation of planet Shapiro delays still biases
TEMPO2 timing by tens of nanoseconds. Therefore the original TEMPO2 has a
timing precision of a few tens of nanoseconds for decade-long
observations, and the original E10 has a timing precision of subseconds.  

\subsection{Radial velocity}{\label{sec:rv}}
We compare the precision of radial velocity modeling by PEXO and
TEMPO2 by calculating the so-called barycentric correction term
$z_{\rm S}$ (see equation \ref{eqn:zb}). Unlike \cite{wright14}, we do
not compare the Doppler shift of pulse frequency with $z_{\rm T}$
because pulse frequency is influenced by aberration delay (E06). We
instead calculate Doppler shift numerically using
\begin{equation}
  1+z_{\rm S}=\frac{\delta \tau_a^{\rm SSB}}{\delta\tau_o}~,
\end{equation}
where $\tau_a^{\rm SSB}$ is $\rm BJD_{TDB}$ and $\tau_o$ is $\rm
JD_{\rm UTC}$. Because the analytical value of the local Doppler shift
$z_{\rm S}$ is not given in TEMPO2, we calculate it numerically by using
${\rm z_{bary}=(BJD_{\rm TDB2}-BJD_{\rm TDB1})/(JD_{\rm UTC2}-JD_{\rm UTC1})-1}$, where UTC2,
UTC1 are separated by 0.02\,day and TDB2 and TDB1 are corresponding TDB times. This UTC time step is chosen such that the rounding error
for both PEXO and TEMPO2 could be as small as possible. However, such
a numerical treatment is only used for comparison. We use the analytical radial velocity model in equation \ref{eqn:vr_obs} for the application of PEXO.

\begin{figure}
  \centering
  \includegraphics[scale=0.4]{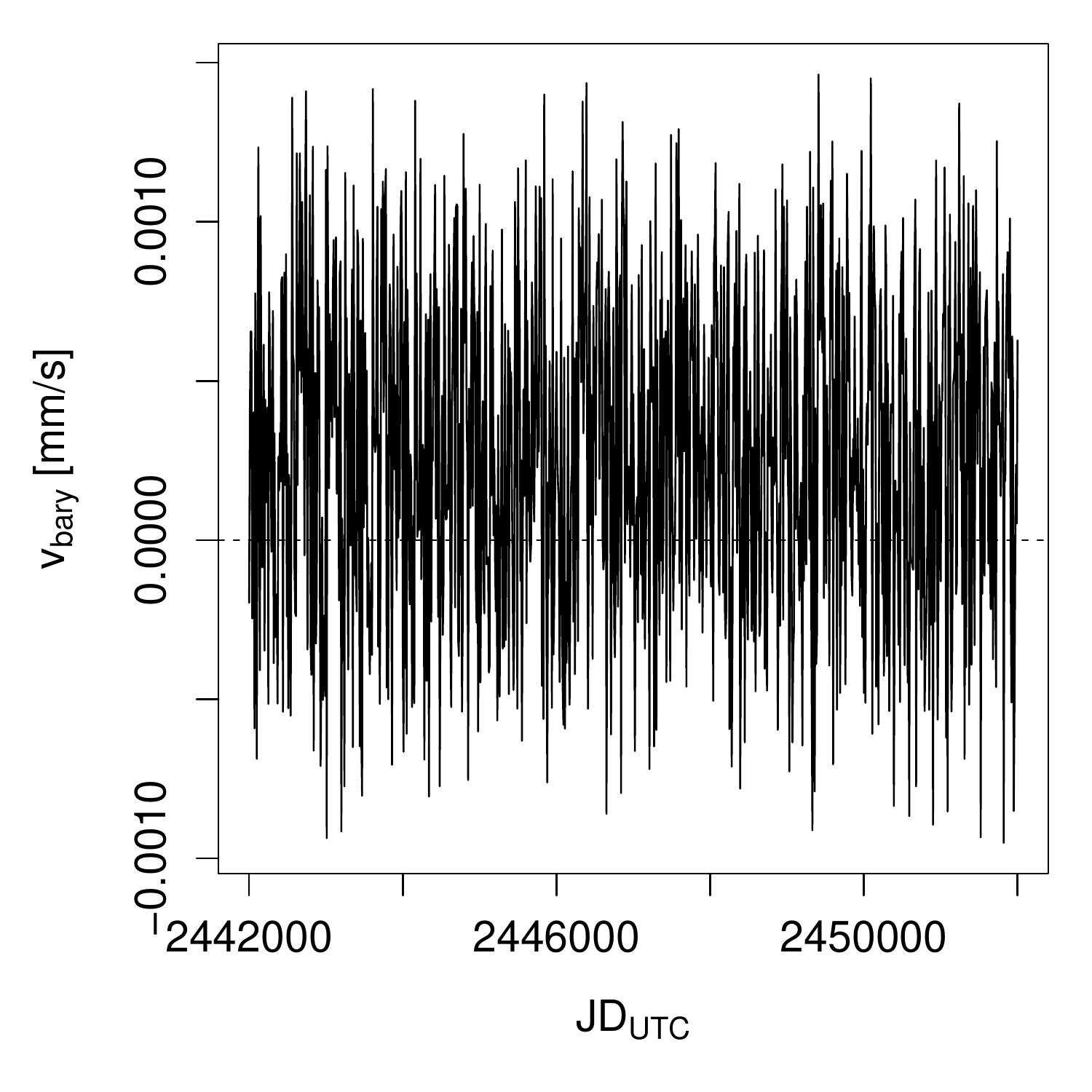}
  \includegraphics[scale=0.4]{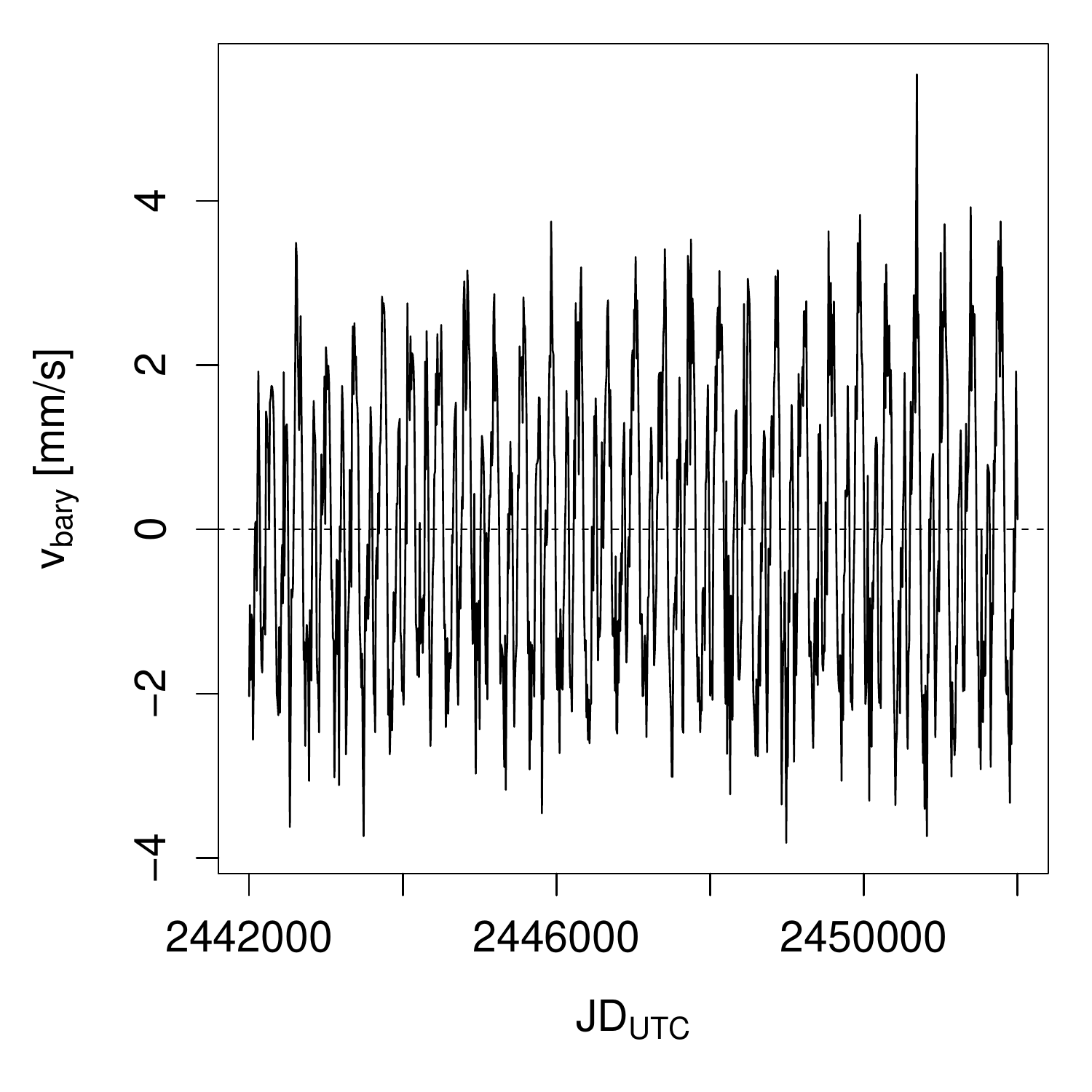}
  \caption{Difference of barycentric correction radial velocity term
    calculated by PEXO and TEMPO2. The left panel shows the
      numerical comparison while the right one shows the analytical
      comparison. }
  \label{fig:rv}
\end{figure}

We take $\tau$ Ceti as a test case and calculate the local Doppler
shift $z_{\rm S}$ numerically for PEXO and TEMPO2. We show the
difference in the corresponding radial velocities over 5\,yr in
Fig. \ref{fig:rv}.
We see that the PEXO radial velocities deviate from the TEMPO2 values with a
peak-to-peak difference of 2\,$\mu$m/s, indicating a radial velocity precision comparable to TEMPO2.

We also compare the analytical PEXO and TEMPO2 models of barycentric
radial velocity defined in equation 28 of \cite{wright14} and show the
results in the right panel of Fig. \ref{fig:rv}. The main error seems to arise from
the annual motion of the Earth, which is projected onto the source
direction to derive kinematic Doppler shift. Because we use DE430
both for TEMPO2 and for PEXO, the annual variation might be caused by
uncertainty in the Earth's rotation model and in the numerical calculation of
TDB-TT. Readers are referred to \cite{kopeikin99b} for a rigorous
treatment of higher-order relativistic Doppler effects.

Therefore, PEXO's radial velocity precision is comfortably beyond the specification of the
current best radial velocity instruments such as ESPRESSO (about a few
cm/s). However, a precision at the mm/s level is only achievable with an
ideal treatment of the atmospheric chromatic aberration
\citep{wright14}. The ESPRESSO instrument's exposure meter provides
three spectral channels, so in time the identification of possible
chromatic effects on exposure time midpoints for radial velocity
measurements can be evaluated. A precision of 1\,mm/s is also
challenged by the acceleration of the barycenters of the target system
and the solar system. The acceleration of the SSB is about 1\,mm/s/yr,
leading to 1\,cm/s bias in the radial velocity model prediction over
one decade. This could provide an opportunity to use radial velocity
data to estimate the acceleration of stars and thus provide
measurements relevant to the Galactic potential analogous to ongoing
efforts to quantify cosmological variations in the fine-structure constant (e.g., \citealt{whitmore14}).

\section{PEXO Simulation of Relativistic Effects in extrasolar systems}\label{sec:effects}
 In this section, we assess various relativistic effects in transit timing,
astrometry, and radial velocity models through comparison of PEXO simulations and real data for example systems. These tests are aimed at roughly assessing the precision of PEXO more than detecting relativistic effects in real data. 
% We also compare our results with analytical solutions to verify the precision of PEXO.
\subsection{Transit Timing}\label{sec:ttv}
TTV \citep{miralda02,holman05} is an
efficient method of constraining the mass and orbital parameters of
transiting planets such as the TRAPPIST-1 system
\citep{gillon16,grimm18}. However, relativistic effects are typically
ignored to simplify the TTV modeling because of their small effects,
such as in {\small EXOFAST} \citep{eastman13}, although these effects
could be detectable with decade-long observations of some systems
\citep{miralda02,jordan08}. To assess the importance of relativistic
effects on transit timing, we use XO-3 b
  \citep{johns-krull08}, a transiting hot Jupiter, as an example
  because it is a hot Jupiter on an eccentric orbit and is also
  recommended by \cite{jordan08} for searching for relativistic precession.

%First, we estimate the GR effects by comparing the
%classical Kepler timing model and the DDGR timing model. To determine
%the Keplerian orbits, we use the orbital parameters for TRAPPIST-1 b
%inferred through TTV by \cite{grimm18}. We use the stellar parameters
%from Gaia DR2 \citep{brown18} and fix the longitude of ascending node
%$\Omega$ at zero. We ignore the perturbations from other planets to
%focus on the relativistic effects. We use equations
%\ref{eqn:omdot}-\ref{eqn:Pdot} and equations
%\ref{eqn:drT-pT}-\ref{eqn:DST} to derive the relativistic time delay
%due to planet motion. For classical Kepler motion, we only consider the Roemer delay according to equation \ref{eqn:drT}.

\begin{figure}
    \centering
    \includegraphics[scale=0.8]{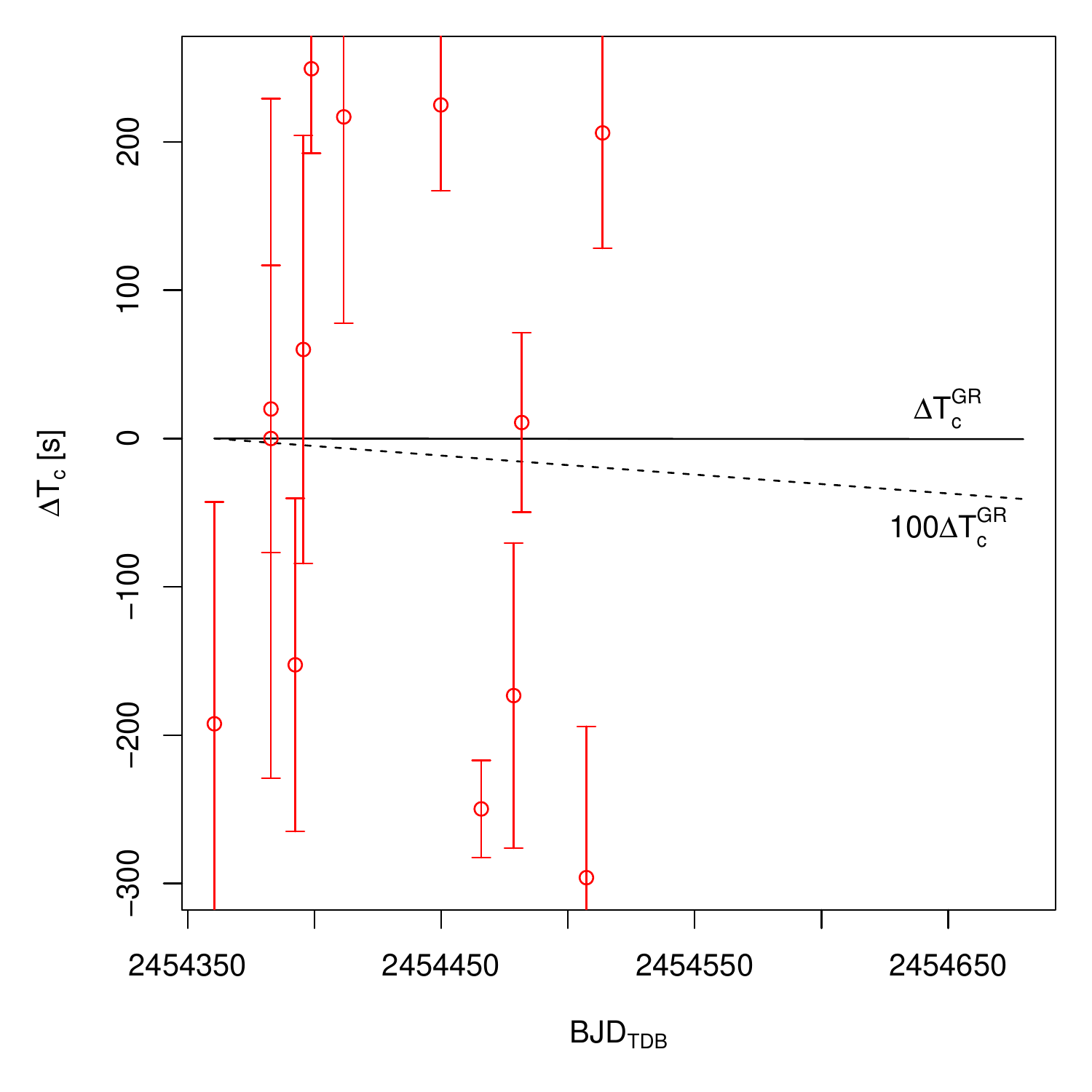}
    \caption{Comparison of the relativistic TTV ($\Delta T_c^{\rm
        GR}$; black solid line) and observed TTV (red circles with
      error bars). To visualize the small relativistic TTV signal, we
      show the amplified relativistic TTV prediction for $100 \Delta T_c^{\rm
        GR}$. }
    \label{fig:xo3}
  \end{figure}

XO-3 b has a mass of 11.79\,$M_{\rm Jup}$ and an
  orbital period of 3.1915239$\pm$0.00023 days
  \citep{johns-krull08,winn08}. We simulate the system over 100
  orbital periods using PEXO, calculate the transit epoch for each orbit, and compare the simulated relativistic TTV with the observed transit timing data
  \citep{winn08} in Fig. \ref{fig:xo3}. The period of primary transits is changed by about
  0.4\,s over 100 orbits (or about 319 days) due to relativistic
  precession. This corresponds to a time derivative of transit period
  of $7.98\times 10^{-5}$, consistent with the prediction using 
  equation 21 of \cite{miralda02}. Such a signal might be
  detectable in transit timing measurements with a precision of about
  one minute for each over one decade. Although such a detection of
  relativistic TTV is not impossible with
  current instruments, it is not as efficient as other methods such as
  Transit Duration Variation (TDV) and the variation of time between primary and
  secondary transit (PSV). This is because TTV is proportional to
  $\dot{\omega}_{\rm GR}^2$  while TDV and PSV are proportional to
  $\dot{\omega}_{\rm GR}$ \citep{miralda02}. 

\begin{figure}
  \centering
  \includegraphics[scale=0.8]{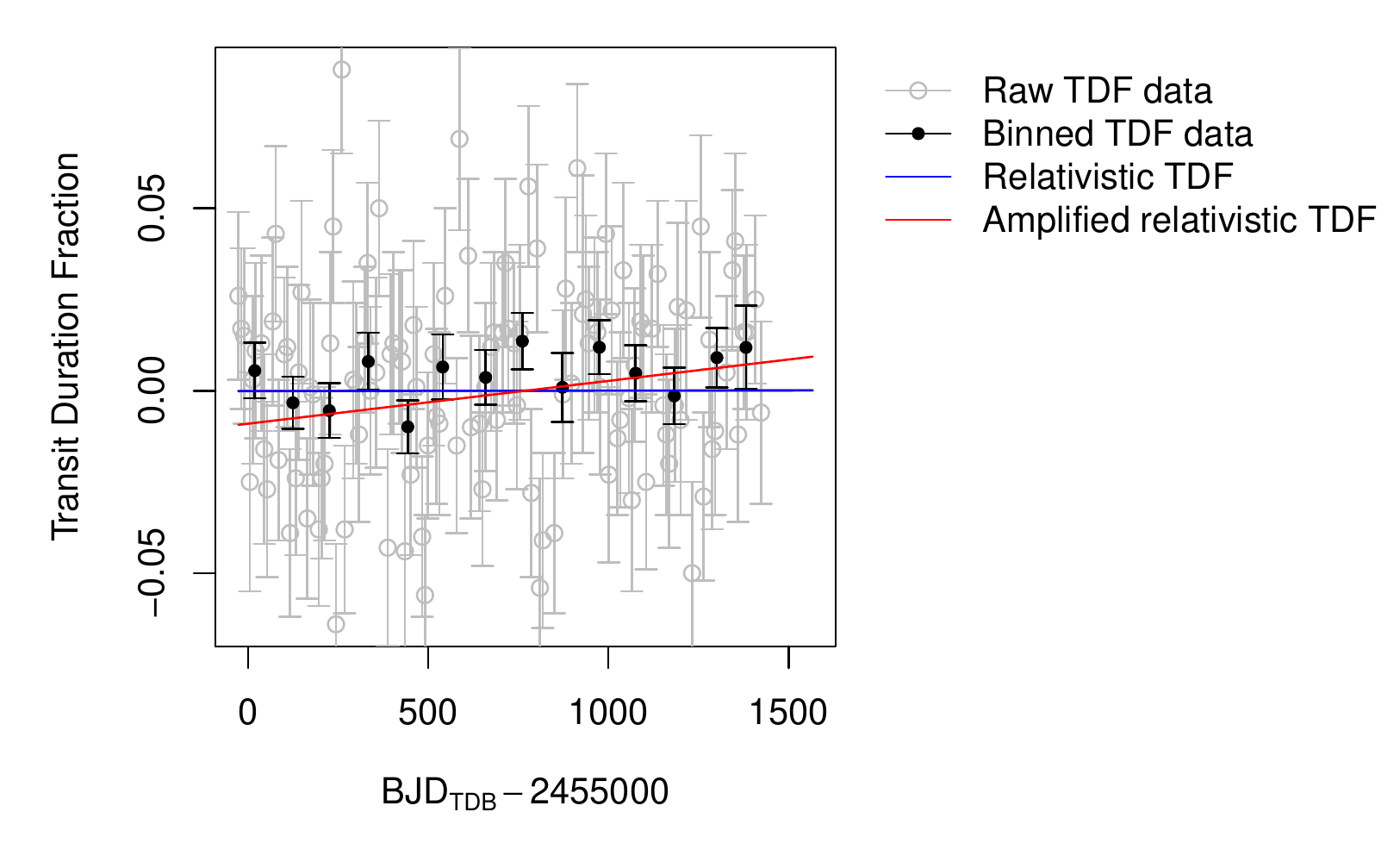}
  \caption{TDF for Kepler-210 over 200 orbits or
  1500 days. The blue line denotes the relativistic TDF, while the red
line represents the relativistic TDF multiplied by 100. The gray error
bars show the raw TDF data, while the black ones are the binned data
with a bin width of 100 days.}
  \label{fig:tdf}
  \end{figure}

  Because the duration for each transit of XO-3 b is not provided in
  previous studies, we select Kepler-210 c \citep{ioannidis14} to assess the
  TDV effect. This planet has an orbital period of 7.9725 days and an
  eccentricity of 0.5 and thus has significant relativistic
  precession. We use the transit duration epoch data provided by
  \cite{holczer16} to compare to PEXO predictions. Treating
  the planet as the target and the star as the companion, we simulate the
  system over 200 orbits and calculate
  the velocity $v_{\rm TC}$. We derive the transit duration ${\rm
    TD}=2\sqrt{(R_{\rm star}+R_{\rm pl})^2-(bR_{\rm star})^2}/v_{\rm
    TC}$, where $b$ is the impact parameter, and $R_{\rm star}$ and $R_{\rm pl}$ are
  respectively the radii of Kepler-210 and Kepler-210 c. Because
  \cite{holczer16} only provide the epoch data for Transit Duration Fraction (${\rm TDF}=({\rm TD}-\overline{\rm TD})/\overline{\rm TD}$), we derive TDF from
$v_{\rm TC}$ using ${\rm TDF}=(\overline{v}_{\rm TC}-v_{\rm
  TC})/v_{\rm TC}$ assuming the impact parameter does not change over
time. We show the TDF data and the PEXO prediction in
Fig. \ref{fig:tdf}. To visualize the relativistic TDF properly, we
also show an amplified TDF in the figure. The relativistic TDF
changes by $1.8\times 10^{-4}$ over 1500 days, equivalent to 1.93\,s
variation in transit duration. This is consistent with the prediction using
equation 23 of \cite{miralda02} or equation 15 of
\cite{jordan08}. Considering that the mean uncertainty of transit
durations is about 4\,min and the relativistic precession is about
$\dot{\omega}_{\rm GR}=0^\circ.61$/century, the relativistic TDF is not
detectable with the current {\it Kepler} data for this system. However, such
an effect becomes detectable if high-precession systems are observed
with high cadence ($<10$\,s) by space- or ground-based telescopes (e.g. \citealt{ivanov11}). 

 % The mean uncertainties of the raw and binned TDF data are 0.023 and
%$8\times 10^{-3}$, respectively. Thus it requires thousands of
%transit timings with such precision to detect relativistic TDF. We
%need at least two decades of continuous observations of transits by at least 10 instruments simultaneously to detect such a small effect.
%However, the TDF due to the changing perspective caused by the annual
%motion of the observer has an amplitude $2.5\times 10^{-3}$ and thus
%is at least one order of magnitude more significant than the
%relativistic TDF. Because this type of TDF is caused by the annual
%motion of the observer around the Sun, we call it ``parallax
%TDF''. Considering that Kepler-210 is about 234\,pc away from the Sun
%\citep{brown18}, we expect stronger parallax TDF for nearby transit
%systems, which are the main targets of the TESS mission \citep{ricker14}. 

In summary, the relativistic precession is unlikely to be detectable
in current transit timing data due to the large timing uncertainty caused
by low-cadence observations. For transit systems such as XO-3, the
relativistic precession is larger than 1 degree per century and is thus
detectable with high-cadence observations with exposure time as small
as a few seconds. 

\subsection{Astrometry}\label{sec:astrometry_binary}
PEXO is similar to the GREM package \citep{klioner03,lindegren18} used
by {\it Gaia} to model single stars. For binaries or stars hosting massive
companions, PEXO also accounts for the gravitational lensing caused by
companions in the target system. Additionally, PEXO models the
  atmospheric refraction in order to account for the differential
  refraction effects in ground-based direct imaging. For decade-long astrometry
data, the parameter uncertainty in an astrometry catalog would lead to
significant deviation of model prediction from the real position. For
example, a proper motion error of 1\,mas/yr would result in 100\,mas position
error over one century. Hence, the astrometry parameters should be
determined {\it a posteriori} in combination with the orbital
parameters of companions through Markov chain Monte Carlo (MCMC) posterior sampling. This is also
an approach adopted by pulsar timing and is more suitable for
precision exoplanet research than the traditional approach, which separates
the barycentric correction from the motions in the target system.

\begin{figure}
  \centering
    \includegraphics[scale=0.6]{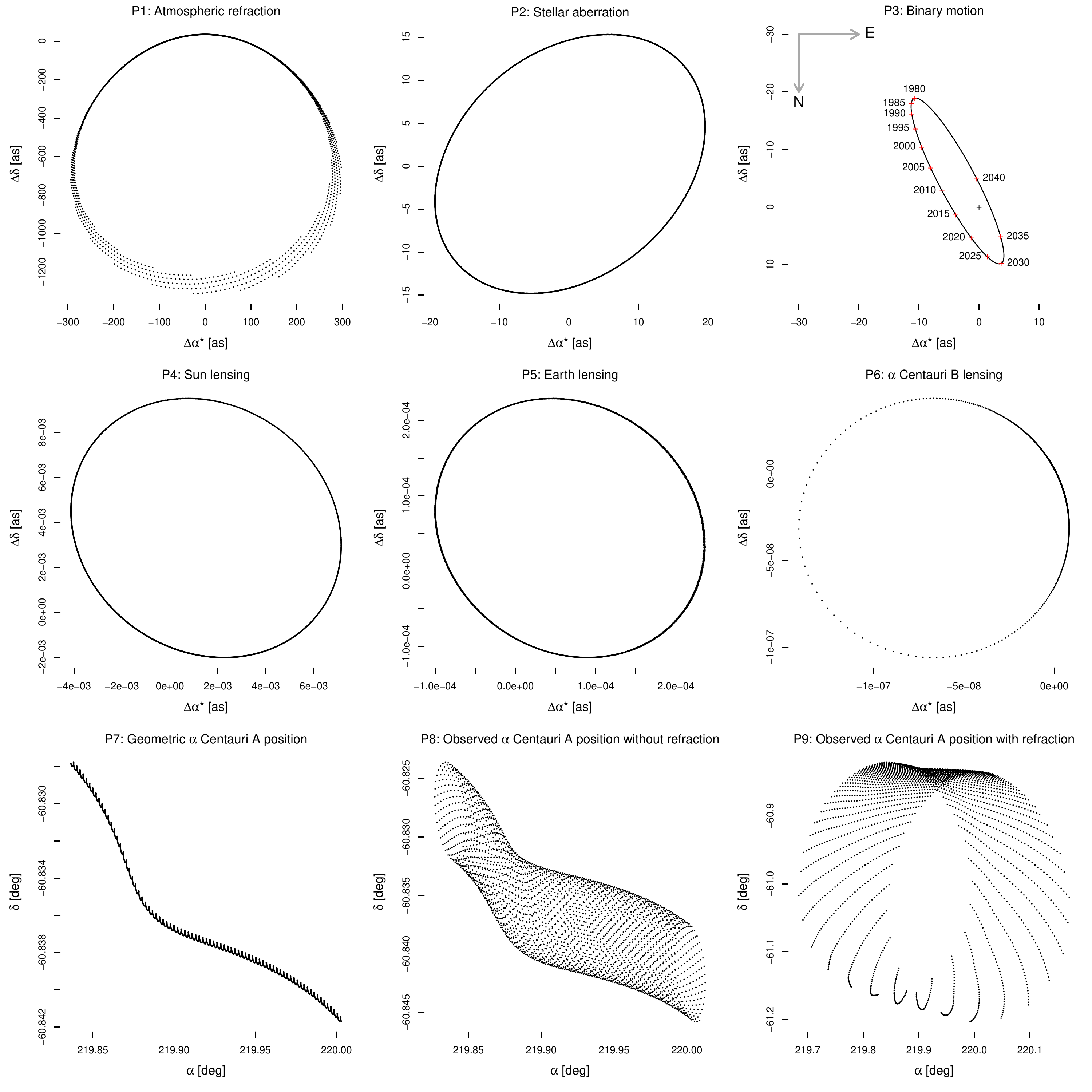}
  \caption{Various effects on the observed position of $\alpha$
    Centauri A over an orbital period. The linear proper motion
      effect is easy to comprehend and thus is not shown here. The
      panels are denoted by ``Pn'' where n is the panel number. The
      names of effects are denoted by the panel titles. For P1 to P6,
      the panels are ordered in the order of decreasing
      significance. The plots P7 to P9 in the bottom row show the absolute
      position of $\alpha$ Centauri A without relativistic and
      atmospheric refraction  effects, without refraction effect, and
      with all effects, respectively. The binary orbit in P3
        shows the motion of $\alpha$ Centauri B with respect to A. It is scaled to be comparable with figure 1 in \cite{pourbaix99}. The north (N) and east directions are shown by arrows, and epochs are denoted by red crosses. The black cross represents $\alpha$ Centauri A.} 
  \label{fig:ac}
\end{figure}

To compare the various effects on the measured astrometry, we consider the
nearest binary system $\alpha$ Centauri as an example. We use
the CTIO observatory as an example observatory site. The
orbital and astrometry parameters of $\alpha$ Centauri
A and B are determined by \cite{kervella16} through a combined radial
velocity and astrometry analysis. Based on a simulation of the
position of $\alpha$ Centauri A over one orbital period with a
  time step of ten days, we compare
various astrometric effects in Fig. \ref{fig:ac}. One of the main position changes is caused by the heliocentric motion of the barycenter of $\alpha$
  Centauri. Because it is linear and easy to comprehend, we do not show
  this effect in the figure. For ground-based observations, the
  atmospheric refraction (P1 in Fig. \ref{fig:ac}) induces a few arcminute position offset. Because this effect can only be modeled to a precision of about 1\,arcsec
  \citep{mangum15}, the ground-based astrometry is not able to achieve
  1\,arcsec absolute precision. The segments of curves in P1 are
  due to the annual variation of the elevation angle. We use the {\it
 slaRefro} routine to calculate the refraction with an effective
wavelength of about 500\,nm, temperature of 278\,K, and relative humidity of 0.1.

The stellar aberration due to the Earth's location (P2) is
another main factor alternating the observed direction of $\alpha$
Centauri A. This effect is linearly proportional to the barycentric
velocity of the observatory for ground-based astrometry. Hence, a
  precise knowledge of the Earth's ephemeris and rotation are required
  to properly model this effect. The third most significant effect is
  caused by the binary motion (P3). Instead of showing the
    barycentric motion of A in P3 of Fig. \ref{fig:ac}, we show the orbit of
    $\alpha$ Centauri B with respect to A and scale the axes such that
    the binary orbit is comparable with the one shown in figure 1 of
    \cite{pourbaix99}. The good match between P3 and the one in
    \cite{pourbaix99} demonstrates the consistency of our convention
    (see Appendix \ref{sec:conv3} for details) with the ones used
    in previous studies of visual binaries. Although the binary
  motion of visual binaries such as $\alpha$ Centauri A and B is significant, it was typically ignored in previous radial velocity modeling due to a decoupling of the solar and target systems. A more rigorous treatment of the stellar motion around the Galactic center is also needed to account for the secular aberration \citep{kopeikin06}.

The other less significant effects are the gravitational lensing in the solar and target systems. The gravitational lensing effects
caused by the Sun (P4) and Earth (P5) are detectable in
astrometric data with mas and sub-mas precision. As see in the bottom
left panel of Fig. \ref{fig:ac}, the annual motion of the Earth is
superposed on the binary motion of $\alpha$ Centauri in the solar
lensing effect. On the other hand, the lensing effect in the target system only changes the apparent position of the target star by less
than 1\,$\mu$as. According to \cite{kopeikin99}, the lensing effect
due to a companion in the target system is only significant for nearly
edge-on systems that host massive companions.

In panels P7 to P9, we show the position of $\alpha$ Centauri A
  with various combinations of effects. In the geometric position  of
  $\alpha$ Centauri A (P7), we only combine the proper motion, parallax, and
  binary motion. We see that the orbit is dominated by a linear trend
  caused by proper motion and a periodic component due to the binary
  motion. The annual parallax is superposed on this long-term
  trend. If we add stellar aberration and lensing effects (P8), we
  find that the aberration adds another dimension to the trend
  and forms a ``tube.'' The diameter of the tube is determined by the
  magnitude of stellar aberration. If we combine all effects (P9), the
  position offset is dominated by the refraction effect. If the elevation angle is large enough, the refraction can be smaller than 1\,arcmin, which is still much more significant than other effects. 

\begin{figure}
  \centering
    \includegraphics[scale=0.6]{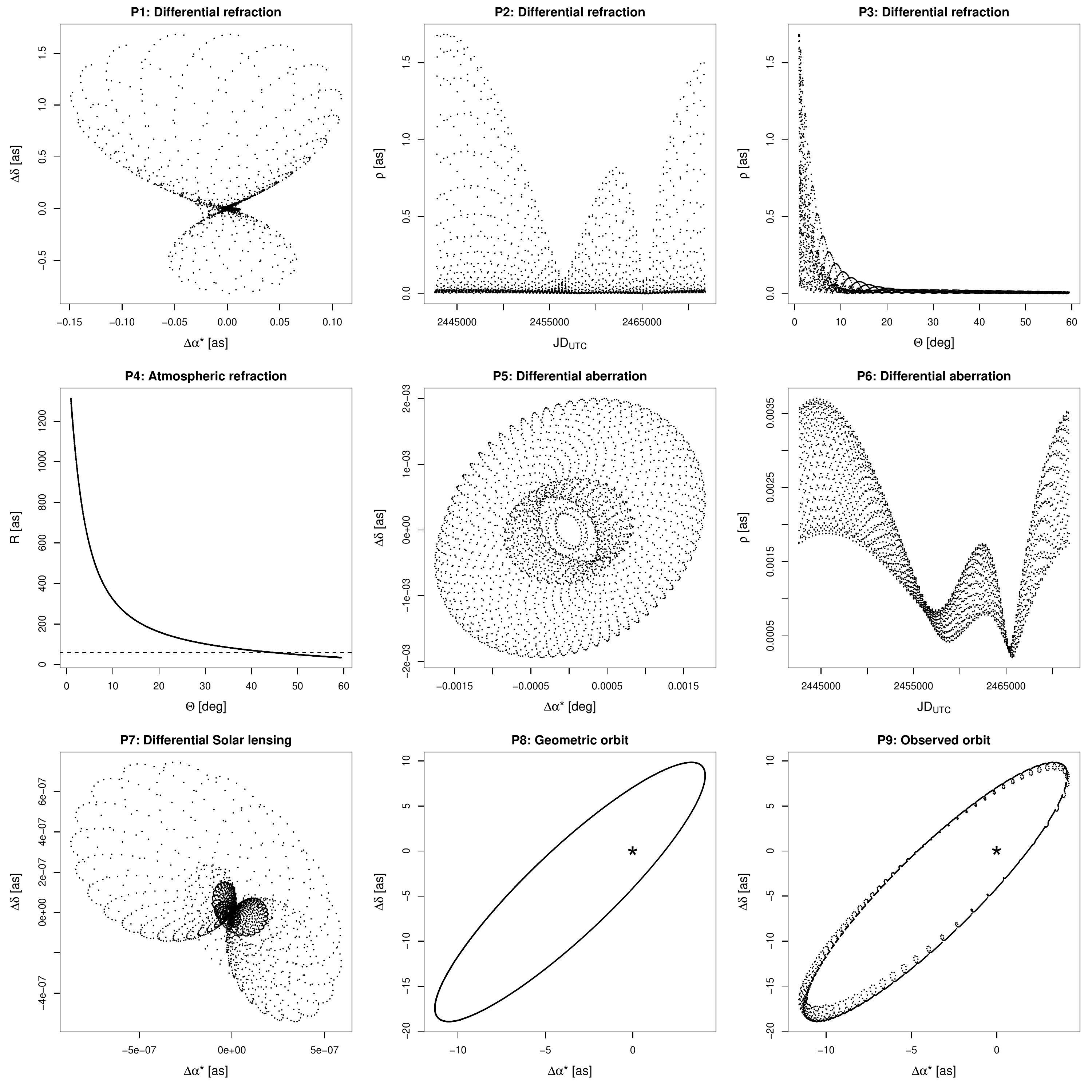}
  \caption{Various effects on the relative position of $\alpha$
    Centauri B with respect to A. The differential
    refraction is shown in the offset coordinates (P1), and as a function
    of time (P2), as a function of elevation angle (P3). The
    refraction as a function of elevation angle is shown in P4. The
    horizontal dashed line indicates a refraction of 1\,arcmin. The
    differential aberrations in the offset coordinates and as a
    function of time are shown in P5 and P6, respectively. P7 shows the
    differential Solar lensing in offset coordinates. The geometric
    orbit of B around A without lensing, aberration, and refraction
    effects is shown in P8. The observed orbit with all effects are
    shown in P9. }
  \label{fig:acrel}
\end{figure}

To see the influence of various effects on relative astrometry, we
show the position of B with respect to A in Fig. \ref{fig:acrel}. The
offset coordinates are defined as
\begin{eqnarray}
  \Delta\alpha^*&=&(\alpha_2-\alpha_1)\cos{\frac{\delta_1+\delta_2}{2}}~,\\
  \Delta\delta&=&\delta_2-\delta_1~,
\end{eqnarray}
where $(\alpha_1,\delta_1)$ and
$(\alpha_2,\delta_2)$ are the equatorial coordinates of two points on
the celestial sphere which are close to each other. We explain the various effects shown in Fig. \ref{fig:acrel} as follows.
%We define $\rho=\sqrt{\Delta\alpha^2+\Delta\delta^2}$ as the angular
%separation between A and B. 
\begin{itemize}
  \item {\bf P1: Differential refraction. }Because we only consider a
    single wavelength in the calculation of refraction, this
    refraction effect is achromatic. Chromatic refraction can be
    calculated simply by applying the {\it slaRefro} routine to different
    wavelengths. Despite much scatter and complexity in the pattern
    observed in the offsets, the differential refraction is less than
    0.05\,arcsec for most time steps when the elevation angle is
    higher than 30$^\circ$. Without properly modeling this
    differential effect to a sub-mas precision, relative astrometry based on direct imaging would be significantly biased in characterizing exoplanets. 
    \item {\bf P2: Differential refraction. }Because we adopt a uniform
      time step of 10 days, the modulation of elevation and refraction
      over time is due to the Earth's motion with respect to the SSB. 
      \item {\bf P3: Differential refraction. }This panel shows the
        differential refraction as a function of elevation angle. The
        upper limit of the differential refraction is determined by
        the elevation angle while the binary motion modulates the
        differential refraction at a given elevation angle.
      \item {\bf P4: Atmospheric refraction. }This panel shows the
        refraction as a function of elevation. For the atmospheric
        parameter adopted in this simulation, the absolute refraction
        is less than 1\,arcminute, and the relative refraction is less
        than 0.05\,arcsec if the elevation angle is larger than
        30$^\circ$ (see section \ref{sec:special_shift}). In principle,
        the current atmospheric model allows a differential refraction
        model precision of 10\,$\mu$as \citep{gubler98}. Parameters such as temperature of the star, local pressure and temperature, and relative humidity may not be well known. As already routinely practiced by some ground-based astrometric programs, it is necessary to observe target systems close to the zenith if sub-mas relative astrometry is required.
      \item {\bf P5 and P6: Differential aberration. }The aberration
        is determined by the component of $\bm{r}_{\rm SO}$ which is
        perpendicular to the target direction. Hence, the aberration is
        modulated by the Earth's rotation and barycentric
        motion. Although this effect contributes a few mas positional
        offset (comparable with the astrometric signal induced by a Jupiter analog) and shows strong variation in time, it is rarely considered in analyses of relative astrometry data. 
        \item {\bf P7: Differential solar lensing. }This effect is
          at most a few $\mu$as and thus is only important for future
          space-based astrometry missions such as the SIM PlanetQuest
          \citep{catanzarite06,unwin08}.
          \item {\bf P8: Geometric orbit.} This is the binary motion
            projected onto the plane of the sky and is frequently used by the community to model relative astrometry.
            \item {\bf P9: Observed orbit.} This is a combination of
              geometric and other effects. The atmospheric refraction
              biases the binary orbit by at most 2\,arcsec,
              producing about 10\% of the total offsets. If the system
              is observed with an elevation angle larger than 30$^\circ$,
              we expect $<$0.05\,arcsec refraction bias,
              equivalent to a 0.2\% uncertainty in the binary orbital
              solution. Without properly removing this bias through
              modeling, it is unlikely to detect astrometric signals
              of exoplanets reliably. 
\end{itemize}
Based on the above analyses of the $\alpha$ Centauri orbit, the
atmospheric and aberration effects are only marginally important for
constraints on its binary orbit based on relative astrometry. However,
these effects are far more significant than potential planetary
signals. For example, an Earth-like planet around $\alpha$ Centauri B
would induce $\sim 1~\mu$as stellar reflex motion, and a Jupiter-like
planet would induce $\sim 1$\,mas reflex motion. Hence, the atmospheric
and aberration effects should be modeled to a high-precision level if
solar system analogs are to be detected through the astrometry
method.

\begin{figure}
  \centering
  \includegraphics[scale=0.8]{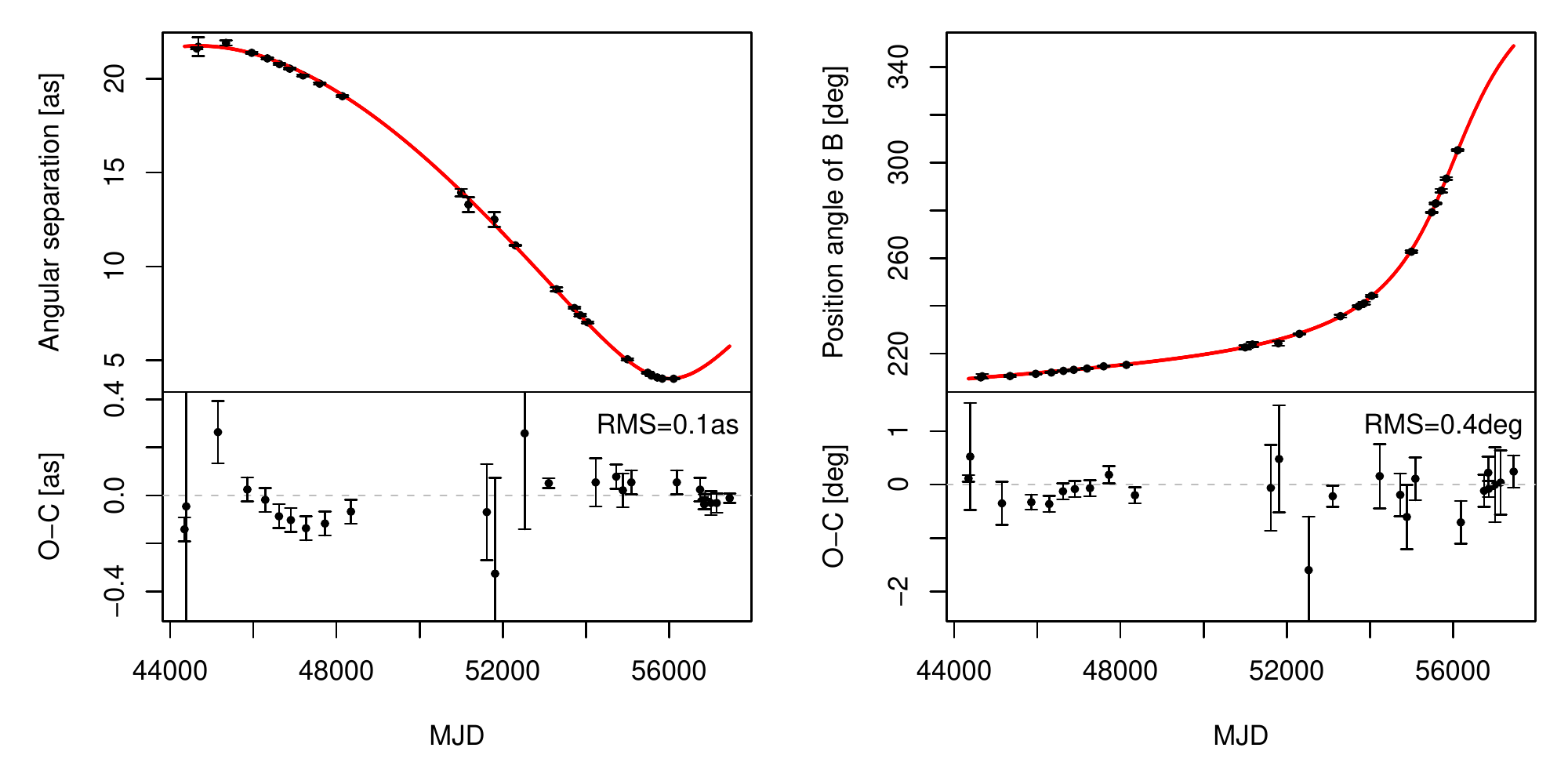}
  \caption{Predicted and observed angular separation (left) and
    position angle (right) of $\alpha$ Centauri B with respect to
    A. The red lines denote the best-fit orbital solution given by
    \cite{kervella16}. The black error bars denote the astrometry data
    that the solution is based on and can be visualized in the lower observed-calculated plot.}
  \label{fig:acmodel}
\end{figure}

To roughly test the precision of PEXO prediction, we compare the
geometric binary orbit (P8 in Fig. \ref{fig:acrel}) with the
astrometry data from \cite{kervella16}. The model and data for the
angular separation and the position angle are shown in
Fig. \ref{fig:acmodel}. The small residual suggests that PEXO is
able to recover previous results. The observational details for each astrometry data point are beyond this work, so we have not considered the aberration and atmospheric effects that could introduce differential positional offsets.

\subsection{Radial Velocity}\label{sec:rv_binary}
Adopting the same observatory coordinates and orbital parameters as in
section \ref{sec:astrometry_binary}, we now consider the relativistic
and classical effects on the radial velocities of $\alpha$ Centauri
below and show the results in Fig. \ref{fig:rv_binary}:
\begin{figure}
  \centering
    \includegraphics[scale=0.45]{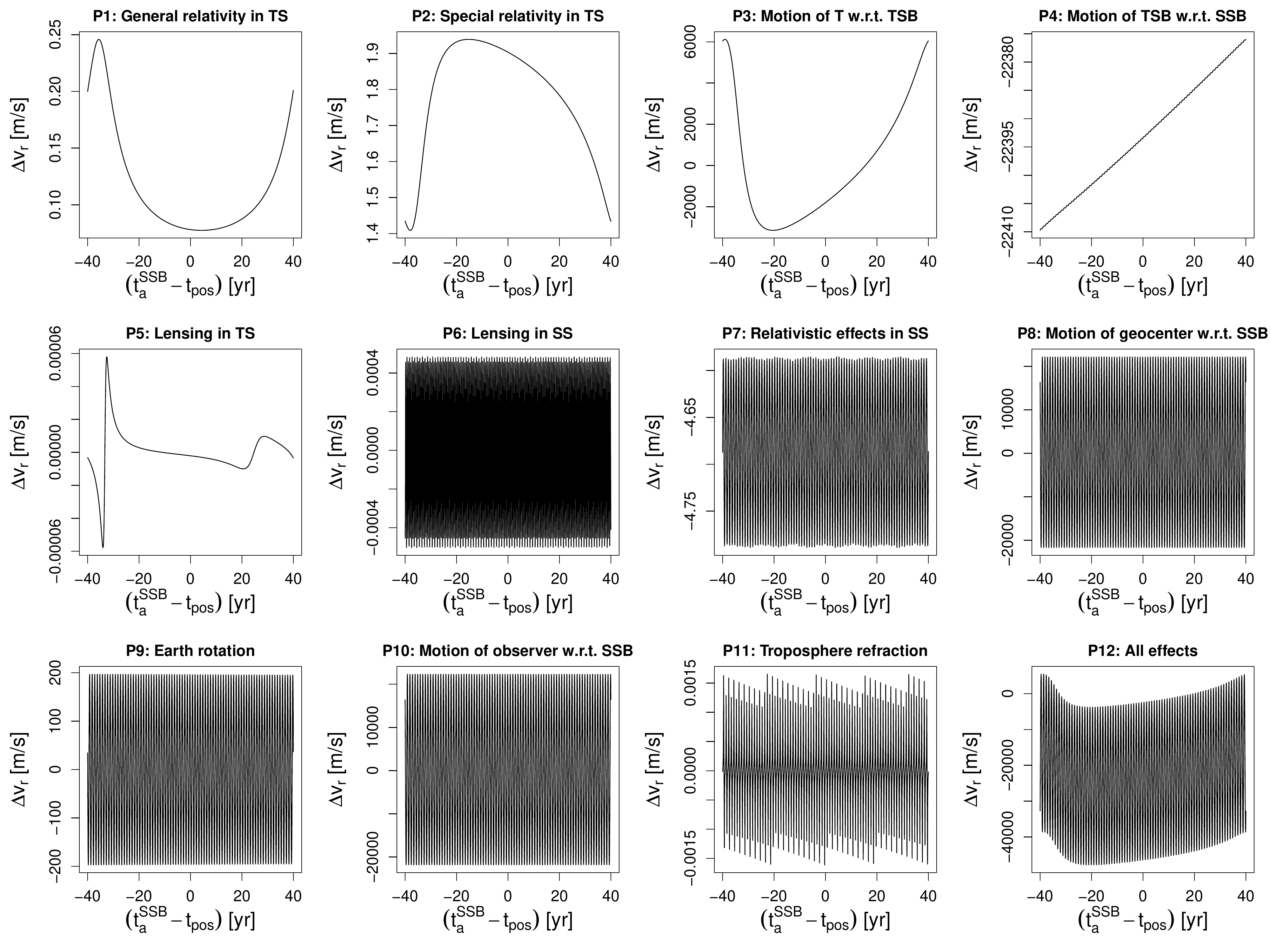}
  \caption{Relativistic and classical effects on the measured radial velocity of
    $\alpha$ Centauri A over an orbital period. }
  \label{fig:rv_binary}
\end{figure}

\begin{itemize}
\item {\bf P1: General relativity in the target system. }This GR effect is caused by the gravitational field of $\alpha$
  Centauri B, which makes the light ray from $\alpha$ Centauri A
  Doppler shifted. This leads to a radial velocity variation of 0.17\,m/s, which is larger than the maximum radial velocity variation of 0.1\,m/s that would be caused by the presence of any Earth-like planets around $\alpha$ Centauri A. To the best of our knowledge, this effect is not considered in current radial velocity analysis packages. 
\item {\bf P2: Special relativity in the target system. }This is a special relativity
  effect due to the motion of $\alpha$ Centauri A around the
  barycenter of the binary. This effect contributes to a radial
  velocity variation of 0.53\,m/s over one orbital period and again does not appear to be included in existing radial velocity packages.
\item {\bf P3: Motion of the target with respect to the TSB.} This radial velocity variation is due to the motion of $\alpha$ Centauri A around the binary barycenter. This motion is typically ignored for the barycentric correction, although it can be determined {\it a priori} if
  the Keplerian parameters of the binary motion are known to a high precision. 
\item {\bf P4: Motion of the TSB with respect to the SSB. }This is a kinematic effect
  due to the relative motion of the TSB with respect to
  the SSB. This motion would change the viewing perspective and lead
  to the so-called ``perspective acceleration'' in radial velocity. This
  perspective acceleration is coupled with the binary motion and the
  observer's motion in the solar system. This is evident in the corresponding
  radial velocity acceleration shown in Fig. \ref{fig:rv_accelerate}. The mean
  acceleration is the perspective acceleration, the short
  periodic variation is due to the Earth's annual motion around the
  Sun, and the long periodic variation is caused by the binary motion
  of $\alpha$ Centauri A and B. This time-varying acceleration casts doubt on the reliability of subtracting a linear trend with a constant perspective acceleration from the radial velocity data (e.g., \citealt{zechmeister13}). 
  \begin{figure}
  \centering
  \includegraphics[scale=0.45]{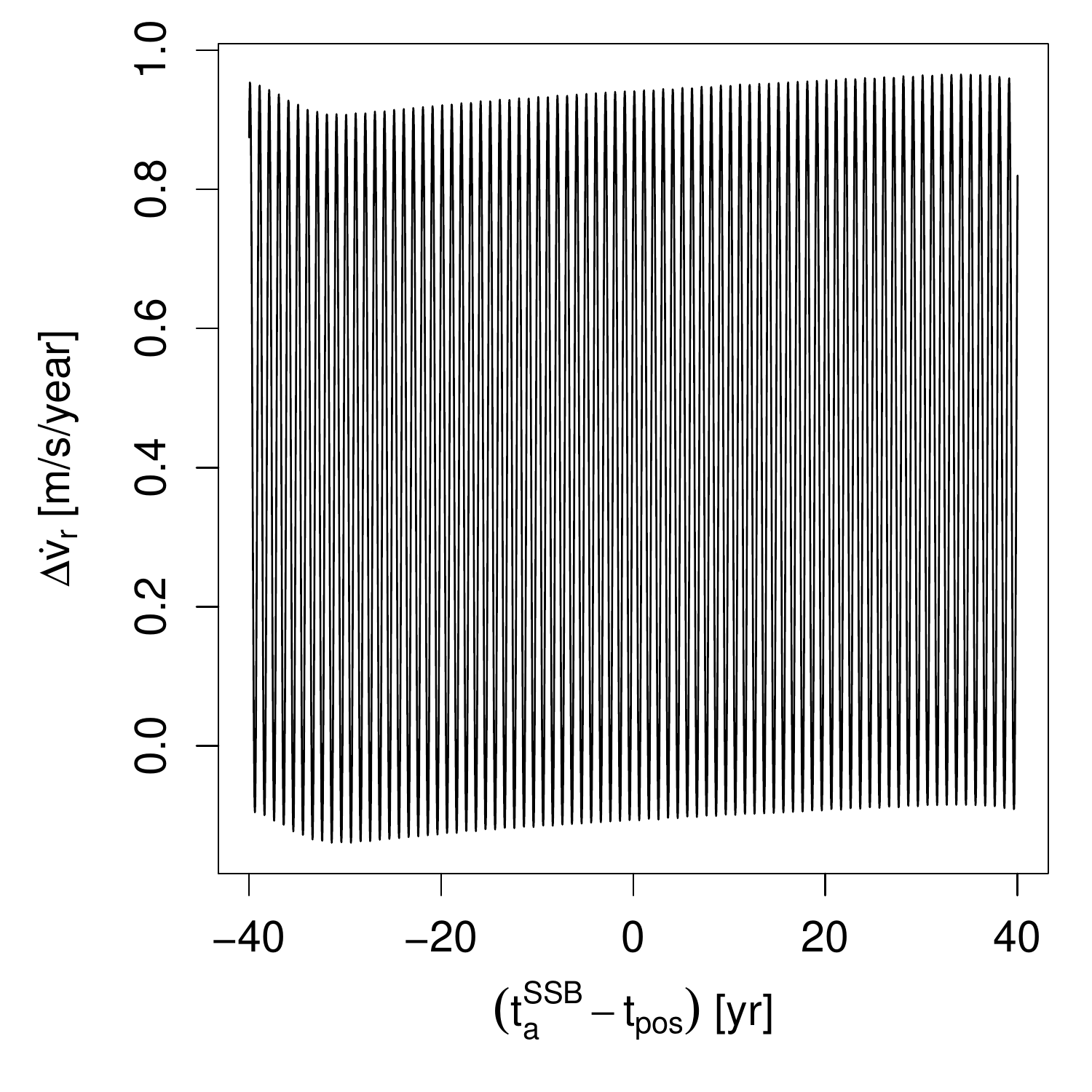}
  \caption{Radial velocity acceleration induced by the motion of the barycenter of
    $\alpha$ Centauri relative to the SSB. }
  \label{fig:rv_accelerate}
\end{figure}
\item {\bf P5: Lensing in the target system.} This effect corresponds to the
  gravitational lensing of the companion, which is $\alpha$ Centauri B
  in this case. This effect contributes at most a 0.1\,mm/s radial velocity
  variation. Considering that this effect is proportional to the inclination
  and the semi-major axis, it might be detectable
  with radial velocity instruments such as ESPRESSO in a nearly edge-on
  binary system, in short-period binaries, and in transiting systems
  hosting massive, short-period planets.
\item {\bf P6: Lensing in the solar system.} This effect is due to the gravitational lensing of the Sun and the Earth, contributing to a 0.1\,mm/s radial velocity variation. This effect is proportional to $\cot{\psi/2}$ \citep{klioner03}, where $\psi$ is the angular distance between the Sun and the target star. Considering that the minimum $\psi$ is 40$^\circ$ for the $\alpha$ Centauri example, the lensing effect would contribute to 1\,cm/s if the target star is less than
  1$^\circ$ from the Sun. 
\item {\bf P7: Relativistic effects in the solar system.} These effects are due to the gravitational field in the solar system and the barycentric motion of the observer. These effects
  also cause the Einstein delay, which transforms TT to TCB. The ratio of the increments of TT and TCB is simply the relativistic Doppler shift. Because the motion of the observatory
  in the solar system is derived from the JPL ephemeris, we can subtract this type of radial velocity variation directly from the measured radial velocity to remove these local effects. However, in the cases
  that the observatory site or the ephemeris is not well determined,
  the Earth's motion as well as the motions of the target star should
  be determined {\it a posteriori}. 
\item {\bf P8:  Motion of geocenter with respect to the SSB.} This effect is due to the
  barycentric motion of the geocenter and contributes about a 20\,km/s radial velocity variation. Such significant kinematic effects are not completely local
  because the corresponding radial velocity variation depends on the direction of
  the target, which changes over time due to the motion of the target
  star.
  \item {\bf P9: Earth rotation.} This effect is due to the
  Earth's rotation and contributes to a 200\,m/s variation. Hence, the radial velocity
  data precision highly depends on the Earth rotation
  model. Specifically, a radial velocity precision of 1\,cm/s requires 1\,cm/s
  modeling precision of the Earth rotation. Alternatively, the Earth's
  rotation can be determined a posterori through a fit of the
  combined model to the data.
\item {\bf P10: Motion of observer with respect to the SSB.} It is a combination
  of the P9 and P10 effects.
\item {\bf P11: Troposphere refraction.} In this panel, we show
    refraction-induced radial velocity variation for elevation angles
    larger than 10$^\circ$ to be representative of most ground-based
    observations. This indicates at most a few mm/s variation in
    radial velocity, and thus refraction-induced effects are
    negligible for the current radial velocity observations.
\item {\bf P12: All effects. }This is the observed radial velocity, which is a
  combination of all effects.
\end{itemize}
We further assess the performance of PEXO by comparing the PEXO model
  prediction of radial velocity with the radial velocity data in the
  literature in Fig. \ref{fig:ACcombined}. For data sets with relative
radial velocities, we add an offset so that the mean predicted and
observed radial velocities are equal. It is obvious that the PEXO
prediction well fits the combined data, leading to 4.7 and 4.2\,m/s standard
deviations of residual radial velocities for $\alpha$ Centauri A and B,
respectively. Nevertheless, we still see significant variations in
the CHIRON and HARPS data sets, indicating potential bias in the
\cite{kervella16} solution and in the barycentric correction. To
mitigate such biases, a comprehensive modeling of the $\alpha$
Centauri system is needed and is beyond the scope of this work. 
\begin{figure}
  \centering 
  \includegraphics[scale=0.5]{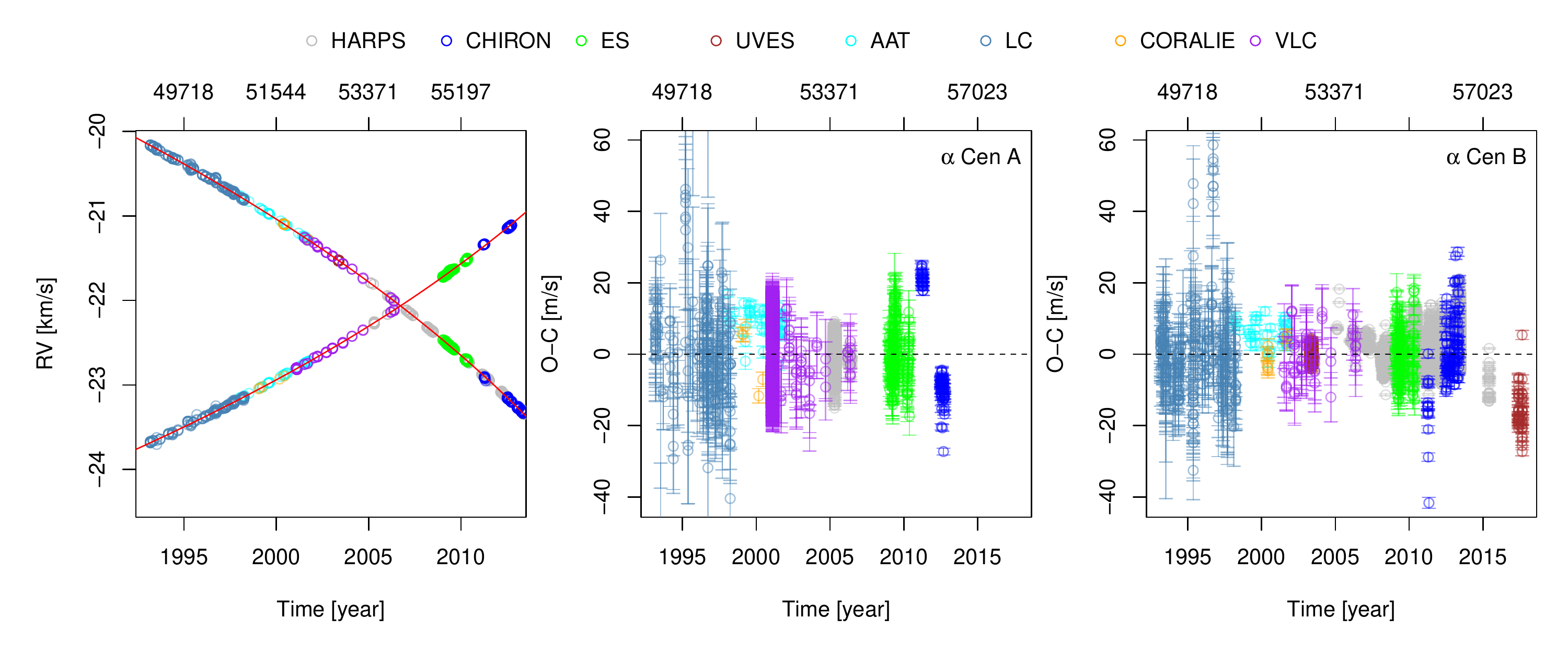}
  \caption{Comparison of the radial velocity model prediction and the
    radial velocity data sets from various sources. The red line shows
    the model prediction based on the solution given by
    \cite{kervella16}. Left panel: model prediction and observed
    radial velocities of $\alpha$ Centauri A and B; middle panel:
    radial velocity residual for $\alpha$ Centauri A; right panel:
    radial velocity residual for $\alpha$ Centauri B. In each panel,
    the top axis shows the modified Julian date (MJD$=$JD$-$2400000.5). The HARPS data is
    from \cite{lisogorskyi19}, the CHIRON
    and ES data sets are from \cite{zhao18}, the UVES data is obtained
    by \cite{kjeldsen05}, the AAT and CORALIE data are from
    \cite{pourbaix02}, and the LC and VLC data are from \cite{endl01}. }
  \label{fig:ACcombined}
\end{figure}

\subsection{A Comparison of Relativistic Effects on Timing,
  Astrometry,  and Radial Velocities between Different Packages}\label{sec:comparison_table}
We apply PEXO to simulate the timing, astrometry, and radial velocity
of $\tau$ Ceti, $\alpha$ Centauri A, and XO-3 over one decade to
assess the significance of various effects. In the XO-3 system,
we treat XO-3 b as the primary and XO-3 as the secondary so that the timing model can be used to predict transit
timing, the astrometry model is applicable to the relative astrometry
of directly imaged planets, and the radial velocity model can be used in the
studies of systems such as the S2-Sgr A* system
\citep{gravity18}. Additionally, we compare the functions in PEXO with
those in TEMPO2, EXOFAST+\footnote{EXOFAST+ represents a
  combination of the timing routines such as {\small utc2bjd.pro}
  developed by E10, EXOFAST developed by \cite{eastman13}, and {\small zbarycorr.pro} developed by \cite{wright14}. The Roemer delay in TS
  applies to EXOFAST, the radial velocity functionalities apply to
  {\small zbarycorr.pro}, and the other timing functionalities apply to {\small utc2bjd.pro}.}, and GREM\footnote{Because we do not have
  access to the software, information about the functions implemented
  comes from \cite{klioner92} and \cite{klioner03}.} and show the results in Table \ref{tab:functions}.

Although TEMPO2 is able to model radial velocity by
differentiating the timing function over time, such a numerical
treatment is likely to result in significant rounding errors due to
the huge scale difference between various quantities. The pulse
frequency in TEMPO2 is not the same as the light frequency; for example, the
light from a star can be gravitationally red shifted while the
rotation frequency of a pulsar cannot. Thus, TEMPO2 is not optimal for radial velocity modeling especially in the case of exoplanet detections. 

As seen in Table \ref{tab:functions}, TEMPO2 provides a precise timing
model by including many high-order effects, EXOFAST+ includes some
relativistic effects in timing and radial velocity, and GREM is able to model
astrometry to a high precision. In comparison with these packages,
PEXO aims to include most high-order effects whilst also providing a 
combined modeling of timing, astrometry, and radial velocity. As shown in the table, the significant time delay not included in
EXOFAST+ and GREM is the Einstein delay in the target system,
which is about 0.01\,s for $\alpha$ Centauri A and XO-3
  b. The Einstein delay in $\alpha$ Centauri over one binary
  orbit is as large as 0.1\,s. This delay variation actually measures
the difference between periastron and apastron for a given Keplerian
orbit. Thus it is significant for an eccentric orbit and can be measured
by the timing difference between the primary and secondary transits for
a long-period transit planet. 

Without considering proper motion effects, EXOFAST+ introduces
  subsecond timing bias for $\tau$ Ceti and $\alpha$ Centauri
  A. This bias increases with the time difference between the reference
  epoch of the astrometry catalog and the mean epoch of the data. On the other hand, the decoupling approach adopted by EXOFAST+ and GREM would introduce 0.02\,s bias
  in their timing model for $\alpha$ Centauri. Because TEMPO2 does not
  consider the third-order Roemer delay, there would be timing biases
  of 200 and 900\,ns for $\tau$ Ceti and $\alpha$ Centauri for one
  decade, respectively. Because the third-order Roemer delay is
  determined by the time difference between simulated epochs and the
  reference epoch, the corresponding 200\,ns timing bias for the
  simulated epochs starting from the {\it Gaia} reference epoch for $\tau$
  Ceti (shown in Table \ref{tab:functions}) is much higher than that
  for earlier epochs shown in the right panel of Fig. \ref{fig:bjd}.

The atmospheric effects contribute a few arcminute position offset
  and are not modeled by GREM, which is aimed at space-based
  astrometry. The second-most significant astrometric effect is the
  stellar aberration. Because the first-order aberration is considered
  in most packages, we only show the second and third-order aberration
  values in Table \ref{tab:functions}. The second-order aberration
  contributes at most 1\,mas position offset and is not sensitive to
  stellar distance. Thus it is important for mas-precision astrometry
  for all stars. Another significant effect is
  gravitational lensing by the Sun, which could contribute more than
  10\,mas position offset. The Earth lensing dominates the planetary
  lensing effects and contributes sub-mas position offset. The third-order geometric effect contributes to a 30\,$\mu$as astrometric
bias for $\alpha$ Centauri A and is not included in GREM. Such an
effect could be significant for some nearby binary systems and needs
to be properly modeled if $\mu$as precision is required. The decoupling effect contributes a few arcsecond offset.

For the comparison of radial velocity models, the total decoupling bias is about 4\,m/s for $\alpha$ Centauri over one decade and thus significantly higher than potential signals in the radial velocity data \citep{zhao18}. The time-varying special relativity effect in the $\alpha$ Centauri system is as much
as 0.3\,m/s and is not included in current radial velocity
packages. The relativistic effects are even more significant in nearly
edge-on systems such as XO-3. In such cases, even the lensing
Doppler effect in the target system contributes up to 2\,m/s and is measurable with current radial velocity instruments. 

It is evident in Table \ref{tab:functions} that no current packages
are able to precisely model different types of data consistently. In contrast, PEXO stands out as a package for modeling
multiple types of data in a precise and consistent fashion. This makes
it a suitable package for synthesizing data from high precision
ground-based or space-based facilities. 
\begin{longrotatetable}
\begin{table}
  \caption{Comparison of various packages and the amplitudes of high-order classical and relativistic effects in timing, astrometry, and
    radial velocity. The amplitudes are determined based on
    simulations over ten years starting from the Gaia DR2 epoch. The solar system is denoted by SS while the target system is denoted by TS. We do not show constant effects such as the frame transformation between TSB and SSB. Because TEMPO2 is not optimal and is not aimed at radial velocity modeling, but in principle might offer the feature, we assign a dash mark to each radial velocity effect for TEMPO2. Because the wet component of
    tropospheric delay is not well modeled, we follow TEMPO2 to only
    model the hydrostatic component. The delay values are for
    elevation angle larger than 10$^\circ$ and thus is representative
    for most astronomical observations. Like TEMPO2, PEXO only models
    the well understood components such as hydrostatic delay in the
    troposphere in these effects and model the other components as
    fittable parameters. The trend decoupling bias is calculated for a
  ten year time span while the period decoupling bias is for an orbital
  period of a given target system. The decoupling effects are recorded
    as a function of a package if all motions are modeled
    simultaneously to avoid decoupling bias and the decoupling effects
    can be separated as a data product (or ``barycentric
    correction''). If a function is included in a package, a tick is
    assigned. Otherwise, a cross is assigned. If there is a coding
    error for a function, we use two crosses to mark it. }
  \label{tab:functions}
\footnotesize{
  \hspace{-1in}
%  \begin{tabular}{*10{c}}
    \begin{tabular}{lp{5cm}cc ccc cccc}
    % \startdata
    \hline
      \hline
  Model&Function&Equations&Unit&$\tau$ Ceti&$\alpha$ Centauri A& XO-3
                                                  b&PEXO&TEMPO2\footnote{TEMPO2
                                                     is not designed
                                                     primarily for radial velocity
                                                     modeling, although
                                                     the pulse
                                                     frequency
                                                     variation can be
                                                     converted into
                                                     Doppler shift. 
                                                     }&EXOFAST+
      & GREM\\
      \hline
Timing&Second-order Roemer delay in SS&\ref{eqn:dpS}&s&$3\times
                                                        10^{-4}$&$5\times
                                                                  10^{-4}$&$4\times
                                                                            10^{-6}$&\cmark&\cmark&\xmark\xmark&\cmark\\
Timing&Third-order Roemer delay in SS&\ref{eqn:offset}&s&$6\times10^{-5}$&$9\times 10^{-7}$&$3\times
                                                                         10^{-12}$&\cmark&\xmark&\xmark&\xmark\\
Timing&Einstein delay in SS&\ref{eqn:einstein}&s&20&20&20&\cmark&\cmark&\cmark&\cmark\\
Timing&Shapiro delay due to the Sun&\ref{eqn:DSO}&s&$3\times
                                                     10^{-5}$&$2\times
                                                               10^{-5}$&$2\times
                                                                         10^{-5}$&\cmark&\cmark&\cmark&\cmark\\
Timing&Shapiro delay due to SS planets&\ref{eqn:DSO}&s&$4\times
                                                              10^{-8}$&$2\times
                                                                        10^{-8}$&$3\times
                                                                                  10^{-8}$&\cmark&\xmark\xmark&\xmark&\xmark\\
Timing&Proper motion of TS&\ref{eqn:dpS}&s&0.2&0.6&$1\times 10^{-4}$&\cmark&\cmark&\xmark&\cmark\\
Timing&Roemer delay in TS&\ref{eqn:drT-pT}&s&0&$3\times 10^3$&40&\cmark&\cmark&\cmark&\cmark\\
Timing&Einstein delay in TS&\ref{eqn:deT}&s&0&0.01&0.01&\cmark&\cmark&\xmark&\xmark\\
Timing&Shapiro delay in TS&\ref{eqn:DST}&s&0&$4\times 10^{-6}$&$2\times 10^{-7}$&\cmark&\cmark&\xmark&\xmark\\
Timing&Atmospheric effects&\ref{eqn:Dtropo2}&s&$8\times 10^{-8}$&$3\times 10^{-8}$&$3\times 10^{-8}$ &\cmark&\cmark&\xmark&\xmark\\
Timing&Trend decoupling effects&\ref{eqn:timing_trend}&s&0&0.02&$4\times 10^{-3}$&\cmark&\cmark&\xmark&\xmark\\
Timing&Period decoupling effects&\ref{eqn:timing_period}&s&0&0.02&$5\times 10^{-7}$&\cmark&\cmark&\xmark&\xmark\\
\hline
Astrometry&Second-order stellar aberration&\ref{eqn:uo}&as&$6\times10^{-4}$&$7\times 10^{-4}$&$7\times 10^{-6}$&\cmark&\xmark&\xmark&\cmark\\
Astrometry&Third-order stellar aberration&\ref{eqn:uo}&as&$2\times 10^{-7}$&$1\times 10^{-7}$&$1\times 10^{-7}$&\cmark&\xmark&\xmark&\cmark\\
Astrometry&Lensing by the Sun&\ref{eqn:lo}&as&0.02&0.009&0.01&\cmark&\xmark&\xmark&\cmark\\
Astrometry&Lensing by SS planets&\ref{eqn:lo}&as&$6\times 10^{-4}$&$2\times
                                                                      10^{-4}$&$9\times10^{-4}$&\cmark&\xmark&\xmark&\cmark\\
Astrometry&Gravitational lensing in TS&\ref{eqn:ll}&as&0&$3\times 10^{-9}$&$1\times
                                                                    10^{-9}$&\cmark&\xmark&\xmark&\xmark\\
Astrometry&Second-order geometric
            effects&\ref{eqn:offset}&as&0.2&4&$2\times 10^{-3}$&\cmark&\xmark&\xmark&\cmark\\
Astrometry&Third-order geometric effects&\ref{eqn:offset}&as&$2\times 10^{-6}$&$3\times
                                                                                10^{-5}$&$2\times10^{-14}$&\cmark&\xmark&\xmark&\xmark\\
Astrometry&Atmospheric effects&\ref{eqn:refro}&as&$4\times 10^3$&$1\times 10^3$&$3\times 10^3$&\cmark&\xmark&\xmark&\xmark\\      
Astrometry&Trend decoupling effects&\ref{eqn:u_trend}&as&0&9&2&\cmark&\xmark&\xmark&\cmark\\
Astrometry&Period decoupling effects&\ref{eqn:u_period}&as&0&0.7&$5\times
                                                 10^{-3}$&\cmark&\xmark&\xmark&\cmark\\
      \hline
Radial velocity&Relativistic effects in the SS&\ref{eqn:zrS}&m/s&0.2&0.2&0.2&\cmark&-&\cmark&\xmark\\
Radial velocity&Lensing by the Sun&\ref{eqn:zlS1}&m/s&$2\times10^{-3}$&$1\times10^{-3}$&$1\times
                                                                            10^{-3}$&\cmark&-&\cmark&\xmark\\
Radial velocity&Lensing by SS planets&\ref{eqn:zlS1}&m/s&$5\times10^{-6}$&$2\times
                                                               10^{-6}$&$2\times
                                                               10^{-6}$&\cmark&-&\cmark&\xmark\\
Radial velocity&Special relativity in TS&\ref{eqn:zsrT}&m/s&0&0.2&40&\cmark&-&\xmark&\xmark\\
Radial velocity&General relativity in TS&\ref{eqn:zgrT}&m/s&0&0.04&50&\cmark&-&\xmark&\xmark\\
Radial velocity&Lensing Doppler shift in TS&\ref{eqn:zlS}&m/s&0&$2\times10^{-5}$&2&\cmark&-&\xmark&\xmark\\
Radial velocity&Second-order geometric
                 effects&\ref{eqn:offset}&m/s&0.07&0.9&$8\times 10^{-3}$&\cmark&-&\cmark&\xmark\\
Radial velocity&Third-order geometric effects&\ref{eqn:offset}&m/s&$6\times 10^{-7}$&$1\times
                                                               10^{-5}$&$2\times
                                                                         10^{-14}$&\cmark&-&\xmark&\xmark\\
Radial velocity&Atmospheric effects&\ref{eqn:ztropo}&m/s&$8\times
                                                         10^{-3}$&$3\times
                                                                   10^{-3}$&$4\times
                                                                             10^{-3}$&\cmark&-&\xmark&\xmark\\
Radial velocity&Trend decoupling effects&\ref{eqn:vr_trend}&m/s&0&2&0.4&\cmark&-&\xmark&\xmark\\
Radial velocity&Period decoupling effects&\ref{eqn:vr_period}&m/s&0&2&$5\times 10^{-5}$&\cmark&-&\xmark&\xmark\\\hline
    \end{tabular}
  }
  \end{table}
\end{longrotatetable}

\section{Conclusion}\label{sec:conclusion}
In this work, we introduce relativistic models of timing, astrometry,
and radial velocity in the PEXO package, which is mainly aimed at data
analysis for exoplanets as well as tests of GR. PEXO
includes the general and special relativistic effects both in the
solar system and in the target system. These relativistic effects lead
to Einstein delays in timing, stellar aberration in astrometry as well
as gravitational Doppler shift in radial velocity. PEXO also models
the gravitational lensing and high-order geometric effects in both
systems. The lensing effects lead to the Shapiro delay in timing,
light deflection in astrometry, and Doppler shift in radial
velocity. Based on our comparison of PEXO with TEMPO2, PEXO is able to
achieve a timing precision of $\sim$1\,ns, an astrometry precision of
$\sim$1\,$\mu$as, and a radial velocity precision of $\sim$1\,$\mu$m/s. 

These figures are comfortably better than what were expected to be achieved by
current facilities. To test the precision of PEXO, we compare it with TEMPO2 and the package developed by E10. The timing precision of PEXO is at least
  comparable with TEMPO2 at the level of a few nanoseconds. It is
  better than TEMPO2 for decade-long timing data for nearby targets
  due to its consideration of the third-order terms of Roemer delay. We find an error in the routine {\small shapiro\_delay.C} of TEMPO2 that could induce tens of nanoseconds timing
  bias in the calculation of Shapiro delay. Considering the popularity
  of TEMPO2 and the potential for coding errors in complex packages, we
  strongly recommend the application of independent packages for
  important discoveries in pulsar timing and exoplanetology as well as
  in other astrophysical applications. We also compare the IDL routine {\small
    utc2bjd.pro} developed by E10 and its corresponding applet with
  TEMPO2. We find that the applet is able to provide a timing precision of a few
  milliseconds if ignoring the proper motion effects. However, we notice an
  error in the calculation of parallax delay in {\small
    utc2bjd.pro}, leading to a timing error of about 0.3\,ms for
  $\tau$ Ceti. Although such a bug is not significant for current
  exoplanet science, it could become significant for high-precision
  applications. The corrected IDL version of E10 is able to model timing to a precision of a few microseconds if the propagation of TSB is provided externally. The errors in high-precision packages such as TEMPO2 and {\small utc2bjd.pro} demonstrate the utility of new packages in minimizing coding errors and their potential spread in other applications.

  The numerical implementations of barycentric correction of radial
  velocity for PEXO and TEMPO2 differ by a few \,$\mu$m/s. Considering the
  timing error in radial velocity data, PEXO is able to provide a
  practical radial velocity precision of 1\,cm/s. The main
  limitation of radial velocity precision comes from the bias in the
  determination of the appropriate mid-point of an exposure caused by effects such as the
  atmospheric chromatic aberration. We do not compare PEXO with the
  known high-precision astrometry package GREM because it is not
  publicly available. Considering the consistency between astrometry and radial velocity
  modeling, we expect the astrometry precision to be $\mu$as.

We test various effects in transit timing by applying PEXO to XO-3 b and Kepler-210 c. The relativistic effects are not significant enough to be detectable in these two cases. High cadence
and long term observations are needed to reliably detect relativistic
precession in short-period and eccentric-transit systems where
planet-induced precession is minimized. Follow-up work on how transit
timing is sensitive to various relativistic effects is needed for the potential
application of transit timing in the test of GR, as
done in pulsar timing. The Einstein delay in the target system
contributes to a 0.2\,s timing variation for $\alpha$ Centauri
over one decade but is not included in many previous packages
such as EXOFAST and GREM. The Einstein delay in some high-eccentricity
transit systems might be detectable in the timing difference between
the primary and secondary transits. We further investigate the
feasibility of using PEXO for relativity tests in binaries. The
gravitational redshift variation caused by the companion of the target
star is as large as a few m/s for 52 binaries with orbital periods
less than 10\,yr. Such tests are able to examine GR
and MOND theories in a new regime of stellar mass and weak gravitational field. 

Using $\alpha$ Centauri as an example, we assess relativistic effects
in the modeling of astrometry. We find that the lensing effect in the
solar system contributes to 9\,mas position variation, while the
lensing effect in $\alpha$ Centauri only contributes to 0.003$\mu$as
position variation over one decade. The barycentric motion of the
binary and the stellar aberration due to the Earth's motion, as well as the atmospheric refraction are the
major effects changing the observed direction of photons. For
ground-based astrometric observations, the atmospheric modeling
uncertainty limits the absolute positional precision to be 1\,arcsec
and the relative positional precision to be tens of $\mu$as. For space-based
observations, an astrometry precision of $\mu$as requires a telescope
ephemeris with 1\,cm/s velocity precision. An alternative approach is
to determine the Earth's motion {\it a posteriori} through a combined fit to the
data. The third-order geometric effect contributes to a few $\mu$as position bias and is not
included in the GREM package, which is used by {\it Gaia} for relativistic
astrometric solutions \citep{lindegren18}. 

We assess various effects in the radial velocity model using the
example of $\alpha$ Centauri A. The general and special relativistic
effects in the $\alpha$ Centauri system affect the radial velocity of
$\alpha$ Centauri A by 0.04 and 0.2\,m/s over one decade,
respectively. Although these effects are essential for sub-m/s radial
velocity modeling of binary systems, they are not accounted for in
radial velocity packages. The special relativity effect due to
$\alpha$ Centauri B changes the radial velocity by nearly 1\,m/s over
one orbital period. The binary motion of the target star would change
the viewing perspective, leading to a change in the projection of various motions
onto the radial velocity direction. Furthermore, errors in astrometry data would
lead to considerable radial velocity variation for nearby
stars. We find that decoupling could introduce $\sim$0.1\,m/s bias in one
  decade of radial velocity observations of a nearby star ($<$10\,pc) with
  hot or cold Jupiters and could introduce $\sim$1\,m/s bias over one year for
  nearby stars with stellar-mass companions. Therefore, a barycentric
correction of the measured radial velocity is not appropriate to
achieve 1\,cm/s. We suggest a combined modeling of stellar reflex motion, stellar proper motion, and the Earth's motion for high-precision radial velocity modeling.

The $\sim$1\,ns timing precision of PEXO and TEMPO2 can be achieved if the uncertainty in the ephemeris
  of solar system bodies is less than 1\,m and the effect of interstellar
  scattering is well understood (E06). On the other hand, the radial velocity precision of
  1\,$\mu$m is the software precision that is achievable only if we could
  model the observational effects to a high precision. For example, to achieve a precision of 1\,mm/s in radial
velocity modeling, we need to determine the midpoint of exposures in
spectroscopic observations to a precision of a few milli-seconds by
properly modeling the atmospheric chromatic aberration of incoming
photons. For future space-based spectrographs such as EarthFinder
  \citep{plavchan18}, the exposure time might be better determined and
  telluric effects would disappear, enabling $<$1\,cm/s radial velocity
  precision if combined with PEXO's precise modeling of astrophysical
  effects. PEXO’s precision can be further improved by the
    appropriate modeling of the Galactic acceleration of stars and the
    dispersion of photons in the interplanetary and interstellar
    medium. The extra packages envisaged for PEXO are (1) the Galactic
    acceleration of stars and the corresponding secular aberration as
    well as cosmological effects (e.g., \citealt{klioner03} and
    \citealt{lindegren03}) and (2) the inclusion of gravitational wave effects to provide an independent package for the detection of gravitational waves using pulsar timing arrays such as NANOGrav (e.g., \citealt{arzoumanian18}).

Based on our investigation of various relativistic effects and
  comparison of various packages, we summarize the main results of
  this paper and give relevant recommendations as follows:
  \begin{itemize}
    \item By accounting for relativistic and high-order geometric
      effects, PEXO is able to model timing to a precision of 1\,ns,
      astrometry to 1\,$\mu$as, and radial velocity to 1\,$\mu$m/s. PEXO
      is able to model multiple types of data precisely and consistently. 
    \item Decoupling of the target system and the solar system introduces
      considerable bias in the modeling of timing, astrometry and
      radial velocity. For stars with stellar mass companions and for
      nearby stars with Jupiter-mass companions, we recommend a
      combined modeling of binary motion, binary barycentric motion,
      and telescope ephemeris. An alternative, efficient approach is to model the
      decoupling trend bias using astrometric offsets and fit all
      motions to the data corrected for barycentric effects.
    \item A detectable relativistic effect in extrasolar systems is the
      gravitational redshift of the light from a target star caused by
      its companion star. This test is feasible for a few binaries
      with long-term and high-precision radial velocity data
      ($\sim$1\,m/s uncertainty), such as $\alpha$ Centauri A and B.
    \item Atmospheric effects limit absolute astrometry to a precision
      of 1\,arcsec and relative astrometry to a precision of tens of
      $\mu$as. Such precisions are only achievable if various
      atmospheric parameters are well measured at the observation site
      and the refraction effects are correctly modeled through
      reliable packages. To avoid high model uncertainty, we recommend an
      elevation angle of at least 30$^\circ$ for direct imaging of
      planets (see section \ref{sec:astrometry_binary}). 
    \item Second-order stellar aberration is needed to model space-based astrometric observations correctly. The third-order
      geometric effects become significant for decade-long observations
      of nearby stars. Because GREM is the core package used for {\it Gaia}'s
      astrometry solution \citep{lindegren18}, we recommend a more
      comprehensive analysis of astrometric epoch data for nearby stars to reveal potential planetary signals.
    \item The consideration of relativistic effects in the target
      system is the missing piece in previous exoplanet
      packages. The Einstein delay and gravitational Doppler shift
      are significant in some binary systems and are detectable with
      current technology.
    \item The coding error in the calculation of planetary Shapiro
      delay would bias TEMPO2 timing modeling by at least tens of nanoseconds
      over one decade. The bug in the calculation of parallax delay in
      {\small utc2bjd.pro} would bias its timing precision by
      $\sim$1\,ms for nearby stars.
    \item PEXO is tested by comparison with TEMPO2 and by recovering
      previous fitting results for the astrometric and radial velocity
      data for $\alpha$ Centauri. 
  \end{itemize}

In summary, PEXO provides a high-precision combined model for timing,
radial velocity, and astrometry data. PEXO is versatile enough to take
binary motions into account and precise enough to consider high-order
classical and relativistic effects. By applying PEXO to the
analysis of high-precision data provided by the state-of-the-art
facilities such as TESS, ESPRESSO, and {\it Gaia}, we expect to have the
ability to make a reliable detection of an Earth twin as well as a
test of GR in extrasolar systems in the near
future. 

\section*{Acknowledgements}
We are indebted to the anonymous referee of this paper for their
inspiring and insightful comments that led to very substantial improvements in the content and clarity of this manuscript and PEXO. M.L. is supported by a
 University of Hertfordshire PhD studentship. F.F. and H.J. acknowledge
 support from the UK Science and Technology Facilities Council
 [ST/M001008/1]. This work has made use of data from the European
 Space Agency (ESA) mission {\it Gaia} ({\url
   https://www.cosmos.esa.int/gaia}), processed by the {\it Gaia} Data
 Processing and Analysis Consortium (DPAC, {\url
   https://www.cosmos.esa.int/web/gaia/dpac/consortium}). Funding for
 the DPAC has been provided by national institutions, in particular the
institutions participating in the {\it Gaia} Multilateral Agreement. We adapt various SOFA routines (\url{http://www.iausofa.org}) to R functions in PEXO. 

\appendix
  \section{ Conventions in binary studies}\label{sec:convention}
The angular parameters for a binary orbit depend on the coordinate
system where it is defined. We introduce three coordinate systems and
corresponding orbital parameters in order to clarify the previous
conventions and to assist the community in using data consistently.

Before defining the coordinate system, we need to be clear what we mean by
rotation matrix, which is essential to transforming between different
coordinate systems. We define a rotation matrix $R$ as an operation to
rotate vector $\bm r_a$ to $\bm r_b$. For example, in a 2D Cartesian
system, we use the following matrix to rotate $\bm r_a$
counterclockwise by $\theta$ to get $\bm r_b$:
\begin{equation}
  {\bm r_b}\equiv
  \begin{bmatrix}
    x_b\\
    y_b\\
      \end{bmatrix}
  =
      \begin{bmatrix}
                \cos{\theta} & -\sin{\theta}\\
                \sin{\theta} &\cos{\theta} \\
              \end{bmatrix}
                \begin{bmatrix}
    x_a\\
    y_a\\
      \end{bmatrix}
      ~.
          \label{eqn:2Drotate}
        \end{equation}

For 3D rotation matrix in Cartesian coordinate system, we define an
counter-clockwise rotation around the x axis by $\theta$ as
\begin{equation}
  R_x(\theta)=
  \begin{bmatrix}
    1 & 0 & 0\\
    0&\cos{\theta} & -\sin{\theta}\\
    0&\sin{\theta} &\cos{\theta} \\
  \end{bmatrix}
  ~,
  \label{eqn:3Dx}
\end{equation}
a counterclockwise rotation around the y axis by $\theta$ as
\begin{equation}
  R_y(\theta)=
  \begin{bmatrix}
\cos{\theta}&0 & \sin{\theta}\\
0&1 &0\\
-\sin{\theta}&0&\cos{\theta}\\
  \end{bmatrix}
  ~,
  \label{eqn:3Dy}
\end{equation}
and a counterclockwise rotation around the z axis by $\theta$ as
\begin{equation}
  R_z(\theta)=
  \begin{bmatrix}
    \cos{\theta} & -\sin{\theta}&0\\
    \sin{\theta} &\cos{\theta}&0 \\
    0 & 0 & 1\\    
  \end{bmatrix}
  ~.
  \label{eqn:3Dz}
\end{equation}
For example, if the rotation sequence is to rotate vector $\bm r_a$ around the x
axis by $\theta_1$, y axis by $\theta_2$, and z axis by $\theta_3$, the
new vector $\bm r_b$ should be
\begin{equation}
  {\bm r_b}=  R_x(\theta_1)R_y(\theta_2)R_z(\theta_3){\bm r_a}~.
  \label{eqn:3Drot}
\end{equation}
If $\bm r_a$ is one of the axes of a coordinate system, the above
operation would lead to a new coordinate system defined by $\bm
r_b$. In other words, two coordinate systems are related to each
other by
\begin{equation}
  \begin{bmatrix}
    x'&y'&z'\\
  \end{bmatrix}
  \begin{bmatrix}
    {\bf e}_x'\\
    {\bf e}_y'\\
    {\bf e}_z'\\
  \end{bmatrix}
  =
  \begin{bmatrix}
    x&y&z\\
  \end{bmatrix}
  \begin{bmatrix}
    {\bf e}_x\\
    {\bf e}_y\\
    {\bf e}_z\\
  \end{bmatrix}
  ~,
\end{equation}
where $\bf e_x'$, $\bf e_y'$, $\bf e_z'$ are axes of the new
coordinate system, $\bf e_x$, $\bf e_y$, $\bf e_z$ are axes of the
old coordinate system, and $[x',y',z']$ and $[x,y,z]$ are respectively
the new and old coordinates of a vector. If the new coordinate system
is transformed from the old one by rotation matrix $R$, which is
orthogonal, the new coordinates would be
\begin{equation}
  \begin{bmatrix}
    x'\\
    y'\\
    z'\\
  \end{bmatrix}
  =R^{\rm T}
  \begin{bmatrix}
    x\\
    y\\
    z\\
  \end{bmatrix}
\end{equation}
For the rotation operations in Eqn. \ref{eqn:3Drot}, the coordinate
transformation would be
\begin{equation}
  \begin{bmatrix}
    x'\\
    y'\\
    z'\\
  \end{bmatrix}
  =R_z(-\theta_3)R_y(-\theta_2)R_x(-\theta_1)
  \begin{bmatrix}
    x\\
    y\\
    z\\
  \end{bmatrix}
~.
\end{equation}
We will apply the definition and conclusion in the above analysis to
the following subsections. We will introduce three conventions and
label the corresponding coordinates by subscripts 1, 2, and 3. In the
following conventions, we model the motion of the target object with
respect to the binary barycenter. It is also known as stellar reflex
motion if the star is the target. 

\subsection{Definition of orbital elements}\label{sec:element}
Before introducing coordinate systems, we define the orbital
  elements independent of the chosen coordinate system. By doing this,
  we emphasize that conventions for orbital elements and coordinate systems are independent. To distinguish
 these two conventions, we use ``definition'' to name the
 former and ``convention'' to name the latter. The orbit of the target
 is in the ``orbit plane'' while the plane perpendicular to
 $\bm{r}_{\rm SB}$ is called the ``sky plane''. To focus on the study of
 stellar reflex motion, we assume that the observer is located at the SSB and the proper motion of TSB
 is zero. According to Eqn. \ref{eqn:offset}, the offsets are
 $\xi\equiv\Delta\alpha*\approx \bm{p}_b\cdot \bm{r}_{\rm
   BT}(t)\tilde{\omega}^b/A$ and $\eta\equiv\Delta\delta\approx
 \bm{q}_b\cdot\bm{r}_{\rm BT}(t)\tilde{\omega}^b/A$. This assumption
 is only used for a convenient definition of orbital elements and is
 not used in PEXO modeling. Because the semi-major axis and eccentricity do not depend on the viewing geometry, we
 focus on the definition of angular orbital elements. 

The inclination $I$ is defined as the angle between the
target-to-observer direction (i.e. $-\bm{u}_b$) and the angular momentum. Thus, for an orbit with
$0^\circ<I<90^\circ$, the motion of the target with respect to the
binary barycenter is counterclockwise if viewed by the observer.
The line of nodes are at the intersection between the
orbit plane and the sky plane. The ascending node is the node
at which the target crosses the sky plane and moves away from the
observer. The longitude of ascending node $\Omega$ is the
counterclockwise angle in the sky plane from the north to the
ascending node viewed from the observer. For the motion of the target with respect to the
  system barycenter or to the companion, we use the argument of periastron $\omega_{\rm T}$ which is the angle in
the orbit plane from the ascending node to the periastron along the
motion of the target. For the orbit of a companion with
  respect to the barycenter or to the target, the argument of
  periastron is $\omega_{\rm C}=\omega_{\rm T}+\pi$. We will use the
  barycentric motion of the target as an example for the following
  introduction and discussions. 

  It is important to note that astrometry data alone cannot distinguish between the ascending and descending nodes because we
  do not know whether the star is moving away from or towards the
  observer. Thus we recommend restricting the range of $\omega$ and $\Omega$ to
  be $[0,\pi]$ for astrometry-only analyses. This restriction should
  not be used for combined analyses of astrometry and radial velocity
  data. We also recommend reporting $\omega+\pi$ and $\Omega+\pi$ as
  an alternative solution. These are practical suggestions for fitting
  that do not change the definitions of the ascending node. If only
  one set of $\omega$ and $\Omega$ values are reported based on an astrometry-only analysis, $\omega+\pi$ and $\Omega+\pi$ should be considered as an alternative solution during a combined analysis of astrometry and radial velocity data.

The orbital elements are illustrated in
Fig. \ref{fig:binary_orbit}. In principle, coordinate systems are not
needed because the orbital elements are defined in independently of
the coordinate system. We can transform the target motion
  from the orbit plane frame to the sky plane frame, which is formed by
  the north (increase of decl.), the East (increasing of R.A.), and the direction from the target to the
  observer. However, for the convenience of formalization, we can
  randomly assign these three axes or their opposite directions as $X$,
  $Y$, and $Z$ axes to form a left-handed or right-handed coordinate
  system for the sky plane frame. We will introduce three of them as follows.

\subsection{Convention I: right-handed coordinate
  system}\label{sec:conv1}
We use a right-handed convention where $+X$ direction is along the north
(or increasing declination), $+Y$ direction is along the East (or
increasing right ascension), $+Z$ axis is from the target to the
observer (or decreasing distance). In the orbit plane, +x axis is the
direction of periastron, $+z$ axis is along the direction of orbital angular
momentum of a binary, $+y$ axis is in the orbit plane and is chosen
such that $x$, $y$ and $z$ axes form a right-handed Cartesian coordinate
system. The rotation matrix transforming the orbital-plane frame to the sky
plane frame is $R_z(-\omega)R_x(I)R_z(-\Omega)$. Thus the coordinate transformation matrix\footnote{The rotation matrices defined in this work are the transposes of the corresponding ones defined in \cite{catanzarite10}.} would be its inverse or transpose, which is
$R_z(\Omega)R_x(-I)R_z(\omega)$.

The orbital-plane coordinates are transformed into the sky plane
coordinates following
\begin{equation}
  \begin{bmatrix}
    X_1\\
    Y_1\\
    Z_1\\
  \end{bmatrix}
  = R_z(\Omega)R_x(-I)R_z(\omega)
  \begin{bmatrix}
    x\\
    y\\
    0\\
  \end{bmatrix}
  ~.
  \label{eqn:conv11}
\end{equation}
The expansion of the above rotation matrices leads to
\begin{equation}
  \begin{bmatrix}
    X_1\\
    Y_1\\
    Z_1\\
  \end{bmatrix}
  =
  \begin{bmatrix}
   \cos{\Omega}\cos{\omega_{\rm T}}-\sin{\Omega}\sin{\omega_{\rm
       T}}\cos{I} &-\cos{\Omega}\sin{\omega_{\rm
       T}}-\sin{\Omega}\cos{\omega_{\rm T}}\cos{I}&-\sin{\Omega}\sin{I}\\
    \sin{\Omega}\cos{\omega_{\rm T}}+\cos{\Omega}\sin{\omega_{\rm
        T}}\cos{I}&-\sin{\Omega}\sin{\omega_{\rm
        T}}+\cos{\Omega}\cos{\omega_{\rm
        T}}\cos{I}&\cos{\Omega}\sin{I}\\
 -\sin{\omega_{\rm T}}\sin{I}&-\cos{\omega_{\rm T}}\sin{I}&\cos{I}\\
  \end{bmatrix}
  \begin{bmatrix}
    x\\
    y\\
    0\\
  \end{bmatrix}
  ~.
  \label{eqn:conv12}
\end{equation}
Ignoring the scaling between the orbital ellipse in the orbital-plane
and that in the sky plane, the elements of the matrix in the above
equation are also known as Thiele-Innes constants \citep{thiele83}:
\begin{eqnarray}
  A'&=&\cos\Omega \cos\omega_{\rm T} - \sin\Omega \sin\omega_{\rm
       T} \cos{I}\nonumber~,\\
  B'&=&\sin\Omega \cos\omega_{\rm T} + \cos\Omega \sin\omega_{\rm
       T} \cos{I}~,\\
  F'&=&-\cos\Omega \sin\omega_{\rm T}-\sin\Omega \cos\omega_{\rm T} \cos{I}\nonumber~,\\
  G'&=&-\sin\Omega \sin\omega_{\rm T} + \cos\Omega
       \cos\omega_{\rm T} \cos{I}\nonumber~.
\end{eqnarray}
Then Eqn. \ref{eqn:conv12} becomes
\begin{equation}
  \begin{bmatrix}
    X_1\\
    Y_1\\
    Z_1\\
  \end{bmatrix}
  =
  \begin{bmatrix}
      A' &F'&-\sin{\Omega}\sin{I}\\
B'&G'&\cos{\Omega}\sin{I}\\
   -\sin\omega_{\rm T} \sin{I}&-\cos\omega_{\rm T} \sin{I}&\cos{I}\\
  \end{bmatrix}
  \begin{bmatrix}
    x\\
    y\\
    0\\
  \end{bmatrix}
  ~.
  \label{eqn:conv13}
\end{equation}

\subsection{Convention II: left-handed coordinate system}\label{sec:conv2}
In the study of binaries, another convention closely
related to convention I is also frequently used. In this convention,
the $+Z$ axis is along the increasing distance, leading to a left-handed
Cartesian coordinate system. To keep the convention that a counterclockwise
orbit on the sky plane viewed from the observer corresponds to an
inclination of $0^\circ - 90^\circ$, we keep the definition of inclination
as well as other orbital elements in convention I. Thus the new coordinates
in this convention are related to the coordinates in convention I by
\begin{equation}
  \begin{bmatrix}
    X_2\\
    Y_2\\
    Z_2\\
  \end{bmatrix}
  =
  \begin{bmatrix}
    X_1\\
    Y_1\\
    -Z_1\\
  \end{bmatrix}
  =  \begin{bmatrix}
      A' &F'&-\sin{\Omega}\sin{I}\\
B'&G'&\cos{\Omega}\sin{I}\\
   \sin\omega_{\rm T} \sin{I}&\cos\omega_{\rm T} \sin{I}&-\cos{I}\\
  \end{bmatrix}
  \begin{bmatrix}
    x\\
    y\\
    0\\
  \end{bmatrix}
  ~.
  \label{eqn:conv2}
  \end{equation}

  \subsection{Convention III: astrometric and precession-compatible convention}\label{sec:conv3}
  In this study, we used the right-handed coordinate system formed by
  the triad $\bm{[p,q,u]}$. Here, $\bm p$ is in the direction of increasing
  R.A., $\bm q$ is in the direction of increasing
  decl., and $\bm u$ is in the direction of increasing
  distance. Thus, $\bm p$ and $\bm q$ axes in this convention
  correspond to the $Y$ and $X$ axes in convention II, respectively. Because
  this $\bm{[p,q,u]}$ triad is used in the astromery models of
  {\it Hipparcos} and {\it Gaia} \citep{esa97,lindegren11}, we call the
  corresponding convention of binary motion the ``astrometric
  convention.'' By using the same definition of orbital elements to
  avoid unnecessary confusion, we get the new coordinates in this convention by
\begin{equation}
  \begin{bmatrix}
    Y_3\\
    X_3\\
    Z_3\\
  \end{bmatrix}
  =
  \begin{bmatrix}
    X_2\\
    Y_2\\
    Z_2\\
  \end{bmatrix}
  ~.
  \label{eqn:conv3}
\end{equation}
From the above transformation, we derive the position of the target
relative to TSB as
\begin{equation}
  {\bf r}_{\rm BT}=
  \begin{bmatrix}
    X_3\\
    Y_3\\
    Z_3\\
  \end{bmatrix}
  ~.
\end{equation}
This convention is also used in TEMPO2 (E06) although they set the
+x direction in the orbital-plane as the ascending node considering
the precession of periastron. We call this convention the
``precession-compatible convention '' due to its consideration of
precession of periastron. Thus the rotation matrix in
Eqn. \ref{eqn:rBT} is the same as the one in Eqn. \ref{eqn:conv2} and
in equation 54 of E06 if setting $\omega=0$ and swapping the first and
second rows.

We emphasize that the conventions for coordinate systems are used
  for transformation from the orbit plane to the sky plane
  frames. Like the choice of TDB and TCB time standards, the choice of coordinate system does not matter as long as the
stellar reflex motion is correctly modeled in offset coordinates
(i.e. $\Delta \alpha*$ and $\Delta \delta$). Considering the variety
of conventions used in binary studies, we recommend a consistent definition and introduction of conventions including orbital elements and coordinate systems in the studies of binaries and exoplanets.

\subsection{Comments on previous conventions}\label{sec:conv1}
 The conventions for binaries and exoplanets are not always
  clearly presented and consistently used in the literature. To help
  the community use conventions consistently, we comment on some of
  the previous conventions that are not consistently defined or presented. 

Although we have used the definition of orbital elements in
\cite{catanzarite10} for all three conventions of coordinate systems,
\cite{catanzarite10} did not present his convention consistently in his
figure 1. He used an image from Wikipedia
(\url{https://en.wikipedia.org/wiki/Longitude_of_the_ascending_node#/media/File:Orbit1.svg})
to illustrate his convention, although this image is more suitable for
the studies of the solar system. In his figure 1, the observer
views the planetary system from above the reference plane. The target moves toward the observer after crossing the
ascending node. Thus the ascending node should be the descending node in the
figure according to the definition used by
\cite{catanzarite10}. Although the ascending node would be correct if
the observer views the system from the south (or $-Z$ direction), the
inclination would be incorrect because the orbit would be retrograde
and thus the inclination should be larger than 90$^\circ$ (i.e. 180$^\circ$
minus the inclination presented in Figure 1). 

In solar system studies, the ascending node is a point in the orbit
and sky planes where an object moves from the south to the north (or
$+Z$ direction) of the reference plane. Because the observer is absent in
the definition, we call this definition the ``observer-independent ascending node''. Nevertheless, the ascending node defined in binary studies
 is the point both in the orbit plane and in the sky plane where the
 target moves away from the observer. Because it depends on the direction of the observer, we call this definition ``observer-dependent ascending
node''. The observer-dependent ascending node is consistent with the
convention that the radial velocity of a target is positive if it is
moving away from us. Therefore, the observer-dependent ascending
node is correctly defined by \cite{catanzarite10} but is presented as
the observer-independent ascending node. The differences between
observer-dependent and observer-independent ascending nodes are intrinsic and do not depend on the choice of coordinate systems.

Similarly, figures 4 and 7 in \cite{murray10} and figure 31 in
\cite{binnendijk60} show that they use the observer-independent ascending
node to study binaries and exoplanets. Although the ascending node is not consistently defined in \cite{murray10} and
\cite{binnendijk60}, they transform the binary motion from the
orbit plane to sky plane (or reference plane) correctly. Their use of the
observer-independent ascending node would lead to a sign flip in the
value of radial velocity (i.e. $-Z_1$; see
Eqn. \ref{eqn:conv12}). This sign flip would be absorbed through the
fit of $\omega_T$ to the radial velocity data, leading to a value of
$\omega_T$ inconsistent with their definition of ascending node. In
particular, it would become problematic for astrometry modeling where
the proper motion of the barycenter and the stellar reflex motion need
to be combined.

Compared with previous illustrations of binary motion,
Fig. \ref{fig:binary_orbit} consistently visualizes the observer-dependent
ascending node. The star is moving away from the
observer after crossing the ascending node. The longitude of the ascending
node $\Omega$ is measured counterclockwise from the north to the
ascending node from the observer's perspective. The inclination is the angle
between the angular momentum and the $-{\bm u}_b$ direction so that the
orbit is prograde for $0^\circ<I<90^\circ$ and retrograde for $I>90^\circ$
viewed by the observer. The requirement of consistency makes
Fig. \ref{fig:binary_orbit} different from the equivalent ones
frequently used in the literature. 

\section{Acronyms and symbols}
\startlongtable
\begin{deluxetable}{llr}
\tablecaption{Meanings of acronyms and the sections where
  they first appear. For vectors with opposite signs, only one
of them is shown here. The magnitude or length of a vector is denoted
by the same symbol but without bold faced font. Many secondary
variables derived from other variables are not shown. Acronyms are shown
first and are followed by English and Greek mathematical symbols. For
each category, the acronyms or symbols are listed in the alphabetic order.}
\tablehead{
  \colhead{Acronym or Symbol}&\colhead{Meaning} & \colhead{Sections}
}
\startdata
BCRS & Barycentric Celestial Reference system  & \ref{sec:geometric}\\
BIPM & International Bureau of Weights and Measures &\ref{sec:delay_solar}\\
BIPMXY & BIPM realization of TT in year 19XY if XY$>$20 or 20XY if XY$<$20 & \ref{sec:delay_solar}\\
BJD & Barycentric Julian date &\ref{sec:delay_solar}\\
C & Companion in the target system & \ref{sec:geometric}\\
DD & Post-Newtonian binary model proposed by \cite{damour86} &\ref{sec:PN}\\
DDGR & General relativity binary model proposed by \cite{damour86} &
\ref{sec:PN}\\
GR & General relativity & \ref{sec:introduction}\\
JD&Julian date&\ref{sec:delay_solar}\\
MJD&Modified Julian date&\ref{sec:rv_binary}\\
NCP & North Celestial Pole&\ref{sec:reflex}\\
PEXO & Precision Exoplanetology & \ref{sec:introduction}\\
PPN& Parameterized post-Newtonian formalism&\ref{sec:abberration}\\
O & Observatory site & \ref{sec:astrometry}\\
RV & Radial velocity & \ref{sec:introduction}\\
SS & Solar system & \ref{sec:astrometry}\\
SSB or S & Solar system barycenter & \ref{sec:geometric}\\
T & Target star& \ref{sec:geometric}\\
TAI & International Atomic Time & \ref{sec:delay_solar}\\
TCB & Barycentric Coordinate Time&\ref{sec:delay_solar}\\
TCG & Geocentric Coordinate Time&\ref{sec:delay_solar}\\
TDB & Barycentric Dynamical Time&\ref{sec:delay_solar}\\
TDF & Transit Duration Fraction&\ref{sec:ttv}\\
TDV & Transit Duration Variation&\ref{sec:ttv}\\
TS & Target system& \ref{sec:astrometry}\\
TSB or B & Target system barycenter& \ref{sec:astrometry}\\
TT & Terrestrial Time or proper time at the geoid&\ref{sec:delay_solar}\\
TTV & Transit Timing Variation&\ref{sec:ttv}\\
UTC & Coordinated Universal Time & \ref{sec:delay_solar}\\
\hline
$A$&1 au&\ref{sec:astrometry}\\
$a$&Semi-major axis of the target star with respect to the
companion&\ref{sec:reflex}\\
$a_{\rm C}$&Semi-major axis of the barycentric orbit of the companion&\ref{sec:reflex}\\
$a_r$& Counterpart of $a_{\rm T}$ in the DD model\\
$a_{\rm T}$&Semi-major axis of the barycentric orbit of the
  target star&\ref{sec:reflex}\\
$c$&Speed of light&\ref{sec:PN}\\
$\dot{e}$&Time derivative $e$ in DD model&\ref{sec:PN}\\
$e_0$&Eccentricity at a reference time in post-Newtonian models&\ref{sec:reflex}\\
$E$&Eccentric anomaly of the barycentric orbit of the target star&\ref{sec:reflex}\\
$f$&True anomaly of the barycentric orbit of the target star&\ref{sec:reflex}\\
$G$ & Gravitational constant&\ref{sec:PN}\\
$g$ & Timing model parameter in the DD model&\ref{sec:PN}\\
$h_{\rm O}$&Altitude of the observer&\ref{sec:tropo}\\
$I$&Orbital inclination of the target star with respect to the TSB&\ref{sec:reflex}\\
${\bm l}_e$&Light ray direction at the emission time&\ref{sec:abberration}\\
${\bm l}_i$&Incident light ray before entering atmosphere&\ref{sec:deflection}\\
${\bm l}_l$&Light ray direction after lensing by the companion&\ref{sec:abberration}\\
${\bm l}_o$&Light ray direction at the observation time&\ref{sec:abberration}\\
$L_B$&Scaling factor for the transformation between TCB and TDB&\ref{sec:delay_solar}\\
$L_G$&Scaling factor for the transformation between TT and TCG&\ref{sec:delay_solar}\\
$\mathcal L$&Light ray path in the atmosphere&\ref{sec:tropo}\\
$m_{\rm C}$&Mass of the companion&\ref{sec:reflex}\\
$m_h$&Mapping function for hydrostatic delay&\ref{sec:reflex}\\
$m_{\rm T}$&Mass of the target star&\ref{sec:reflex}\\
$m_w$&Mapping function for wet delay&\ref{sec:reflex}\\
$m_\odot$ &Mass of the Sun &\ref{sec:reflex}\\
$n$&Mean motion of the target binary orbit&\ref{sec:reflex}\\
$n_{\rm O}$&Refraction index at the telescope&\ref{sec:refraction}\\
$N_{\rm wet}$&Refractivity of wet component&\ref{sec:tropo}\\
$N_{\rm hydro}$& Refractivity of hydrostatic component&\ref{sec:tropo}\\
$P$&Orbital period of the target system&\ref{sec:relativity_test}\\
$P_0$&Orbital period at a reference time in post-Newtonian models&\ref{sec:reflex}\\
$\dot{P}$&Time derivative of $P$ in the DD model&\ref{sec:reflex}\\
$p_{\rm O}$&Air pressure at the observation site&\ref{sec:tropo}\\
${\bm p}_b$ &  Unit vector in the directions of increasing R.A. $\alpha$ at the reference time $t_0$& \ref{sec:astrometry}\\
${\bm q}_b$ &  Unit vector in the directions of increasing decl. $\delta$ at the reference time $t_0$& \ref{sec:astrometry}\\
$r_s$& Range parameter of Shapiro delay&\ref{sec:astrometry}\\
$r_s^{\rm GR}$& Range parameter of Shapiro delay assuming general relativity&\ref{sec:astrometry}\\
${\bm r}_{\rm BT}$ & Vector from B to T & \ref{sec:astrometry}\\
${\bm r}_{\rm CT}$ & Vector from C to T & \ref{sec:special_shift}\\
${\bm r}_{\rm OB}$ & Vector from O to B & \ref{sec:astrometry}\\
${\bm r}_{\rm OC}$ & Vector from O to C & \ref{sec:special_shift}\\
${\bm r}_{\rm orb}$& Barycentric position of the target in the orbital plane&\ref{sec:reflex}\\
${\bm r}_{\rm OS}$ & Vector from O to S & \ref{sec:astrometry}\\
${\bm r}_{\rm OT}$ & Vector from O to T & \ref{sec:astrometry}\\
${\bm r}_{\rm SB}$ & Vector from S to B & \ref{sec:astrometry}\\
${\bm r}_{\rm ST}$ & Vector from S to T & \ref{sec:delay_solar}\\
${\bm r}_\oplus$ &  Barycentric position of the geocenter&\ref{sec:delay_solar}\\
${\bm R}$&$\left[{\bm r}_{\rm BT}(t) -{\bm r}_{\rm
    SO}(t)\right]\widetilde\omega^b/A$&\ref{sec:astrometry}\\
$\bm{\mathcal{R}}$&Refraction vector&\ref{sec:astrometry}\\
${\bm s}$& Position of observatory with respect to the geocenter&\ref{sec:astrometry}\\
$s_s$& Shape parameter of Shapiro delay&\ref{sec:astrometry}\\
$s_s^{\rm GR}$& Shape parameter of Shapiro delay assuming general relativity&\ref{sec:astrometry}\\
$t$& Arbitrary coordinate time&\ref{sec:astrometry}\\
$t_0$& Arbitrary reference coordinate time, $t_0=t_{\rm pos}$ by default&\ref{sec:astrometry}\\
$t_a^{\rm SSB}$& Light arrival time at SSB&\ref{sec:astrometry}\\
$t_a^{\rm TSB}$& Light arrival time at TSB&\ref{sec:astrometry}\\
$t_e$& Coordinate time of light emission&\ref{sec:astrometry}\\
$t_i$ &Arrival time of incident light ray without atmospheric refraction& \ref{sec:special_shift}\\
$t_{\mathcal G}$&Vacuum light propagation time for straight light ray& \ref{sec:tropo}\\
$t_{\mathcal L}$&Vacuum light propagation time for deflected light ray& \ref{sec:tropo}\\
$t_o$& Coordinate time of light arrival at the
observatory&\ref{sec:astrometry}\\
$t_{\rm pos}$& Reference epoch when the position or astrometry of the target star is measured&\ref{sec:relativistic}\\
$T_0$&Proper time of periastron in post-Newtonian
models&\ref{sec:reflex}\\
$T_c$&Midtransit time&\ref{sec:ttv}\\
$T_\odot$&$Gm_\odot/c^3$; Half the light travel time across the solar Schwarzschild radius& \ref{sec:reflex}\\
$U$ &  Relativistic eccentric anomaly &  \ref{sec:reflex}\\
$U_\oplus$ &  Gravitational potential of all solar system objects apart from the Earth at the observatory&  \ref{sec:delay_solar}\\
${\bm u}_b$ & Unit vector from S to B at the reference time $t_0$&\ref{sec:astrometry}\\
${\bm u}_o$&Observed direction of arriving light ray&\ref{sec:abberration}\\
$\hat{\bm u}_o$&Model prediction of the observed direction of arriving
light ray&\ref{sec:abberration}\\
${\bm u}_{\rm OT}$ & Unit vector from O to T & \ref{sec:rv}\\
${\bm u}_{\rm SB}$ & Unit vector from S to B & \ref{sec:delay_solar}\\
${\bm u}_{\rm ST}$ & Unit vector from S to T & \ref{sec:astrometry}\\
${\bm u}_{\rm Z}$ & Unit vector in the zenith direction & \ref{sec:astrometry}\\
$v_e$ & Escape velocity &\ref{sec:VG_shift}\\
${\bm v}_{\rm BT}$ & Velocity of T with respect to B & \ref{sec:delay_target}\\
${\bm v}_{\rm CT}$ & Velocity of T with respect to C & \ref{sec:special_shift}\\
${\bm v}_{\rm OS}$ & Velocity of S with respect to O & \ref{sec:rv}\\
${\bm v}_{\rm OT}$ & Velocity of T with respect to O & \ref{sec:rv}\\
$\bar{\bm v}_{\rm reflex}$&Average reflex motion of the target star &\ref{sec:decoupling}\\
${\bm v}_r^b$ &${\bm v}_{\rm OT}$ at the reference epoch $t_0$ &\ref{sec:astrometry}\\
${\bm v}_r^{\rm obs}$ &Observed absolute radial velocity& \ref{sec:astrometry}\\
${\bm v}_{\rm SB}$ & Velocity of B with respect to S & \ref{sec:rv}\\
${\bm v}_{\rm ST}$ & Velocity of T with respect to S &\ref{sec:VG_shift}\\
${\bm v}_{\rm TC}$ & Velocity of C with respect to T &\ref{sec:ttv}\\
$v_{\rm tot}$ & Characteristic radial velocity of a star with respect to the observer&\ref{sec:decoupling}\\
$v_\oplus$ &  Barycentric velocity of the geocenter&\ref{sec:delay_solar}\\
$W_0$&Gravitational and spin potential of the Earth at the
geoid&\ref{sec:delay_solar}\\
$x$& $x$ Coordinate in the orbital plane&\ref{sec:reflex}\\
$y$& $y$ Coordinate in the orbital plane&\ref{sec:reflex}\\
$x_a$&$a$sin$I$/c; Light travel time across the projected semi-major axis& \ref{sec:PN}\\
$\dot{x}_a$&Time derivative of $x_a$ in the DD
model&\ref{sec:reflex}\\
$Z_i$&Zenith angle of incident light ray&\ref{sec:refraction}\\
$Z_o$&Observed zenith angle&\ref{sec:refraction}\\
$z_{\rm kS}$&Doppler shift due to kinematic effects in the solar system&\ref{sec:VG_shift}\\
$z_{\rm kT}$&Doppler shift due to kinematic effects in the target system&\ref{sec:VG_shift}\\
$z_{\rm lS}$&Doppler shift due to kinematic effects in the solar system&\ref{sec:VG_shift}\\
$z_{\rm lT}$&Doppler shift due to kinematic effects in the target system&\ref{sec:VG_shift}\\
$z_{\rm grS}$&Doppler shift due to general relativity effects in the solar system&\ref{sec:VG_shift}\\
$z_{\rm grT}$&Doppler shift due to general relativity effects in the target system&\ref{sec:VG_shift}\\
$z_{\rm srS}$&Doppler shift due to special relativity effects in the solar system&\ref{sec:VG_shift}\\
$z_{\rm srT}$&Doppler shift due to special relativity effects in the target system&\ref{sec:VG_shift}\\
$z_{\rm S}$&Doppler shift due to all effects in the solar system&\ref{sec:VG_shift}\\
$z_{\rm T}$&Doppler shift due to all effects in the target system&\ref{sec:VG_shift}\\
\hline
$\alpha^b$ & BCRS R.A. of the TSB at the reference time $t_0$& \ref{sec:astrometry}\\
$\gamma$&One PPN parameter&\ref{sec:abberration}\\
$\phi_{\rm O}$&Latitude of the observer&\ref{sec:tropo}\\
$\Phi_{\rm T}$&Gravitational potential at the target star center&\ref{sec:relativity_test}\\
$\delta^b$ & BCRS decl. of the TSB at the reference time $t_0$&\ref{sec:astrometry}\\
$\delta t$&Arbitrary time span &\ref{sec:decoupling}\\
$\delta T$& Difference between the BIPM and TAI realizations of TT
&\ref{sec:delay_solar}\\
$\delta v_r^{\rm grT}$&Radial velocity variation due to general relativity in TS &\ref{sec:VG_shift}\\
$\delta v_r^{\rm srT}$& Radial velocity variation due to special relativity in TS &\ref{sec:VG_shift}\\
$\delta v_r^{\rm trend}$& Trend decoupling bias in radial velocity &\ref{sec:decoupling}\\
$\delta v_r^{\rm period}$& Periodic decoupling bias in radial velocity &\ref{sec:decoupling}\\
$\delta u$& Positional bias &\ref{sec:decoupling}\\
$\delta u_{\rm trend}$& Trend decoupling bias in target position &\ref{sec:decoupling}\\
$\delta u_{\rm period}$& Periodic decoupling bias in target position &\ref{sec:decoupling}\\
$\delta \Delta^{\rm trend}$& Trend decoupling bias in timing &\ref{sec:decoupling}\\
$\delta \Delta^{\rm period}$& Periodic decoupling bias in timing &\ref{sec:decoupling}\\
$\delta z_{\rm grT}$&Doppler shift variation due to general relativity in TS &\ref{sec:VG_shift}\\
$\delta z_{\rm srT}$& Doppler shift variation due to special relativity in TS &\ref{sec:VG_shift}\\
$\delta \mu$& Proper motion bias caused by ignoring the target reflex motion &\ref{sec:decoupling}\\
$\Delta\alpha*$ & Offset in the R.A. direction&\ref{sec:astrometry_binary}\\
$\Delta\delta$ & Offset in the decl. direction &\ref{sec:astrometry_binary}\\
$\Delta\xi$ & $\xi$ of the secondary with respect to the primary &\ref{sec:astrometry}\\
$\Delta\eta$ & $\eta$ of the secondary 1 with respect to the primary&\ref{sec:astrometry}\\
$\Delta{\bm r}$&${\bm r}_{\rm BT2}(t) -{\bm r}_{\rm BT1}(t)$; Position
of the secondary with respect to the primary&\ref{sec:astrometry}\\
$\Delta T_c^{\rm GR}$&Transit timing variation due to general relativity&\ref{sec:ttv}\\
$\Delta v_r$&Relative radial velocity&\ref{sec:rv}\\
$\Delta \hat{v}_r$&Model estimation of relative radial velocity&\ref{sec:rv}\\
$\Delta v_0$&Radial velocity offset&\ref{sec:rv}\\
$\Delta_{\rm ei}$& Einstein delay due to the relative velocity between the SSB to the TSB&\ref{sec:timing}\\
$\Delta_{\rm eS}$& Einstein delay in the solar system&\ref{sec:timing}\\
$\Delta_{\rm eT}$& Einstein delay in the target system&\ref{sec:timing}\\
$\Delta_{\rm geo}$& Time delay due to geometric effects&\ref{sec:timing}\\
$\Delta_{\rm hydro}$& Hydrostatic component in tropospheric delay &\ref{sec:timing}\\
$\Delta_{\rm is}$& Time delay in interstellar medium &\ref{sec:timing}\\
$\Delta_{\rm pS}$& Parallax delay in the solar system&\ref{sec:timing}\\
$\Delta_{\rm pT}$& Parallax delay in the target system&\ref{sec:timing}\\
$\Delta_{\rm sS}$& Shapiro delay in the solar system &\ref{sec:timing}\\
$\Delta_{\rm sT}$& Shapiro delay in the target system &\ref{sec:timing}\\
$\Delta_{\rm rS}$& Roemer delay in the solar system&\ref{sec:timing}\\
$\Delta_{\rm rT}$& Roemer delay in the target system&\ref{sec:timing}\\
$\Delta_{\rm S}$& Time delay in the solar system &\ref{sec:timing}\\
$\Delta_{\rm T}$& Time delay in the target system &\ref{sec:timing}\\
$\Delta_{\rm tropo}$& Tropospheric delay &\ref{sec:timing}\\
$\Delta_{\rm vp}$& Vacuum propagation delay due to the light travel between TSB and SSB&\ref{sec:timing}\\
$\Delta_{\rm wet}$& Wet component in tropospheric delay &\ref{sec:timing}\\
$\lambda_e$ &Wavelength of light when emitted&\ref{sec:VG_shift}\\
$\lambda_o$ &Wavelength of light when observed&\ref{sec:VG_shift}\\
$\eta$ & Projection of ${\bm r}_{\rm OT}$ on ${\bm
  q}_b$&\ref{sec:astrometry}\\
$\Theta$ & Elevation angle&\ref{sec:tropo}\\
$\bm\mu$ &${\bm p}_b\mu_\alpha^b+{\bm q}_b\mu_\delta^b$; Total proper motion of the TSB at the reference time $t_0$ & \ref{sec:astrometry}\\
$\mu_\alpha^b$ & Proper motion in R.A. of the TSB at the reference time $t_0$& \ref{sec:astrometry}\\
$\mu_\delta^b$ & Proper motion in decl. of the TSB at the reference time $t_0$& \ref{sec:astrometry}\\
$\mu_r^b$  & $v_r^b \widetilde\omega^b/A$; Proper motion equivalent of radial velocity at the reference time $t_0$&\ref{sec:astrometry}\\
$\nu_e$ &Frequency of light when emitted &\ref{sec:VG_shift}\\
$\nu_o$ &Frequency of light when observed &\ref{sec:VG_shift}\\
$\xi$ & Projection of ${\bm r}_{\rm OT}$ on ${\bm p}_b$&\ref{sec:astrometry}\\
$\tau_e$& Proper time of light emission &\ref{sec:timing}\\
$\tau_o$& Proper time of light arrival at the
observatory&\ref{sec:timing}\\
$\tau_p$& Proper time of periastron&\ref{sec:timing}\\
$\omega$&Argument of periastron of the target star with respect to the TSB&\ref{sec:reflex}\\
$\omega_0$&Argument of periastron at a reference time in post-Newtonina models&\ref{sec:reflex}\\
$\tilde{\omega}^b$ & Annual parallax of the TSB at the reference time $t_0$& \ref{sec:astrometry}\\
$\omega_{\rm C}$&Argument of periastron of the barycentric orbit of the companion star&\ref{sec:reflex}\\
$\omega_{\rm T}$&Argument of periastron of the barycentric orbit of the target star&\ref{sec:reflex}\\
$\dot{\omega}$&Time derivative of $\omega$ in the DD model&\ref{sec:reflex}\\
$\Omega$&Longitude of ascending node of the barycentric orbit of the target star&\ref{sec:reflex}\\
\enddata
\end{deluxetable}

\bibliographystyle{aasjournal}
\bibliography{nm}

\begin{thebibliography}{}
\expandafter\ifx\csname natexlab\endcsname\relax\def\natexlab#1{#1}\fi

\bibitem[{{Abbott} {et~al.}(2016){Abbott}, {Abbott}, {Abbott}, {Abernathy},
  {Acernese}, {Ackley}, {Adams}, {Adams}, {Addesso}, {Adhikari}, \&
  et~al.}]{gw150914}
{Abbott}, B.~P., {Abbott}, R., {Abbott}, T.~D., {et~al.} 2016, Physical Review
  Letters, 116, 061102

\bibitem[{{Abbott} {et~al.}(2017){Abbott}, {Abbott}, {Abbott}, {Acernese},
  {Ackley}, {Adams}, {Adams}, {Addesso}, {Adhikari}, {Adya}, \&
  et~al.}]{gw170817}
---. 2017, Physical Review Letters, 119, 161101

\bibitem[{Arzoumanian {et~al.}(2018)Arzoumanian, Baker, Brazier, Burke-Spolaor,
  Chamberlin, Chatterjee, Christy, Cordes, Cornish, Crawford,
  {et~al.}}]{arzoumanian18}
Arzoumanian, Z., Baker, P., Brazier, A., {et~al.} 2018, The Astrophysical
  Journal, 859, 47

\bibitem[{{Auer} \& {Standish}(2000)}]{auer00}
{Auer}, L.~H., \& {Standish}, E.~M. 2000, \aj, 119, 2472

\bibitem[{{Banik} \& {Zhao}(2018)}]{banik18}
{Banik}, I., \& {Zhao}, H. 2018, \mnras, 480, 2660

\bibitem[{{Barstow} {et~al.}(2005){Barstow}, {Bond}, {Holberg}, {Burleigh},
  {Hubeny}, \& {Koester}}]{barstow05}
{Barstow}, M.~A., {Bond}, H.~E., {Holberg}, J.~B., {et~al.} 2005, \mnras, 362,
  1134

\bibitem[{{Beichman} {et~al.}(2014){Beichman}, {Benneke}, {Knutson}, {Smith},
  {Dressing}, {Latham}, {Deming}, {Lunine}, {Lagage}, {Sozzetti}, {Beichman},
  {Sing}, {Kempton}, {Ricker}, {Bean}, {Kreidberg}, {Bouwman}, {Crossfield},
  {Christiansen}, {Ciardi}, {Fortney}, {Albert}, {Doyon}, {Rieke}, {Rieke},
  {Clampin}, {Greenhouse}, {Goudfrooij}, {Hines}, {Keyes}, {Lee}, {McCullough},
  {Robberto}, {Stansberry}, {Valenti}, {Deroo}, {Mand ell}, {Ressler},
  {Shporer}, {Swain}, {Vasisht}, {Carey}, {Krick}, {Birkmann}, {Ferruit},
  {Giardino}, {Greene}, \& {Howell}}]{beichman14}
{Beichman}, C., {Benneke}, B., {Knutson}, H., {et~al.} 2014, arXiv e-prints,
  arXiv:1411.1754

\bibitem[{{Binnendijk}(1960)}]{binnendijk60}
{Binnendijk}, L. 1960, {Properties of double stars; a survey of parallaxes and
  orbits.}

\bibitem[{{Blandford} \& {Teukolsky}(1976)}]{blandford76}
{Blandford}, R., \& {Teukolsky}, S.~A. 1976, \apj, 205, 580

\bibitem[{{Butkevich} \& {Lindegren}(2014)}]{butkevich14}
{Butkevich}, A.~G., \& {Lindegren}, L. 2014, \aap, 570, A62

\bibitem[{{Capitaine} \& {Wallace}(2006)}]{capitaine06}
{Capitaine}, N., \& {Wallace}, P.~T. 2006, \aap, 450, 855

\bibitem[{Capitaine {et~al.}(2003)Capitaine, Wallace, \&
  Chapront}]{capitaine03}
Capitaine, N., Wallace, P.~T., \& Chapront, J. 2003, Astronomy \& Astrophysics,
  412, 567

\bibitem[{{Catanzarite} {et~al.}(2006){Catanzarite}, {Shao}, {Tanner}, {Unwin},
  \& {Yu}}]{catanzarite06}
{Catanzarite}, J., {Shao}, M., {Tanner}, A., {Unwin}, S., \& {Yu}, J. 2006,
  \pasp, 118, 1319

\bibitem[{{Catanzarite}(2010)}]{catanzarite10}
{Catanzarite}, J.~H. 2010, ArXiv e-prints, arXiv:1008.3416

\bibitem[{{CHIME/FRB Collaboration} {et~al.}(2019){CHIME/FRB Collaboration},
  {Amiri}, {Bandura}, {Bhardwaj}, {Boubel}, {Boyce}, {Boyle}, {.~Brar},
  {Burhanpurkar}, {Cassanelli}, {Chawla}, {Cliche}, {Cubranic}, {Deng},
  {Denman}, {Dobbs}, {Fandino}, {Fonseca}, {Gaensler}, {Gilbert}, {Gill},
  {Giri}, {Good}, {Halpern}, {Hanna}, {Hill}, {Hinshaw}, {H{\"o}fer},
  {Josephy}, {Kaspi}, {Landecker}, {Lang}, {Lin}, {Masui}, {Mckinven},
  {Mena-Parra}, {Merryfield}, {Michilli}, {Milutinovic}, {Moatti}, {Naidu},
  {Newburgh}, {Ng}, {Patel}, {Pen}, {Pinsonneault-Marotte}, {Pleunis},
  {Rafiei-Ravandi}, {Rahman}, {Ransom}, {Renard}, {Scholz}, {Shaw}, {Siegel},
  {Smith}, {Stairs}, {Tendulkar}, {Tretyakov}, {Vanderlinde}, \&
  {Yadav}}]{chime19}
{CHIME/FRB Collaboration}, {Amiri}, M., {Bandura}, K., {et~al.} 2019, \nat,
  566, 235

\bibitem[{{Cosine-100 Collaboration} {et~al.}(2018){Cosine-100 Collaboration},
  {Adhikari}, {Adhikari}, {Barbosa de Souza}, {Carlin}, {Choi}, {Djamal},
  {Ezeribe}, {Ha}, {Hahn}, {Hubbard}, {Jeon}, {Jo}, {Joo}, {Kang}, {Kang},
  {Kauer}, {Kim}, {Kim}, {Kim}, {Kim}, {Kim}, {Kim}, {Kim}, {Kim}, {Ko},
  {Kudryavtsev}, {Lee}, {Lee}, {Lee}, {Lee}, {Leonard}, {Lynch}, {Maruyama},
  {Mouton}, {Olson}, {Park}, {Park}, {Park}, {Park}, {Park}, {Pettus},
  {Prihtiadi}, {Ra}, {Rott}, {Scarff}, {Shin}, {Spooner}, {Thompson}, {Yang},
  \& {Yong}}]{cosine100}
{Cosine-100 Collaboration}, {Adhikari}, G., {Adhikari}, P., {et~al.} 2018,
  \nat, 564, 83

\bibitem[{{Damour} \& {Deruelle}(1986)}]{damour86}
{Damour}, T., \& {Deruelle}, N. 1986, Ann.~Inst.~Henri Poincar{\'e}
  Phys.~Th{\'e}or., Vol.~44, No.~3, p.~263 - 292, 44, 263

\bibitem[{{Eastman} {et~al.}(2013){Eastman}, {Gaudi}, \& {Agol}}]{eastman13}
{Eastman}, J., {Gaudi}, B.~S., \& {Agol}, E. 2013, \pasp, 125, 83

\bibitem[{{Eastman} {et~al.}(2010){Eastman}, {Siverd}, \& {Gaudi}}]{eastman10}
{Eastman}, J., {Siverd}, R., \& {Gaudi}, B.~S. 2010, \pasp, 122, 935

\bibitem[{{Edwards} {et~al.}(2006){Edwards}, {Hobbs}, \&
  {Manchester}}]{edwards06}
{Edwards}, R.~T., {Hobbs}, G.~B., \& {Manchester}, R.~N. 2006, \mnras, 372,
  1549

\bibitem[{{Einstein}(1916)}]{einstein16}
{Einstein}, A. 1916, Annalen der Physik, 354, 769

\bibitem[{{Endl} {et~al.}(2001){Endl}, {K{\"u}rster}, {Els}, {Hatzes}, \&
  {Cochran}}]{endl01}
{Endl}, M., {K{\"u}rster}, M., {Els}, S., {Hatzes}, A.~P., \& {Cochran}, W.~D.
  2001, \aap, 374, 675

\bibitem[{{ESA}(1997)}]{esa97}
{ESA}, ed. 1997, ESA Special Publication, Vol. 1200, {The HIPPARCOS and TYCHO
  catalogues. Astrometric and photometric star catalogues derived from the ESA
  HIPPARCOS Space Astrometry Mission}

\bibitem[{{Event Horizon Telescope Collaboration} {et~al.}(2019){Event Horizon
  Telescope Collaboration}, {Akiyama}, {Alberdi}, {Alef}, {Asada}, {Azulay},
  {Baczko}, {Ball}, {Balokovi{\'c}}, {Barrett}, \& et~al.}]{eht1}
{Event Horizon Telescope Collaboration}, {Akiyama}, K., {Alberdi}, A., {et~al.}
  2019, \apjl, 875, L1

\bibitem[{{Fabian} {et~al.}(2000){Fabian}, {Iwasawa}, {Reynolds}, \&
  {Young}}]{fabian00}
{Fabian}, A.~C., {Iwasawa}, K., {Reynolds}, C.~S., \& {Young}, A.~J. 2000,
  \pasp, 112, 1145

\bibitem[{{Fairhead} \& {Bretagnon}(1990)}]{fairhead90}
{Fairhead}, L., \& {Bretagnon}, P. 1990, \aap, 229, 240

\bibitem[{{Feng} {et~al.}(2017){Feng}, {Tuomi}, \& {Jones}}]{feng17a}
{Feng}, F., {Tuomi}, M., \& {Jones}, H.~R.~A. 2017, \mnras, 470, 4794

\bibitem[{Folkner {et~al.}(2016)Folkner, Park, \& Jacobson}]{folkner16}
Folkner, W.~M., Park, R.~S., \& Jacobson, R.~A. 2016, JPL IOM 392R-16-003

\bibitem[{Folkner {et~al.}(2008)Folkner, Williams, \& Boggs}]{folkner08}
Folkner, W.~M., Williams, J.~G., \& Boggs, D.~H. 2008, JPL IOM 343R-08-003

\bibitem[{{Folkner} {et~al.}(2014){Folkner}, {Williams}, {Boggs}, {Park}, \&
  {Kuchynka}}]{folker14}
{Folkner}, W.~M., {Williams}, J.~G., {Boggs}, D.~H., {Park}, R.~S., \&
  {Kuchynka}, P. 2014, Interplanetary Network Progress Report, 196, 1

\bibitem[{{Fomalont} {et~al.}(2009){Fomalont}, {Kopeikin}, {Titov}, \&
  {Honma}}]{fomalont09}
{Fomalont}, E.~B., {Kopeikin}, S., {Titov}, O., \& {Honma}, M. 2009, in IAU
  Symposium \#261, American Astronomical Society, Vol.~41, 890

\bibitem[{{Fomalont} \& {Kopeikin}(2003)}]{fomalont03}
{Fomalont}, E.~B., \& {Kopeikin}, S.~M. 2003, \apj, 598, 704

\bibitem[{{Gaia Collaboration} {et~al.}(2018){Gaia Collaboration}, {Brown},
  {Vallenari}, {Prusti}, {de Bruijne}, {Babusiaux}, {Bailer-Jones}, {Biermann},
  {Evans}, {Eyer}, \& et~al.}]{brown18}
{Gaia Collaboration}, {Brown}, A.~G.~A., {Vallenari}, A., {et~al.} 2018, \aap,
  616, A1

\bibitem[{Gillon {et~al.}(2016)Gillon, Jehin, Lederer, Delrez, de~Wit,
  Burdanov, Van~Grootel, Burgasser, Triaud, Opitom, {et~al.}}]{gillon16}
Gillon, M., Jehin, E., Lederer, S.~M., {et~al.} 2016, Nature, 533, 221

\bibitem[{{Gonz{\'a}lez Hern{\'a}ndez} {et~al.}(2017){Gonz{\'a}lez
  Hern{\'a}ndez}, {Pepe}, {Molaro}, \& {Santos}}]{hernandez17}
{Gonz{\'a}lez Hern{\'a}ndez}, J.~I., {Pepe}, F., {Molaro}, P., \& {Santos}, N.
  2017, ArXiv e-prints, arXiv:1711.05250

\bibitem[{{Gravity Collaboration} {et~al.}(2018){Gravity Collaboration},
  {Abuter}, {Amorim}, {Anugu}, {Baub{\"o}ck}, {Benisty}, {Berger}, {Blind},
  {Bonnet}, {Brandner}, {Buron}, {Collin}, {Chapron}, {Cl{\'e}net}, {Coud{\'e}
  Du Foresto}, {de Zeeuw}, {Deen}, {Delplancke-Str{\"o}bele}, {Dembet},
  {Dexter}, {Duvert}, {Eckart}, {Eisenhauer}, {Finger}, {F{\"o}rster
  Schreiber}, {F{\'e}dou}, {Garcia}, {Garcia Lopez}, {Gao}, {Gendron},
  {Genzel}, {Gillessen}, {Gordo}, {Habibi}, {Haubois}, {Haug}, {Hau{\ss}mann},
  {Henning}, {Hippler}, {Horrobin}, {Hubert}, {Hubin}, {Jimenez Rosales},
  {Jochum}, {Jocou}, {Kaufer}, {Kellner}, {Kendrew}, {Kervella}, {Kok},
  {Kulas}, {Lacour}, {Lapeyr{\`e}re}, {Lazareff}, {Le Bouquin}, {L{\'e}na},
  {Lippa}, {Lenzen}, {M{\'e}rand}, {M{\"u}ler}, {Neumann}, {Ott}, {Palanca},
  {Paumard}, {Pasquini}, {Perraut}, {Perrin}, {Pfuhl}, {Plewa}, {Rabien},
  {Ram{\'\i}rez}, {Ramos}, {Rau}, {Rodr{\'\i}guez-Coira}, {Rohloff}, {Rousset},
  {Sanchez-Bermudez}, {Scheithauer}, {Sch{\"o}ller}, {Schuler}, {Spyromilio},
  {Straub}, {Straubmeier}, {Sturm}, {Tacconi}, {Tristram}, {Vincent}, {von
  Fellenberg}, {Wank}, {Waisberg}, {Widmann}, {Wieprecht}, {Wiest},
  {Wiezorrek}, {Woillez}, {Yazici}, {Ziegler}, \& {Zins}}]{gravity18}
{Gravity Collaboration}, {Abuter}, R., {Amorim}, A., {et~al.} 2018, \aap, 615,
  L15

\bibitem[{{Greenstein} {et~al.}(1971){Greenstein}, {Oke}, \&
  {Shipman}}]{greenstein71}
{Greenstein}, J.~L., {Oke}, J.~B., \& {Shipman}, H.~L. 1971, \apj, 169, 563

\bibitem[{{Grimm} {et~al.}(2018){Grimm}, {Demory}, {Gillon}, {Dorn}, {Agol},
  {Burdanov}, {Delrez}, {Sestovic}, {Triaud}, {Turbet}, {Bolmont}, {Caldas},
  {Wit}, {Jehin}, {Leconte}, {Raymond}, {Grootel}, {Burgasser}, {Carey},
  {Fabrycky}, {Heng}, {Hernandez}, {Ingalls}, {Lederer}, {Selsis}, \&
  {Queloz}}]{grimm18}
{Grimm}, S.~L., {Demory}, B.-O., {Gillon}, M., {et~al.} 2018, \aap, 613, A68

\bibitem[{Gubler \& Tytler(1998)}]{gubler98}
Gubler, J., \& Tytler, D. 1998, Publications of the Astronomical Society of the
  Pacific, 110, 738

\bibitem[{{Hobbs} {et~al.}(2006){Hobbs}, {Edwards}, \& {Manchester}}]{hobbs06}
{Hobbs}, G.~B., {Edwards}, R.~T., \& {Manchester}, R.~N. 2006, \mnras, 369, 655

\bibitem[{Hohenkerk \& Sinclair(1985)}]{hohenkerk85}
Hohenkerk, C., \& Sinclair, A. 1985, The computation of angular atmospheric
  refraction at large zenith angles, NAO Tech, Tech. rep., Note 63, HM Nautical
  Almanac Office, Royal Greenwhich Observatory, Greenwich

\bibitem[{{Holczer} {et~al.}(2016){Holczer}, {Mazeh}, {Nachmani},
  {Jontof-Hutter}, {Ford}, {Fabrycky}, {Ragozzine}, {Kane}, \&
  {Steffen}}]{holczer16}
{Holczer}, T., {Mazeh}, T., {Nachmani}, G., {et~al.} 2016, \apjs, 225, 9

\bibitem[{{Holman} \& {Murray}(2005)}]{holman05}
{Holman}, M.~J., \& {Murray}, N.~W. 2005, Science, 307, 1288

\bibitem[{{Hulse} \& {Taylor}(1975)}]{hulse75}
{Hulse}, R.~A., \& {Taylor}, J.~H. 1975, \apjl, 195, L51

\bibitem[{{Ioannidis} {et~al.}(2014){Ioannidis}, {Schmitt}, {Avdellidou}, {von
  Essen}, \& {Agol}}]{ioannidis14}
{Ioannidis}, P., {Schmitt}, J.~H.~M.~M., {Avdellidou}, C., {von Essen}, C., \&
  {Agol}, E. 2014, \aap, 564, A33

\bibitem[{{Irwin} \& {Fukushima}(1999)}]{irwin99}
{Irwin}, A.~W., \& {Fukushima}, T. 1999, \aap, 348, 642

\bibitem[{{Ivanov} {et~al.}(2011){Ivanov}, {C{\'a}ceres}, {Minniti}, {Selman},
  {Melo}, {Naef}, {Mason}, \& {Pietrzynski}}]{ivanov11}
{Ivanov}, V.~D., {C{\'a}ceres}, C., {Minniti}, D., {et~al.} 2011, in European
  Physical Journal Web of Conferences, Vol.~11, European Physical Journal Web
  of Conferences, 05008

\bibitem[{{Johns-Krull} {et~al.}(2008){Johns-Krull}, {McCullough}, {Burke},
  {Valenti}, {Janes}, {Heasley}, {Prato}, {Bissinger}, {Fleenor}, \&
  {Foote}}]{johns-krull08}
{Johns-Krull}, C.~M., {McCullough}, P.~R., {Burke}, C.~J., {et~al.} 2008, \apj,
  677, 657

\bibitem[{{Jord{\'a}n} \& {Bakos}(2008)}]{jordan08}
{Jord{\'a}n}, A., \& {Bakos}, G.~{\'A}. 2008, \apj, 685, 543

\bibitem[{{Kervella} {et~al.}(2016){Kervella}, {Mignard}, {M{\'e}rand}, \&
  {Th{\'e}venin}}]{kervella16}
{Kervella}, P., {Mignard}, F., {M{\'e}rand}, A., \& {Th{\'e}venin}, F. 2016,
  \aap, 594, A107

\bibitem[{Kjeldsen {et~al.}(2005)Kjeldsen, Bedding, Butler,
  Christensen-Dalsgaard, Kiss, McCarthy, Marcy, Tinney, \& Wright}]{kjeldsen05}
Kjeldsen, H., Bedding, T.~R., Butler, R.~P., {et~al.} 2005, The Astrophysical
  Journal, 635, 1281

\bibitem[{Klioner(2003)}]{klioner03}
Klioner, S.~A. 2003, The Astronomical Journal, 125, 1580

\bibitem[{{Klioner} \& {Kopeikin}(1992)}]{klioner92}
{Klioner}, S.~A., \& {Kopeikin}, S.~M. 1992, \aj, 104, 897

\bibitem[{{Klioner} {et~al.}(2010){Klioner}, {Capitaine}, {Folkner}, {Guinot},
  {Huang}, {Kopeikin}, {Pitjeva}, {Seidelmann}, \& {Soffel}}]{klioner10}
{Klioner}, S.~A., {Capitaine}, N., {Folkner}, W.~M., {et~al.} 2010, in IAU
  Symposium, Vol. 261, Relativity in Fundamental Astronomy: Dynamics, Reference
  Frames, and Data Analysis, ed. S.~A. {Klioner}, P.~K. {Seidelmann}, \& M.~H.
  {Soffel}, 79--84

\bibitem[{{Kopeikin}(1996)}]{kopeikin96}
{Kopeikin}, S.~M. 1996, \apjl, 467, L93

\bibitem[{{Kopeikin} \& {Makarov}(2006)}]{kopeikin06}
{Kopeikin}, S.~M., \& {Makarov}, V.~V. 2006, \aj, 131, 1471

\bibitem[{{Kopeikin} \& {Makarov}(2007)}]{kopeikin07}
---. 2007, \prd, 75, 062002

\bibitem[{{Kopeikin} \& {Ozernoy}(1999)}]{kopeikin99b}
{Kopeikin}, S.~M., \& {Ozernoy}, L.~M. 1999, \apj, 523, 771

\bibitem[{{Kopeikin} \& {Sch{\"a}fer}(1999)}]{kopeikin99}
{Kopeikin}, S.~M., \& {Sch{\"a}fer}, G. 1999, \prd, 60, 124002

\bibitem[{{Lelli} {et~al.}(2017){Lelli}, {McGaugh}, {Schombert}, \&
  {Pawlowski}}]{lelli17}
{Lelli}, F., {McGaugh}, S.~S., {Schombert}, J.~M., \& {Pawlowski}, M.~S. 2017,
  \apj, 836, 152

\bibitem[{{Lindegren} \& {Dravins}(2003)}]{lindegren03}
{Lindegren}, L., \& {Dravins}, D. 2003, \aap, 401, 1185

\bibitem[{{Lindegren} {et~al.}(2012){Lindegren}, {Lammers}, {Hobbs},
  {O'Mullane}, {Bastian}, \& {Hern{\'a}ndez}}]{lindegren11}
{Lindegren}, L., {Lammers}, U., {Hobbs}, D., {et~al.} 2012, \aap, 538, A78

\bibitem[{{Lindegren} {et~al.}(2018){Lindegren}, {Hern{\'a}ndez}, {Bombrun},
  {Klioner}, {Bastian}, {Ramos-Lerate}, {de Torres}, {Steidelm{\"u}ller},
  {Stephenson}, {Hobbs}, {Lammers}, {Biermann}, {Geyer}, {Hilger}, {Michalik},
  {Stampa}, {McMillan}, {Casta{\~n}eda}, {Clotet}, {Comoretto}, {Davidson},
  {Fabricius}, {Gracia}, {Hambly}, {Hutton}, {Mora}, {Portell}, {van Leeuwen},
  {Abbas}, {Abreu}, {Altmann}, {Andrei}, {Anglada}, {Balaguer-N{\'u}{\~n}ez},
  {Barache}, {Becciani}, {Bertone}, {Bianchi}, {Bouquillon}, {Bourda},
  {Br{\"u}semeister}, {Bucciarelli}, {Busonero}, {Buzzi}, {Cancelliere},
  {Carlucci}, {Charlot}, {Cheek}, {Crosta}, {Crowley}, {de Bruijne}, {de
  Felice}, {Drimmel}, {Esquej}, {Fienga}, {Fraile}, {Gai}, {Garralda},
  {Gonz{\'a}lez-Vidal}, {Guerra}, {Hauser}, {Hofmann}, {Holl}, {Jordan},
  {Lattanzi}, {Lenhardt}, {Liao}, {Licata}, {Lister}, {L{\"o}ffler},
  {Marchant}, {Martin-Fleitas}, {Messineo}, {Mignard}, {Morbidelli}, {Poggio},
  {Riva}, {Rowell}, {Salguero}, {Sarasso}, {Sciacca}, {Siddiqui}, {Smart},
  {Spagna}, {Steele}, {Taris}, {Torra}, {van Elteren}, {van Reeven}, \&
  {Vecchiato}}]{lindegren18}
{Lindegren}, L., {Hern{\'a}ndez}, J., {Bombrun}, A., {et~al.} 2018, \aap, 616,
  A2

\bibitem[{{Lisogorskyi} {et~al.}(2019){Lisogorskyi}, {Jones}, \&
  {Feng}}]{lisogorskyi19}
{Lisogorskyi}, M., {Jones}, H.~R.~A., \& {Feng}, F. 2019, \mnras, 485, 4804

\bibitem[{{Malkov} {et~al.}(2012){Malkov}, {Tamazian}, {Docobo}, \&
  {Chulkov}}]{malkov12}
{Malkov}, O.~Y., {Tamazian}, V.~S., {Docobo}, J.~A., \& {Chulkov}, D.~A. 2012,
  \aap, 546, A69

\bibitem[{{Mamajek}(2012)}]{mamajek12}
{Mamajek}, E.~E. 2012, arXiv e-prints, arXiv:1210.1616

\bibitem[{{Mangum} \& {Wallace}(2015)}]{mangum15}
{Mangum}, J.~G., \& {Wallace}, P. 2015, \pasp, 127, 74

\bibitem[{McCarthy \& Luzum(2003)}]{mccarthy03}
McCarthy, D.~D., \& Luzum, B.~J. 2003, Celestial Mechanics and Dynamical
  Astronomy, 85, 37

\bibitem[{McCarthy \& Petit(2004)}]{mccarthy04}
McCarthy, D.~D., \& Petit, G. 2004, IERS conventions (2003), Tech. rep.,
  International Earth Rotation And Reference Systems Service (IERS)(Germany)

\bibitem[{{Mignard} \& {Klioner}(2010)}]{mignard10}
{Mignard}, F., \& {Klioner}, S.~A. 2010, in IAU Symposium, Vol. 261, Relativity
  in Fundamental Astronomy: Dynamics, Reference Frames, and Data Analysis, ed.
  S.~A. {Klioner}, P.~K. {Seidelmann}, \& M.~H. {Soffel}, 306--314

\bibitem[{{Milgrom}(1983)}]{milgrom83}
{Milgrom}, M. 1983, \apj, 270, 365

\bibitem[{{Miralda-Escud{\'e}}(2002)}]{miralda02}
{Miralda-Escud{\'e}}, J. 2002, \apj, 564, 1019

\bibitem[{Misner {et~al.}(1973)Misner, Thorne, Wheeler, \& Kaiser}]{misner73}
Misner, C.~W., Thorne, K.~S., Wheeler, J.~A., \& Kaiser, D.~I. 1973,
  Gravitation (Princeton University Press)

\bibitem[{{Moe} {et~al.}(2019){Moe}, {Kratter}, \& {Badenes}}]{moe18}
{Moe}, M., {Kratter}, K.~M., \& {Badenes}, C. 2019, \apj, 875, 61

\bibitem[{{Murray} \& {Correia}(2010)}]{murray10}
{Murray}, C.~D., \& {Correia}, A.~C.~M. 2010, {Keplerian Orbits and Dynamics of
  Exoplanets}, ed. S.~{Seager}, 15--23

\bibitem[{{Niell}(1996)}]{niell96}
{Niell}, A.~E. 1996, \jgr, 101, 3227

\bibitem[{Nilsson {et~al.}(2013)Nilsson, B{\"o}hm, Wijaya, Tresch, Nafisi, \&
  Schuh}]{nilsson13}
Nilsson, T., B{\"o}hm, J., Wijaya, D.~D., {et~al.} 2013, in Atmospheric Effects
  in Space Geodesy (Springer), 73--136

\bibitem[{{Nordtvedt} \& {Will}(1972)}]{nordtvedt72}
{Nordtvedt}, Jr., K., \& {Will}, C.~M. 1972, \apj, 177, 775

\bibitem[{{Perryman} {et~al.}(2014){Perryman}, {Hartman}, {Bakos}, \&
  {Lindegren}}]{perryman14}
{Perryman}, M., {Hartman}, J., {Bakos}, G.~{\'A}., \& {Lindegren}, L. 2014,
  \apj, 797, 14

\bibitem[{{Petit}(2004)}]{petit04}
{Petit}, G. 2004, in Proceedings of the Journ{\'e}es 2003 ``Syst{\`e}mes
  der{\'e}f{\'e}rence spatio-temporels'': Astrometry, Geodynamics and Solar
  System Dynamics: from milliarcseconds to microarcseconds, held at IAA,
  St.Petersburg, Russia, 22-25 September 2003, A. Finkelstein \& N.Capitaine
  (eds.), ISBN 5-93197-019-3 \& ISBN 2-901057-50-0, p. 314-317, ed.
  A.~{Finkelstein} \& N.~{Capitaine}, 314--317

\bibitem[{Petit \& Luzum(2010{\natexlab{a}})}]{iers10}
Petit, G., \& Luzum, B. 2010{\natexlab{a}}, IERS conventions (2010), Tech.
  rep., BUREAU INTERNATIONAL DES POIDS ET MESURES SEVRES (FRANCE)

\bibitem[{Petit \& Luzum(2010{\natexlab{b}})}]{petit10}
---. 2010{\natexlab{b}}, IERS Techn. Note No. 36

\bibitem[{{Plavchan} {et~al.}(2018){Plavchan}, {Cale}, {Newman}, {Hamze},
  {Latouf}, {Matzko}, {Beichman}, {Ciardi}, {Purcell}, {Lightsey}, {Cegla},
  {Dumusque}, {Bourrier}, {Dressing}, {Gao}, {Vasisht}, {Leifer}, {Wang},
  {Gagne}, {Thompson}, {Crass}, {Bechter}, {Bechter}, {Blake}, {Halverson},
  {Mayo}, {Beatty}, {Wright}, {Wise}, {Tanner}, {Eastman}, {Quinn}, {Fischer},
  {Basu}, {Sanchez-Maes}, {Howard}, {Vahala}, {Wang}, {Diddams}, {Papp},
  {Pope}, {Martin}, \& {Murphy}}]{plavchan18}
{Plavchan}, P., {Cale}, B., {Newman}, P., {et~al.} 2018, arXiv e-prints,
  arXiv:1803.03960

\bibitem[{Pound \& Rebka~Jr(1959)}]{pound59}
Pound, R.~V., \& Rebka~Jr, G.~A. 1959, Physical Review Letters, 3, 439

\bibitem[{{Pourbaix} \& {Boffin}(2016)}]{pourbaix16}
{Pourbaix}, D., \& {Boffin}, H.~M.~J. 2016, \aap, 586, A90

\bibitem[{{Pourbaix} {et~al.}(1999){Pourbaix}, {Neuforge-Verheecke}, \&
  {Noels}}]{pourbaix99}
{Pourbaix}, D., {Neuforge-Verheecke}, C., \& {Noels}, A. 1999, \aap, 344, 172

\bibitem[{{Pourbaix} {et~al.}(2002){Pourbaix}, {Nidever}, {McCarthy}, {Butler},
  {Tinney}, {Marcy}, {Jones}, {Penny}, {Carter}, {Bouchy}, {Pepe}, {Hearnshaw},
  {Skuljan}, {Ramm}, \& {Kent}}]{pourbaix02}
{Pourbaix}, D., {Nidever}, D., {McCarthy}, C., {et~al.} 2002, \aap, 386, 280

\bibitem[{{Psaltis}(2004)}]{psaltis04}
{Psaltis}, D. 2004, in American Institute of Physics Conference Series, Vol.
  714, X-ray Timing 2003: Rossi and Beyond, ed. P.~{Kaaret}, F.~K. {Lamb}, \&
  J.~H. {Swank}, 29--35

\bibitem[{Ricker {et~al.}(2014)Ricker, Winn, Vanderspek, Latham, Bakos, Bean,
  Berta-Thompson, Brown, Buchhave, Butler, {et~al.}}]{ricker14}
Ricker, G.~R., Winn, J.~N., Vanderspek, R., {et~al.} 2014, Journal of
  Astronomical Telescopes, Instruments, and Systems, 1, 014003

\bibitem[{{Rickman}(2001)}]{rickman01}
{Rickman}, H. 2001, Transactions of the International Astronomical Union,
  Series B, 24

\bibitem[{R{\"u}eger(2002)}]{rueger02}
R{\"u}eger, J. 2002, Unisurv Report S-68, School of Surveying and Spatial
  Information Systems, University of New South Wales, UNSW SYDNEY NSW, 2052, 1

\bibitem[{{Sana} \& {Evans}(2011)}]{sana11}
{Sana}, H., \& {Evans}, C.~J. 2011, in IAU Symposium, Vol. 272, Active OB
  Stars: Structure, Evolution, Mass Loss, and Critical Limits, ed. C.~{Neiner},
  G.~{Wade}, G.~{Meynet}, \& G.~{Peters}, 474--485

\bibitem[{{Shapiro}(1964)}]{shapiro64}
{Shapiro}, I.~I. 1964, Physical Review Letters, 13, 789

\bibitem[{Standish(2006)}]{standish06}
Standish, E. 2006, JPL IOM 343R-06-002

\bibitem[{Standish(1998)}]{standish98}
Standish, E.~M. 1998, F-98\_048

\bibitem[{{Tanaka} {et~al.}(1995){Tanaka}, {Nandra}, {Fabian}, {Inoue},
  {Otani}, {Dotani}, {Hayashida}, {Iwasawa}, {Kii}, {Kunieda}, {Makino}, \&
  {Matsuoka}}]{tanaka95}
{Tanaka}, Y., {Nandra}, K., {Fabian}, A.~C., {et~al.} 1995, \nat, 375, 659

\bibitem[{{Taylor} \& {Weisberg}(1982)}]{taylor82}
{Taylor}, J.~H., \& {Weisberg}, J.~M. 1982, \apj, 253, 908

\bibitem[{{Taylor} \& {Weisberg}(1989)}]{taylor89}
---. 1989, \apj, 345, 434

\bibitem[{{Taylor} {et~al.}(1992){Taylor}, {Wolszczan}, {Damour}, \&
  {Weisberg}}]{taylor92}
{Taylor}, J.~H., {Wolszczan}, A., {Damour}, T., \& {Weisberg}, J.~M. 1992,
  \nat, 355, 132

\bibitem[{{Thiele}(1883)}]{thiele83}
{Thiele}, T.~N. 1883, Astronomische Nachrichten, 104, 245

\bibitem[{{Unwin} {et~al.}(2008){Unwin}, {Shao}, {Tanner}, {Allen}, {Beichman},
  {Boboltz}, {Catanzarite}, {Chaboyer}, {Ciardi}, {Edberg}, {Fey}, {Fischer},
  {Gelino}, {Gould}, {Grillmair}, {Henry}, {Johnston}, {Johnston}, {Jones},
  {Kulkarni}, {Law}, {Majewski}, {Makarov}, {Marcy}, {Meier}, {Olling}, {Pan},
  {Patterson}, {Pitesky}, {Quirrenbach}, {Shaklan}, {Shaya}, {Strigari},
  {Tomsick}, {Wehrle}, \& {Worthey}}]{unwin08}
{Unwin}, S.~C., {Shao}, M., {Tanner}, A.~M., {et~al.} 2008, \pasp, 120, 38

\bibitem[{Viswanathan {et~al.}(2017)Viswanathan, Fienga, Gastineau, \&
  Laskar}]{viswanathan17}
Viswanathan, V., Fienga, A., Gastineau, M., \& Laskar, J. 2017, Notes
  Scientifiques et Techniques de l’Institut de M{\'e}canique C{\'e}leste, 108

\bibitem[{{Wallace} \& {Capitaine}(2006)}]{wallace06}
{Wallace}, P.~T., \& {Capitaine}, N. 2006, \aap, 459, 981

\bibitem[{Wang {et~al.}(2017)Wang, Coles, Hobbs, Shannon, Manchester, Kerr,
  Yuan, Wang, Bailes, Bhat, {et~al.}}]{wang17}
Wang, J., Coles, W., Hobbs, G., {et~al.} 2017, Monthly Notices of the Royal
  Astronomical Society, 469, 425

\bibitem[{{Weisberg} \& {Taylor}(2005)}]{weisberg05}
{Weisberg}, J.~M., \& {Taylor}, J.~H. 2005, in Astronomical Society of the
  Pacific Conference Series, Vol. 328, Binary Radio Pulsars, ed. F.~A. {Rasio}
  \& I.~H. {Stairs}, 25

\bibitem[{{Wex}(2014)}]{wex14}
{Wex}, N. 2014, arXiv e-prints, arXiv:1402.5594

\bibitem[{Whitmore \& Murphy(2014)}]{whitmore14}
Whitmore, J.~B., \& Murphy, M.~T. 2014, Monthly Notices of the Royal
  Astronomical Society, 447, 446

\bibitem[{{Winn} {et~al.}(2008){Winn}, {Holman}, {Torres}, {McCullough},
  {Johns-Krull}, {Latham}, {Shporer}, {Mazeh}, {Garcia-Melendo}, \&
  {Foote}}]{winn08}
{Winn}, J.~N., {Holman}, M.~J., {Torres}, G., {et~al.} 2008, \apj, 683, 1076

\bibitem[{{Wright} \& {Eastman}(2014)}]{wright14}
{Wright}, J.~T., \& {Eastman}, J.~D. 2014, \pasp, 126, 838

\bibitem[{{Zechmeister} {et~al.}(2013){Zechmeister}, {K{\"u}rster}, {Endl}, {Lo
  Curto}, {Hartman}, {Nilsson}, {Henning}, {Hatzes}, \&
  {Cochran}}]{zechmeister13}
{Zechmeister}, M., {K{\"u}rster}, M., {Endl}, M., {et~al.} 2013, \aap, 552, A78

\bibitem[{{Zhao} {et~al.}(2018){Zhao}, {Fischer}, {Brewer}, {Giguere}, \&
  {Rojas-Ayala}}]{zhao18}
{Zhao}, L., {Fischer}, D.~A., {Brewer}, J., {Giguere}, M., \& {Rojas-Ayala}, B.
  2018, \aj, 155, 24

\end{thebibliography}
% \end{CJK*}
\end{document}